\documentclass[reprint,amsmath,amssymb,aps,prb]{revtex4-2}

\usepackage{hyperref} 
\usepackage{amssymb,amsbsy,amsmath,mathrsfs}
\usepackage{epsfig,psfrag,subfigure}
\usepackage{graphicx,float}
\usepackage{times}
\usepackage{color}
\usepackage{subfigure}
\usepackage{setspace}
\usepackage{bm} % bold math.
\usepackage{xr}
\usepackage{xcolor}
\usepackage{etoolbox}

\usepackage{stackengine}
\usepackage{natbib,url,hyperref}
\hypersetup{colorlinks,allcolors=blue}
\usepackage{filecontents}
\usepackage{comment}
\usepackage{hyperref}
\usepackage{tikz}
\usepackage{tikz}
\usepackage{tikz}
\newcommand*\circled[1]{\tikz[baseline=(char.base)]{
            \node[shape=circle,draw,inner sep=2pt] (char) {#1};}}
\newcommand\bea{\begin{eqnarray}}
\newcommand\eea{\end{eqnarray}}
\newcommand\beq{\begin{equation}}
\newcommand\eeq{\end{equation}}

\newcommand{\noi}{\noindent}
\newcommand{\non}{\nonumber}

\newcommand{\la}{\langle}
\newcommand{\ra}{\rangle}

\def\blue{\textcolor{blue}}
\def\red{\textcolor{red}}

\newcommand{\ket}[1]{|#1\rangle}

\begin{document}
\title{Diagnostics of Hilbert space fragmentation, freezing transition, and its effects in the family of quantum East models involving varying range of constraints}
%Absence of thermalization in a correlated-hopping model 
%due to strong Hilbert space fragmentation and its characterization by irreducible strings \\

\author{Sreemayee Aditya$^{1,2}$}
\affiliation{$^1$Center for High Energy Physics, Indian Institute of Science, Bengaluru 560012, India\\
$^2$ Institute for Theoretical Physics, Cluster of Excellence ML4Q, 
University of Cologne,
Z\"ulpicher Str. 77a , 50937, Cologne, Germany}
\begin{abstract}
This paper explores the effect of the strong-to-weak fragmentation transition, namely the freezing transition, and
its rich characteristics in a family of one-dimensional spinless fermionic models with facilitated hoppings involving short-to-long-range East constraints. Focusing on this family of models with range-$q$ constraints, our
investigation furnishes an exhaustive understanding of the fragmented Hilbert space utilizing the enumerative
combinatorics and transfer matrix methods. This further allows us to get insight into the freezing transition in
this family of models with the help of the generalization of Catalan numbers introduced by Frey and Sellers for $q>1$, further revealing that increasing the range of constraints drives the transition to transpire at lower filling fractions as $n_c = 1/(q+1)$. This
distinct fragmentation structure also yields the emergence of ground states at multiple fillings, which can be explained from the root structure of the largest fragment within the full Hilbert space. Further, the ground state exhibits signatures of criticality with logarithmic scaling of entanglement entropy. Thereafter, our investigation exemplifies that the above transition has a profound impact on the
thermalization of bulk and boundary autocorrelators at long times, which includes intricate filling-dependent inhomogeneous long-time autocorrelation profiles across the chain in open boundary conditions (OBCs). Finally, we probe the effect of the strong-to-weak fragmentation transition on the transport at intermediate times in periodic boundary conditions (PBCs), restricting ourselves to models up to range-3 constraints. This investigation discloses a vast range of anomalous transport possibilities, ranging from size-stretched exponential relaxation through
superdiffusive to subdiffusive behaviors, akin to the fragmentation structure supported by the filling fraction and range of constraints. In brevity, our paper reveals intriguing possibilities conspired by constraints with varying ranges, and those-induced Hilbert space fragmentation and freezing transition, thus offering exciting avenues
for future explorations.
\end{abstract}

\maketitle
\section{Introduction}
The quest for unraveling intriguing non-equilibrium quantum many-body dynamics has attracted a lot of interest in recent years as it can offer tantalizing opportunities for uncovering rich phases of matter with no equilibrium counterpart. In this direction, how quantum many-body systems evade thermalization~\cite{sredniki_1994,rigol_2007,rigol_2008} has always been a central question of investigation due to its close connection to quantum information processing~\cite{Preskill2018quantumcomputingin,Google_SC1}. Further, recent advancements in various experimental platforms~\cite{trapped_ion_2012,Rydberg_2020,optical_atoms_2017} have enabled us to make significant progress in this area over the past decade.

Current developments in kinetically constrained quantum systems have paved pathways to uncover novel ergodicity-breaking mechanisms that go beyond the well-known mechanisms for breakdown of thermalization, such as many-body localization~\cite{vosk_2013,pal_2010,mbl_review} and quantum integrability~\cite{bethe1,bethe2,bethe3,inte_rev1, int_rev2}. These systems circumvent thermalization due to various mechanisms mediated by kinetic hindrance, among which Hilbert space fragmentation (HSF)~\cite{moudgalya_memorial_2021,Moudgalya_review_2022} is one of the recent developments; further, it is also the primary focus of this paper. Similar mechanisms have been exhaustively studied in the context of classical glasses~\cite{cg1,cg2,cg3}, which have lately regained considerable attention in kinetically-constrained quantum many-body systems. HSF fractures the full Hilbert space into exponentially many fragments within a conventional symmetry-resolved sector, thus offering a wide range of possibilities, ranging from the Krylov-restricted thermalization~\cite{Moudgalya_review_2022,Moudgalya_pairhopping_2020,Deepak_HSF,tracer} through statistical edge localization~\cite{sala_ergo_2020} to anomalous glassy transport properties~\cite{balasubramanian2024glassy,singh_2021,brighi_2023,Lenart_hetero,pal_hydro}. Furthermore, this mechanism has been studied in a broad range of static models ~\cite{lenart_folded_XXZ,
Moudgalya_review_2022,Deepak_HSF,two_site_East_model,Moudgalya_pairhopping_2020,khemani_2020,
mukherjee_2021}. In addition, the signature of prethermal HSF has recently been witnessed in periodically driven quantum systems~\cite{ghosh_2023,aditya_2023}.

HSF~\cite{moudgalya_memorial_2021,Moudgalya_review_2022} can primarily be classified into two broad classes depending on the dimensional growth of the largest fragment, $D_{frag}$, with respect to the growth of the conventional symmetry-resolved sector, $D_{sum}$. If the growth of $D_{frag}$ is exponentially smaller than $D_{sum}$ in $L\to\infty$, it is dubbed strong HSF~\cite{Moudgalya_pairhopping_2020,Moudgalya_review_2022,Deepak_HSF}, while in the case of weak HSF, $D_{frag}$ approaches $D_{sum}$ in the thermodynamic limit~\cite{East_Sreemayee,Morningstar_2020,wang_2023,brian_thermal}. Although several models exhibiting strongly fragmented Hilbert space~\cite{Moudgalya_pairhopping_2020,Moudgalya_review_2022,Deepak_HSF} have been comprehensively examined before, there are very few instances where models with weak HSF~\cite{Morningstar_2020,East_Sreemayee,wang_2023,brian_thermal} have been investigated until now. In this context, one fascinating problem involves constraint models offering filling dependent dynamical phase transitions between strongly and weakly fragmented Hilbert spaces, called freezing transition, which has been hardly explored except for a few specific cases concerning short-range constraints~\cite{Morningstar_2020,wang_2023,East_Sreemayee,brian_thermal}. 
%Nevertheless, it has been hardly investigated except for a few specific cases~\cite{Morningstar_2020,wang_2023,East_Sreemayee,brian_thermal}; furthermore, most of the studies in this context have also been limited to models including short-range constraints (limited to three to four-site terms).
Therefore, how the increasing range of constraints affects the freezing transition demands a thorough investigation. Focusing on a family of models with a varying range of kinetic constraints,  comprehending its fragmentation-induced freezing transition~~\cite{Morningstar_2020,wang_2023,East_Sreemayee,brian_thermal,abhisodh2024}, and its other impacts thus sets the primary objective of this current paper.

In this paper, we examine a family of one-dimensional facilitated particle-number-conserving quantum East models of spinless fermions with range-$q$ constraints~\cite{brighi_2023}, which breaks the inversion symmetry. The simplest variant of this family has been exhaustively investigated in one of the very recent studies~\cite{East_Sreemayee}, where the author of the present paper is also a co-author. In this previous study, we have shown that this simple variant demonstrates a freezing transition at half-filling~\cite{East_Sreemayee}, which can be captured with the help of Dyck combinatorics~\cite{catalan}. Through this present analysis, we thus scrutinize the fragmentation structure of the full family of East models with range-$q$ constraints utilizing the enumerative combinatorics~\cite{generating1994} and transfer matrix methods~\cite{Deepak_HSF,Dhar_1993,HariMenon_1995}. Further, our investigation indicates that all the variants exhibit a similar strong-to-weak fragmentation transition~\cite{East_Sreemayee,wang_2023,Morningstar_2020} at a critical filling, $n_{c}=1/(q+1)$. Moreover, this can be exactly captured using the generalization of the Catalan numbers introduced by Frey and Sellers for $q\geq2$~\cite{FS_article}. Our result thus suggests that increasing the range of constraints allows the transition to occur at a lower filling fraction. In this context, we want to emphasize that an analytical understanding of the fragmentation structure~\cite{moudgalya_PRX_2022,sala_ergo_2020,pozsgay_2021,pozsgay_2023,Dhar_1993,zadnik_hydro_2021} is, in principle, an involved task as the fragments can not be labeled by quantum numbers of local and quasilocal symmetries, thus beseeching the concept of unconventional symmetries~\cite{moudgalya_PRX_2022,Deepak_HSF,sala_ergo_2020,pozsgay_2023, Dhar_1993,Barma_1994,menon_1997}. This complexity grows further with an increasing range of constraints. However, our study, for the very first time, facilitates a comprehensive analytical understanding of a class of models with a varying range of constraints, further allowing the first-ever appearance of the generalized Catalan family~\cite{FS_article,Asinowski_2022} in a spinless fermionic model. This opens up further possibilities for future exploration, looking at the possibilities offered by the widely studied Motzkin~\cite{Movassagh_2017,motzkin2} and Fredkin spin chains~\cite{khagebdra2_fredkin,salberger2016fredkinspinchain,Garahhan_fredkin} in the literature (albeit the fact that the model under consideration is not frustration-free unlike the Fredkin and Motzkin spin chains).

 Next, we turn to examine how the fragmentation structure~\cite{Moudgalya_review_2022} affects ground state properties~\cite{two_site_East_model,khagebdra2_fredkin,salberger2016fredkinspinchain,Movassagh_2017} (the largest fragment where typically lies the ground state) in this class of models. Our investigation shows that although the filling fraction for ground state in the range-1 case~\cite{East_Sreemayee} shifts from half-filling as $L/2+a$, $a=\sqrt{L}/2$ in $L\to\infty$ as a consequence of Dyck combinatorics, the generalized Catalan family~\cite{FS_article} in range-$q$ models ($q>1$) does not support such shift for large enough system sizes. They rather host multiple ground states at different fillings (precisely at $q$ number of fillings) with a total degeneracy of $2^{q-1}$ in the full Hilbert space in open boundary conditions (OBCs). Also, we reveal that this feature is a byproduct of the root structure of the disjoint sectors and the dimensional growth of fragments in terms of the generalized Catalan sequences~\cite{FS_article}. Further, the ground states in this family of models are found to be critical with entanglement entropy following logarithmic scaling with system size~\cite{Sachdev_2011,PasqualeCalabrese_2004}. Utilizing the Frey-Sellers sequence~\cite{FS_article}, we also show that the minimum filling for the largest fragment and the ground state for the range-$q$ constraint appears at $N_{f}\simeq (L-q)/2$ in OBCs, thus implying that increasing the range entitles the ground state to occur at a lower filling. 

 Next, we investigate how freezing transition~\cite{East_Sreemayee,wang_2023,Morningstar_2020} controls the long-time saturation value of bulk and boundary autocorrelators starting from a typical random initial state in OBCs. It has been reported that the presence of non-local conserved quantities in fragmented systems gives rise to an inhomogeneous profile of long-time autocorrelators~\cite{Moudgalya_pairhopping_2020,sala_ergo_2020,Deepak_HSF} across the chain, where boundaries harbor a strong signature of the breakdown of thermalization in the presence of weakly thermalizing bulk~\cite{sala_ergo_2020}. A similar examination, in this case, shows that the freezing transition~\cite{East_Sreemayee,Morningstar_2020,wang_2023} and unavailability of inversion symmetry further enrich this behavior. We observe that non-thermal active boundaries and bulk autocorrelators on the strongly fragmented side switch to the non-thermal leftmost active boundary and both thermal bulk and rightmost boundary on the side of weak fragmentation. 
 %Moreover, this behavior has been validated with the help of Mazur-Suzuki inequality~\cite{MAZUR1969533,SUZUKI1971277}. 

\begin{figure*}
\fbox{\includegraphics[width=0.8\hsize]{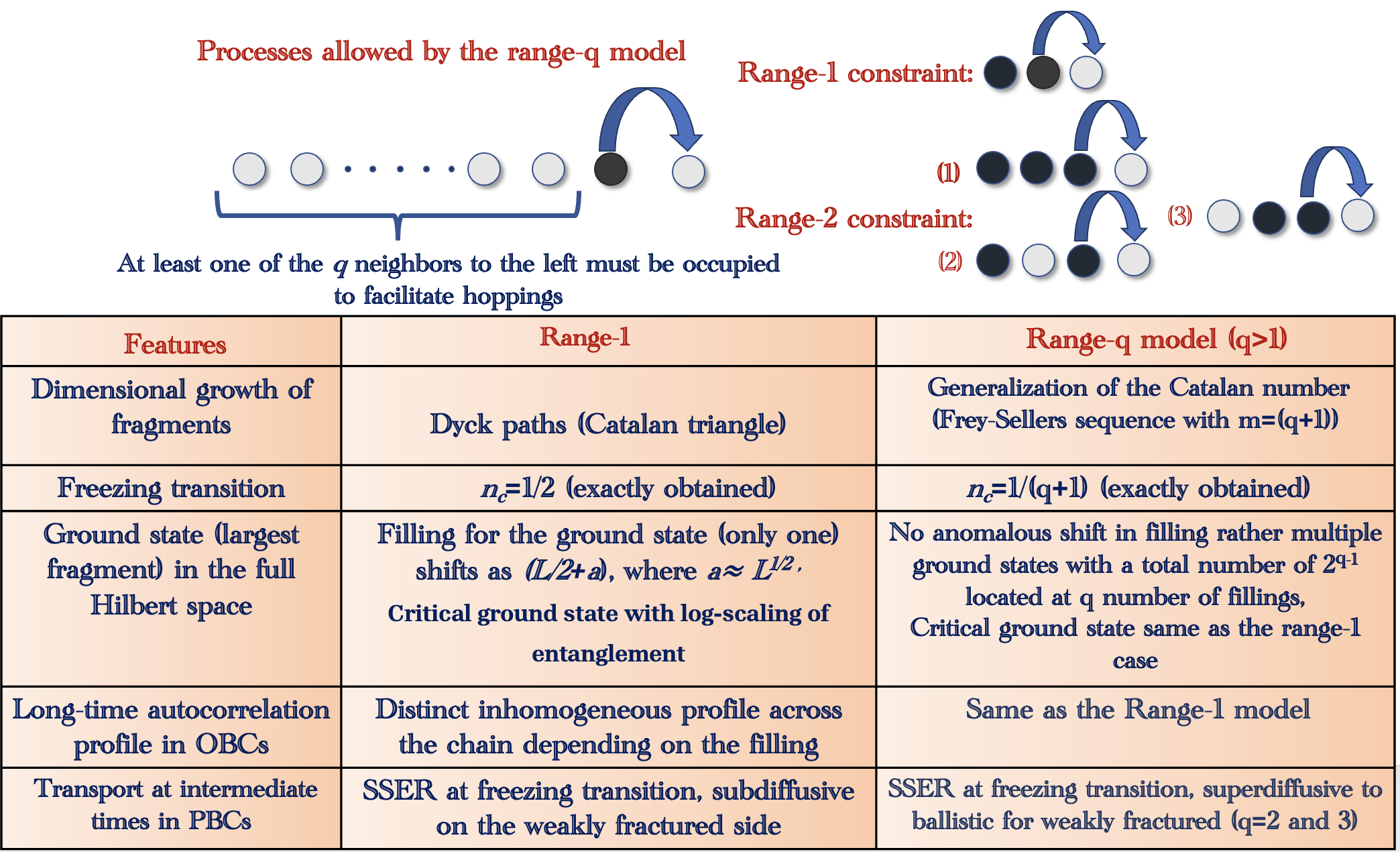}}
\caption{Schematic of the main results obtained from our analysis.}
\label{main}
\end{figure*}

Finally, we examine the behaviors of the infinite temperature transport properties~\cite{singh_2021,brighi_2023,zadnik_hydro_2021,Lenart_hetero,two_site_East_model} in periodic boundary conditions (PBCs) for range-1, range-2, and range-3 constraints, with the help of dynamical quantum typicality~\cite{QDT1,QDT2}, to understand how the freezing transition~\cite{Morningstar_2020,wang_2023,East_Sreemayee,abhisodh2024,brian_thermal} impacts the transport at intermediate times. Our examination of unequal-time autocorrelation function showcases that while the behaviors in all three cases at the freezing transition point show the best numerical fitting with size-stretched exponential relaxation (SSER)~\cite{berma_system_stretched}, it can vary from transient subdiffusive~\cite{singh_2021,pal_hydro} (range-1) through transient superdiffusive~\cite{brighi_2023} (range-3) to transient ballistic~\cite{brighi_anomoulous} power-law decay (range-2) on the weakly fragmented side~\cite{Moudgalya_review_2022,moudgalya_memorial_2021} of the Hilbert space. This thus points toward rich transport properties due to an intricate interplay between the freezing transition and range of constraints. The schematic of the main results obtained from our analysis are illustrated in Fig. \ref{main}.

The rest of the paper is organized as follows. We first begin with describing the family of facilitated quantum East models in Sec. \ref{modelHam}. Thereafter, we outline the strategy to identify the unique root states representing the disjoint fragments in this class of models in Sec. \ref{root}. Afterward, we characterize the features of the fragmented Hilbert spaces as well as describe some of the fragments, including the integrable and largest fragments, using enumerative combinatorics and transfer matrix methods in Secs. \ref{charac}, \ref{int}, and \ref{largest}, respectively. Next, we discuss the robustness of freezing transition in this class of models against the change in boundary conditions in Sec. \ref{freezing}. Following the above, we examine the anomalous ground state behaviors in Sec. \ref{ground state}. Thereafter, we elucidate the effect of freezing transition on the thermalization of bulk and boundary autocorrelators at long times in OBCs and bulk transport in PBCs at intermediate times in Secs. \ref{bulk-boundary} and \ref{inftran}, respectively. Finally, we conclude by recapitulating the primary results obtained from our analysis and offering exciting avenues pertinent to this problem for future exploration in Sec. \ref{dis}.
\begin{section}{Model Hamiltonian}
\label{modelHam}
We will first describe the model under consideration, which is the so-called facilitated quantum East model with range-$q$ constraints~\cite{brighi_2023}. The model is depicted by the following Hamiltonian
\bea
H_{q}~=~\sum_{i=q+1}^{L-1}\hat{K}_{i,q}\,\left(c_{i}^{\dagger}c_{i+1}+{\rm H.c.}\right), \label{Hamgen}
\eea
where $K_{i,q}$ signifies the kinetic constraint of range-$q$, which can be constructed using the projectors with $n_{i-l}=1$ and the region $\left[i-l+1,i-1\right]$ empty, i.e., $\hat{K}_{i,q}=\sum_{l=1}^{q}t_{l}\hat{P}_{i,l}$ where $\hat{P}_{i,l}=\hat{n}_{i-l}\prod_{j=i-l+1}^{i-1} (1-\hat{n}_{j})$ for $l\geq2$ and $P_{i,l}=\hat{n}_{i-l}$ for $l=1$. Here, $t_{l}$ denotes the strength of hoppings enabled by the particle residing at the $l$th site on the left, and $c_{j},c_{j}^{\dagger}$ and $n_{j}$ denote the fermionic annihilation, creation, and number operator on $j$th site, respectively. Further, we consider open boundary conditions (OBCs) while writing Eq. \ref{Hamgen}. As examples, the models with $q=1,2$ and $3$ can be recast as follows with the help of Eq. \ref{Hamgen}
\bea
H_{q=1}&~=~&t_{1}\sum_{i=2}^{L-1}n_{i-1}\left(c_{i}^{\dagger}c_{i+1}+{\rm H.c.}\right),\non\\
H_{q=2}&~=~&\sum_{i=3}^{L-1}\left(t_{1}\,n_{i-1}+t_{2} n_{i-2}(1-n_{i-1})\right)\non\\&&\times\left(c_{i}^{\dagger}c_{i+1}+{\rm H.c.}\right),\non\\
H_{q=3}&~=~&\sum_{i=4}^{L-1} \big(t_{1}\,n_{i-1}\,+\,t_{2}\,n_{i-2}(1-n_{i-1})\non\\
&&+t_{3}\,n_{i-3}(1-n_{i-2})(1-n_{i-1})\big)\non\\&&\times(c_{i}^{\dagger}c_{i+1}+{\rm H.c.}).\non\\ \label{Hamr123}
\eea
Furthermore, this family of models conserve the total particle number, $\hat{N}=\sum_{i}\hat{c}_{i}^{\dagger}\hat{c}_{i}$, but does not remain invariant under inversion symmetry transformation, given by $\hat{c}_{i}\rightarrow \hat{c}_{L-1-i}$. In addition, the energy eigenvalues appear in $E$ and -$E$ pairs due to the sublattice symmetry transformation~\cite{East_Sreemayee}, given by $c_{j}\rightarrow (-1)^{j}c_{j}$, which transforms $H\to-H$. Also, we assume $t_{l}=1$ for most of the analysis unless mentioned explicitly. One should note that the classical fragmentation~\cite{moudgalya_PRX_2022,Moudgalya_review_2022} structure, i.e., the fragmentation structure defined in terms of product basis states discussed shortly after this, remains unaltered irrespective of the values of $t_{l}$.
\end{section}
\section{Root identification method representing classical fragments in facilitated East models}
\label{root}
Due to kinetic constraint imposed by $K_{i,l}$, these models demonstrate Hilbert space fragmentation~\cite{moudgalya_memorial_2021,Moudgalya_review_2022,sala_ergo_2020}, which fractures the full Hilbert space into exponentially many fragments. Although the numerical evidence of fragmentation for these models has been witnessed earlier~\cite{brighi_2023}, the complete analytical characterization of the same remains hardly explored except for the constraint with range $q=1$ case~\cite{East_Sreemayee,wang_2023}. One of the primary purposes is thus to first unravel this for the full family of these models analytically. This further allows us to get insight into the fragmentation-induced freezing transition and ground state properties with an increasing range of constraints. 

\subsection{range-1 model}
We now begin by highlighting the main results obtained for the model with range-1 constraint~\cite{East_Sreemayee} that has been investigated in a very recent paper, where the author of the present paper is also a co-author. As this is the simplest variant involving only three-site terms~\cite{brighi_2023,East_Sreemayee,wang_2023}, reviewing the strategy to unveil the fragmentation structure in this case will facilitate the readers to comprehend the same thing in its longer-range variants in an easier manner. It should be noted that finding canonical representations~\cite{khagebdra2_fredkin,salberger2016fredkinspinchain,East_Sreemayee} for root states uniquely labeling the fragments is the key step toward characterizing the fragmented Hilbert spaces~\cite{moudgalya_memorial_2021,Moudgalya_review_2022} in this class of models. For the $q=1$ case, it can be readily seen that the allowed transitions are $110\leftrightarrow101$, which can be represented in an alternate parentheses notation as $(()\leftrightarrow()($, where open and closed parentheses, $($ and $)$ denote $1$ and $0$ occupation states. Further, this alternate notation furnishes an insight that the number of matched and mismatched parentheses on both sides of the transition is the same, implying the fact the dimensional growth of the fragments has a close connection to the widely studied combinatorics sequence of the Dyck words/ Dyck paths~\cite{catalan}. Dyck paths~\cite{catalan} are lattice paths in the two-dimensional $x-y$ plane with diagonal upsteps and downsteps of $(1,1)$ and $(1,-1)$ units, respectively. Also, it has been revealed that starting from an arbitrary binary string configuration lying in a fully connected fragment (fragments incorporating no blockades comprising frozen states), it is always possible to find a unique root state representing the fragment under consideration through successively substituting $110$'s by $101$'s. In addition, these unique representative states describe all the equivalent classes (fragments) and, thus, unravel the structure of the fragmented Hilbert space entirely~\cite{East_Sreemayee,khagebdra2_fredkin,salberger2016fredkinspinchain}. While the successive replacement of $ 101$'s fails to produce distinctive representations for root states. This local removal rule further furnishes the construction of the transfer matrix to compute the total number of fragments analytically~\cite{HariMenon_1995,menon_1997,Dhar_1993}, which we will discuss later. With the help of this rule, it can be readily shown that the root states labeling each fragment can be recast in a separable form~\cite{East_Sreemayee}, i.e., $\psi_{L}\otimes\psi_{M}\otimes\psi_{R}$,
where $\psi_{L}$
can be a null string ($\phi$) or a substring of $0$'s represented by $)\cdots)$ in the parenthesis notation, $\psi_{M}$ is given by substrings of $10$'s or in an alternative notation $()\cdots()$, and $\psi_{R}$ can be either a null string ($\phi$) or a substring of $1$'s, i.e.,  $(\cdots($ or a substring of $0$'s, $)\cdots)$ or a substring of 0's followed by substring of 1's, i.e., $)\cdots)(\cdots($. Subsequently, the dimension of a fully connected simple fragment can be computed utilizing the rules given below.
\vspace{0.2cm}

\noi (i) One must remove $\psi_{L}$ from the counting problem as this binary string remains dynamically inactive during time evolution. Also, this statement satisfies trivially for $\psi_{L}$ being $\phi$.
\vspace{0.2cm}

\noi (ii) Following the above, if $\psi_R$ is a substring comprising 0's, i.e., $)\cdots)$ or any number of 0's followed by any number of 1's, $)\cdots($, such a  $\psi_R$ also stays frozen under dynamical evolution. This statement again trivially holds when $\psi_R$ is $\phi$. These two scenarios fall under case I. While if $\psi_R$ consists of only $1$'s, i.e., $(\cdots($, it must be incorporated in the counting problem. 
After enforcing these rules, the dimension 
$D(L)$ of a fully connected simple fragment can be readily estimated with the help of the Catalan triangle of order, $l=1$, which is the number of sequences entitled by the well-studied Dyck path/ Dyck words~\cite{catalan} as
\vspace{0.2cm}

\noi (i) For the first case when $\psi_{R}$ falls under case I, we acquire $D(L)=C(n,n)$, where $C(n,n)$ is the $n$-th Catalan number~\cite{catalan} (diagonal elements of the Catalan triangle), i.e., $\frac{1}{n+1}\left(\begin{matrix}2n\\n\end{matrix}\right)$, and $n$ stands for the number of $10$'s.
\vspace{0.3cm}

\noi (ii) For the second case where $\psi_{R}$ lies in case II, the dimension can be computed as $D(L)=C(n+k,n)$, where $C(n+k,n)$ is the element of the Catalan triangle of order $l=1$ ~\cite{catalan} corresponding to $(n+k)$-th row and $n$-th column, where $n$ and $k$ stand for the total number of $10$'s and $1$'s in $\psi_{M}$ and $\psi_{R}$, respectively. The element $C(p,q)$ corresponding to the
$p$-th row and $q$-th column of the Catalan triangle~\cite{catalan} of order $l=1$ is $\left(\begin{matrix}p+q\\q\end{matrix}\right)\,-\,\left(\begin{matrix}p+q\\q-1\end{matrix}\right)$.
\vspace{0.3cm}

This method can be generalized further to fragments where active regions are isolated by blockades~\cite{moudgalya_memorial_2021,Moudgalya_review_2022}, including frozen states. Such fragments can thus be uniquely labeled by similar representative states, i.e., $\psi_{L}\otimes\psi_{m_1}\otimes\psi_{I_1}\otimes\psi_{m_2}\otimes\psi_{I_2}\otimes\cdots\otimes\psi_R$ , where $\psi_L$ is a null string $(\phi)$ or a substring made of any number of 0's, $\psi_{m_i}$ is a substring made of 10's, $\psi_{I_i}$ is a blockade region comprising 0's, and, finally, $\psi_{R}$ includes
only 0's or any number of 0's followed by any number of 1's or all 1's. The dimensions of such fragments with blockades can be calculated with the help of similar rules as discussed above:

\noi (i) For the first case with $ \psi_R$ falling under case I as discussed earlier, $ D(L)~ =~ \prod_{i = 1}^{L-1} C(n_i, n_i) $ where $n_i$ denotes the number of $10$'s in $\psi_{m_i}$, and $C(n_{i},n_{i})$ stands for the $n_{i}$-th Catalan number.
\vspace{0.2cm}

\noi (ii) For $ \psi_R$ with the properties of case II, $ D(L)~ =~ \prod_{i = 1}^{L-1} C(n_i, n_i) \, C(n_L + k, 
n_L)$, where $n_i$ represents the number of 10's in $\psi_{m_i}$ and $k$ denotes the number of $1$'s in $\psi_R$. Further, $C(n_{L}+k,n_{L})$ stands for the element of the Catalan triangle corresponding to the $(n_{L}+k)$-th row and $n_{L}$ column, and $C(n_{i},n_{i})$ is the $n_{i}$-th Catalan number.

\begin{comment}
An equivalent representation of the Dyck words is achievable in terms of Dyck paths. A Dyck path is a path that begins at the origin (0,0) and ends at $(N,0)$ in the two-dimensional $x-y$ plane; additionally, the path contains only two moves, namely, diagonal up steps $(1,1)$ and diagonal down steps
$(1,-1)$. Moreover, Dyck paths meet the condition that they should always be weakly above $y \geq 0$. Following such a prescription,  one can reveal that the number of permitted Dyck paths with $N$ moves is given by the Catalan number $D_{N,0,0}=\frac{1}{N+1}\begin{pmatrix}N\\ N/2\end{pmatrix}$, where $N$ is even. Furthermore, if we consider a more general case, where the paths start at $(0,0)$ and end at $(N,m)$ with $m\geq0$, the number of allowed Dyck paths in the upper half plane is given by
\bea
D_{N,0,m}=\frac{m+1}{N+1}\begin{pmatrix}N+1\\(N-m)/2\end{pmatrix}.
\eea
This is linked to the element of the Catalan triangle of order $l=1$ corresponding to the $n$-th row and $k$-th column, where $n=(N+m)/2$ and $k=(N-m)/2$.
\end{comment}
\subsection{range-2 model}
We will now proceed to the longer-range model with range-2 constraint~\cite{brighi_2023}, where finding the canonical representation for root states~\cite{East_Sreemayee,salberger2016fredkinspinchain,khagebdra2_fredkin} is a more involved task due to increased possibilities of allowed local moves. Our primary objective through this analysis is to capture various features of this fragmented Hilbert space and, most importantly, the filling-dependent strong-to-weak fragmentation transition~\cite{East_Sreemayee,wang_2023,Morningstar_2020} in these longer-range variants in an analytically tractable manner. In this paper, we will perform a detailed analysis of this class of models; however, validation of our analytical results with numerics is limited to range-3 constraints due to numerical restrictions posed by longer range constraints. We will also provide a detailed analytical understanding of the freezing transition~\cite{East_Sreemayee,wang_2023,Morningstar_2020} for the range-$q$ constraint later in the discussion.

To begin with, it should be noted that dynamical transitions in this case do not conserve the number of balanced $()$ and unbalanced parentheses $)($ numbers , unlike the range-1 case. This can be readily confirmed by taking a look at the allowed transitions, i.e.,  $1010\leftrightarrow1001$, $0110\leftrightarrow0101$ and $1110\leftrightarrow1101$, and subsequently employing the parentheses notation, i.e., $1\equiv ($ and $0\equiv)$ (The first process violates the above conservation). This also implies that the growth of the fragments cannot be captured utilizing the conventional Dyck paths~\cite{catalan} with diagonal up and down steps of $(1,1)$ and $(1,-1)$, which preserves the total number of balanced and unbalanced parentheses number for all the allowed transitions. Nevertheless, to find the unique root states labeling each fragment~\cite{East_Sreemayee,khagebdra2_fredkin,salberger2016fredkinspinchain}, we again employ a similar strategy as recapitulated in the earlier case~\cite{East_Sreemayee}. First, we see that for a given binary string in a fully-connected classical fragment, one has to successively substitute all $1110$'s, $1010$'s, and $0110$'s by $1101$'s, $1001$'s, and $0101$'s, respectively, which offers the most dilute string configurations (particles have been maximally spread~\cite{brighi_2023,wang_2023,East_Sreemayee}). Further, we have seen that the imposition of these rules facilitates the correct representative root states in the range-2 model; this has been validated using the transfer matrix method~\cite{generating1994,Deepak_HSF,East_Sreemayee} and numerical basis enumeration method, as elucidated later in the discussion. Accordingly, the root state again takes a separable form as $\psi_{L}\otimes\psi_{M}\otimes\psi_{R}$ similar to the range-1 case where

\noi (i) $\psi_{L}$ can be either a null string ($\phi$), or a 1 or a string of 0's. In the last case, all the 0's except the rightmost one must be subtracted from the counting problem as they remain dynamically frozen. In addition, they do not contribute to the facilitated transitions in any manner.   

\noi (ii) $\psi_{M}$ is units of 100's,

\noi (iii) $\psi_{R}$ can be either $\phi$, 0's, 0's followed by (1's/a 10), all 1's, or a single 10 followed by 1's. For a given $\psi_{R}$, all the strings comprising 0's or 0's followed by 1's should be removed from the dimension counting problem since they remain dynamically inactive. This procedure can be further generalized to more complicated fragments where dynamically active regions are separated from one another by blockades consisting of $0$'s. 

One should note that the dimension of a subclass of simple fully connected classical fragments (when $\psi_{L}$ being all 0's or a single 1) can be computed exploiting the concept of generalization of the Catalan sequence introduced by Frey and Sellers~\cite{FS_article}. The details of this combinatorial sequence are given in the Appendix-\ref{combo}. For this subclass of root states, the dimensional growth of the fragments after imposing the rules given in (i)-(iii) for active $\psi_{R}$'s (i.e., comprised of all 1's or a single 10 followed by 1's) can be shown to be
$D_{frag}(L)=\left\{\begin{array}{cc}n\\ r\end{array}\right\}_{2}$, where
\bea
&&\left\{\begin{array}{cc} n\\ r\end{array}\right\}_{2}=\left(\begin{array}{cc}n+r\\n\end{array}\right)-\left\la\begin{array}{cc}n\\r \end{array}\right\ra_{2},\non\\
&&\left\la\begin{array}{cc} n\\r\end{array}\right\ra_{2}=\sum_{k=0}^{P}\frac{1}{2k+1}\left(\begin{array}{cc} 3k\\k\end{array}\right)\non\\&&
\times\left(\begin{array}{cc}n+r-3k-1\\n-k\end{array}\right),~~\text{where}~~~P=\large\lceil\frac{r}{2}\large\rceil-1,\non\\
\label{FS2}
\eea
where $n$ and $r$ are the total numbers of $1$'s and $0$'s, respectively, in $\psi_{M}\otimes\psi_{R}$ (active $\psi_{R}$ which include 1's or 10 followed by 1's). There is another subclass of root states, where $\psi_{L}$ is $\phi$, and the dimensional growth of these fragments cannot be comprehended utilizing the Frey-Sellers (FS) sequence~\cite{FS_article}. Nevertheless, we demonstrate an example of a root state representing a fragment lying in the non-FS subclass in Appendix-\ref{nonFS}, whose dimension can be captured using the combinatorial sequence of $2_{t}$-Dyck paths~\cite{Asinowski_2022}. This is a special lattice path in a two-dimensional plane with diagonal upsteps and downsteps of $(1, 2)$ and $(1, -1)$ steps, respectively. In addition, $t$ denotes a constraint that all the lattice paths allowed by this sequence must be weakly above the $y=-t$ line, where $t\in Z$ and $0\leq t  \leq2$. Interestingly, this is the first time a generalized Catalan sequence~\cite{Asinowski_2022,FS_article} has appeared in any spinless fermionic models to the best of our knowledge, which opens up further possibilities for future explorations.

\subsection{range-3 model}
An identical procedure is also applicable to apprehend all the canonical representations for the range-3 model, which allows seven transitions given below
\bea
&&11110\leftrightarrow11101,~~01110\leftrightarrow01101,~~00110\leftrightarrow00101,\non\\
&&11010\leftrightarrow 11001,~~10110\leftrightarrow10101,~~10010\leftrightarrow10001,\non\\
&&01010\leftrightarrow01001.
\eea

In this case, we again observe that the most dilute string configuration~\cite{East_Sreemayee} is the representative state labeling each fragment. This can be obtained by substituting all $11110$'s, $01110$'s, $10110$'s, $00110$'s, $10010$'s, $01010$'s, $11010$'s starting from a given binary string configuration. We validate this by employing the numerical basis enumeration method as well as the transfer matrix method\cite{generating1994}, as discussed later. In this case, the representative states again can be recast in a separable form as $\psi_{L}\otimes\psi_{M}\otimes\psi_{R}$, where 
\vspace{0.1cm}

\noi (i) $\psi_{L}$ is either a null string ($\phi$), a single 0 or a single 1, all 0's ($\geq2$), two 1's, and a 10/01. When $\psi_{L}$ is a string of 0's ($\geq2$), all the 0's except the rightmost two 0's have to be removed as they do not contribute to any facilitated hopping processes. However, all other choices of $\psi_{L}$'s (except $\phi$) are capable of allowing more correlated hopping processes, thus causing more binary string configurations to be members of the given fragment. 

\noi (ii) $\psi_{M}$ is a substring of 1000's. 

\noi (iii) $\psi_{R}$ is either a null string($\phi$), all 0's, 0's followed by (1's/a 10/ a 100), a 10 followed by 1's, or a 100 followed by 1's or all 1's. For $\psi_{R}$, being all 0's or 0's followed by 1's or a null string ($\phi$) again does not contribute to the allowed transitions and thus must be deducted while counting the dimension. 

One can thereafter check following the similar procedure shown in the Appendix. \ref{rootr2} that there exists again a subclass of root states even for the range-3 constraint involving $\psi_{L}$ being all 0's ($\geq2$) or $11/ 10/ 01$, which follows the FS sequence with $m=4$~\cite{FS_article}. In this case, one can further validate utilizing the Appendix. \ref{combo} that the dimensional growth of such fragments represented by these root states with active $\psi_{R}$ (i.e., including 1's or $(10)1^{m}$ or $(100)1^{m}$, $m\geq1$) can be shown to be
$D_{frag}(L)=\left\{\begin{array}{cc}n\\ r\end{array}\right\}_{3}$, where $n$ and $r$ are the total number of $1$'s and $0$'s in $\psi_{M}\otimes\psi_{R}$ (in case $\psi_{R}$ is active); further, $\left\{\begin{array}{cc}n\\ r\end{array}\right\}_{m-1}$ can be computed readily with the help of Eq. \eqref{FS}. There is another subclass of root states with $\psi_{L}$, being $\phi$/1/0, which cannot be captured with the help of the FS sequence~\cite{FS_article} in this case as well. However, similar to the range-2 case, there are specific examples of root states in the non-FS subclass, which generate fragments with dimensional growth following the combinatorial sequence of $3_{t}$ Dyck paths~\cite{Asinowski_2022} (see Appendix-\ref{nonFS}). This is again a special lattice path in a two-dimensional plane with diagonal upsteps and downsteps of (1, 3) and (1, -1) lattice steps, respectively, with the constraint that all the allowed paths following the sequence must be weakly above the $y=-t$ line, where $t\in Z$ and $0\leq t  \leq3$. 

\subsection{range-$q$ model}
The above observations thus lead to a general inference about the root structure of the range-$q$ constraint~\cite{brighi_2023}, i.e., it takes an equivalent separable form shown in the previous cases, where $\psi_{m}=10^{q}$, $\psi_{R}$ can be $0^{m_1}1^{m_2}/10^{m_3}1^{m_4}$, where $0\leq m_3\leq (q-1)$, and $\psi_{L}$ would be $0^{k_{1}}$/$1^{k_{2}}/0^{k_3}1^{k_4}/1^{k_3}0^{k_4}$, where $k_1\geq (q-1)$, $k_2=(q-1)$, $k_3+k_4=(q-1)$, respectively (for FS class of root states).
The dimensional growth of such fragments represented by these root states for active $\psi_{R}$ ( $10^{m_1}1^{m_2}$, where $0\leq m_{1}\leq (q-1)$) can therefore be captured with the help of the FS sequence~\cite{FS_article} with $m=q+1$. (One should note that root states with $\psi_{L}$ being a substring of length $l<(q-1)$ fail to generate all possible transitions allowed by the FS sequence. These root states fall under the non-FS class). Hence, the fragmentation structure of this family of models specified by a dominant class of root states is completely extractable using analytical treatments even for the range-$q$ constraint. This has rarely been noticed earlier in the case of models involving comparatively longer-range constraints. 

\section{Characterization of the fragmented Hilbert space}
\label{charac}
\subsection{Number of fragments}
With this root structure in hand, one can now implement the transfer matrix method~\cite{generating1994,Deepak_HSF,East_Sreemayee,Dhar_1993,menon_1997,HariMenon_1995} to count the total number of fragments, which we now discuss for $q=1,2$ and $3$ cases. We will further consider open boundary conditions (OBCs) for future convenience. However, consideration of periodic boundary conditions will not make any qualitative difference in results; nevertheless, the setup of transfer matrix~\cite{generating1994,Deepak_HSF,East_Sreemayee} is a bit tricky to implement in the latter case. In the earlier study~\cite{East_Sreemayee}, we have already shown that the number of fragments in the model with range-1 constraint
 grows as $\tau^{L}$ for large-$L$, where $\tau=\frac{\sqrt{5}+1}{2}$ using the transfer matrix method. In this paper, we will compute the growth of $N_{frag}$ for the range-2 and range-3 cases in a detailed manner.
 
 To begin with, we now consider the range-2 model, which includes transitions involving four consecutive sites. This implies that one requires a $8\times8$ transfer matrix to compute $N_{frag}$. Further, we have already shown that unique identification of root states labeling each fragment requires successive substitutions of $1110$'s, $0110$'s, and $ 1010$'s from a given binary string configuration. This implies that the resultant transfer matrix must not include any $1110$, $0110$, and $1010$. The details of the transfer matrix ($T_{1}$) calculation are shown in Appendix \ref{frag}. Following the above procedure, we note that $N_{frag}$ in this model grows as $1.466^{L}$ asymptotically; this further agrees with the numerically obtained results perfectly. Interestingly, the same asymptotic growth also appears in the case of the number of frozen states for the range-1 constraint, as we have noticed in our previous paper~\cite{East_Sreemayee}. 

 A similar method likewise facilitates the counting of $N_{frag}$ in the range-3 case. Nevertheless, as this model allows transitions including five consecutive sites, it thus requires a $16\times16$ transfer matrix ($T_{2}$) to compute $N_{frag}$. Moreover, as illustrated earlier, the root identification method for the range-3 case requires substitutions of $11110$, $11010$, $01110$, $10110$, $10010$, $01010$, $00110$, and the resultant transfer matrix $T_{4}$ thus should not involve any of this string configurations while computing $N_{frag}$. The details of the transfer matrix method~\cite{generating1994,Deepak_HSF,East_Sreemayee} are displayed in Appendix \ref{frag}, which reveals a growth of $1.38^{L}$ in $L\to\infty$ limit, and is again in perfect agreement with the numerical basis enumeration method. We note that this is precisely the growth of frozen states for the range-2 case, as elucidated in the next part of the discussion.

 This thus infers that the growth of $N_{frag}$ decreases with an increasing range of constraints in this family of models.

\subsection{Frozen fragments}
An intriguing feature of fragmented Hilbert spaces is the presence of frozen state configurations, which are the fragments comprising a single state~\cite{moudgalya_memorial_2021,Moudgalya_review_2022}. These states are zero-energy eigenstates of the Hamiltonian that do not participate in the dynamics. Also, the number of such states ($N_{froz}$) generally grows exponentially with system size due to the kinetic constraints, which can again be readily captured using transfer matrix methods~\cite{generating1994,Deepak_HSF,East_Sreemayee,Dhar_1993,menon_1997,HariMenon_1995}. 

We will now discuss the growth of frozen states with system size with the help of the transfer matrix method~\cite{generating1994,Deepak_HSF,East_Sreemayee}. In the previous study~\cite{East_Sreemayee}, we have already revealed that $N_{froz}$ in the range-1 model grows as $1.466^{L}$ in $L\to\infty$ limit. In this paper, we will, therefore, elaborate on the models with range-2 and range-3 constraints.

As mentioned earlier, the model with range-2 constraints involves terms that include four consecutive sites, which requires us to construct a $8\times 8$ transfer matrix to calculate $N_{froz}$. Furthermore, as these eigenstates are dynamically inactive during transitions, any resultant $1110$, $1010$, $0110$, $1001$, $0101$, and $1101$ have to be removed from the resultant transfer matrix~\cite{generating1994,Deepak_HSF,East_Sreemayee}. The details of the transfer matrix method have been elucidated in Appendix \ref{frozena}. This method~\cite{generating1994,Deepak_HSF,East_Sreemayee} discloses that $N_{froz}$ in this model grows with system size as $1.38^{L}$ asymptotically, which also entirely agrees with our numerical result.

A similar method can be implemented for the range-3 model as well to count the same quantity. However, now, like the case of $N_{frag}$, one needs to construct a $16\times16$ transfer matrix as this model involves transitions, including five consecutive sites. Moreover, for this case, the resultant transfer matrix~\cite{generating1994,Deepak_HSF,East_Sreemayee} should have any of these 14 configurations: $11110$, $11101$, $01110$, $01101$, $11010$, $11001$, $10110$, $10101$, $10010$, $10001$, $01010$, $01001$, $00110$, and $00101$. The details of the construction are again shown in Appendix \ref{frozena}. This reveals that $N_{froz}$ asymptotically grows as $1.325^{L}$, which we also validate utilizing our numerical basis enumeration method.

This analysis thus provides the insight that the growth of $N_{froz}$ reduces with increasing the range of constraints, similar to the growth of $N_{frag}$. 

\section{Descriptions of some of the simple fragments in the models}
\label{int}
We now examine some of the integrable fragments in this class of models that can be easily described in terms of non-interacting tight-binding models.
\subsection{range-1}
The simplest integrable fragments in the range-1 model can be represented by the root state with $\psi_{L}$, $\psi_{M}$ and $\psi_{R}$ being all 0's, a single 10 and all 1's, respectively. As discussed by the rule mentioned earlier in Sec. \ref{root}, $\psi_{L}$ never participates in the dynamics and can thus be removed from the effective description. Thereafter, the effective Hamiltonian with the fragment reduces to a simple tight-binding model with nearest-neighbor hoppings in OBCs where a single 10 hops in the background of 1's. Therefore, the dimension of such a fragment is given by $(N_{A}+1)$, where $N_{A}$ is the number of 1's in $\psi_{R}$. The dispersion of such an effective Hamiltonian can easily be computed as $E_{p}=-2\cos(\pi\,p/(N_{A}+2))$, where $p=1,2,\cdots, (N_{A}+1)$.
\subsection{range-2}
A similar analog of integrable fragments can also be extended to longer-range constraints.
In the range-2 case, an extremely simple integrable fragment is represented by the following root states, $\psi_{L}\otimes\psi_{M}\otimes\psi_{R}$, with (i) $\psi_{L}=1$ or $0$, $\psi_{M}=\phi$, and $\psi_{R}$ incorporating $(10)1^{\otimes N_{A}}$ or (ii) $\psi_{L}=\phi$, $\psi_{M}=100$, and $\psi_{R}=1^{\otimes N_{A}}$ (Here $x^{\otimes y}$ implies $y$ number of $x$'s). 
 
 In the former case, the effective Hamiltonian will be a nearest-neighbor non-interacting tight-binding model, where a single 10 hops in the background of 1's in OBCs~\cite{East_Sreemayee}. In the latter case, the first 10 in $\psi_{M}$ facilitate more hopping process, yet they do not actively participate in the dynamics. Hence, this $10$ should be removed first from the counting problem. Thereafter, the effective Hamiltonian within such fragments again effectively reduces to a tight-binding model where a single hole (0) performs a nearest-neighbor hopping in the background of 1's in OBCs~\cite{East_Sreemayee}. The dimension of such fragments for both cases thus becomes $D(L)=(N_{A}+1)$, where $N_{A}$ is the number of 1's in $\psi_{R}$. In addition, the dispersion can be computed as $E_{p}=-2\cos(\pi\,p/(N_{A}+2))$, where $p=1,2,\cdots, (N_{A}+1)$.

%In the latter case, the effective Hamiltonian with the fragment can be again described by the nearest-neighbor tight-binding model of a single hole in the background of 1's (in $\psi_{R}$). The first 10 always remain dynamically inactive; however, it facilitates the nearest-neighbor hoppings. Therefore, the leftmost 10 should be removed from the dimension counting problem. The dimension of the fragment of such a fragment is thus given by $(N_{A}+1)$, where $N_{A}$ is the number of 1's in $\psi_{R}$. The dispersion of the fragment again can be captured by the same as the previous case, i.e., $E_{p}=-2\cos(\pi\,p/(N_{A}+2))$, where $p=1,2,\cdots, (N_{A}+1)$.

\subsection{range-3}
 In the range-3 case, a similar integrable fragment can be represented by the following root states $\psi_{L}\otimes\psi_{M}\otimes\psi_{R}$, with (i) $\psi_{L}=11/10/01/00$, $\psi_{M}=\phi$, and $\psi_{R}$ a single $(10)1^{\otimes N_{A}}$, (ii) $\psi_{L}=1/0$, $\psi_{M}=\phi$, and $\psi_{R}=(100)1^{\otimes N_{A}}$ and (iii) $\psi_{L}=\phi$, $\psi_{M}=1000$, and $\psi_{R}=1^{\otimes N_{A}}$. 
 
 For the first two cases, $\psi_{L}$ and the leftmost 1 and the leftmost 10 from the root states given in (i) and (ii), respectively, should be removed first from the counting problems as they facilitate more hopping process yet remain unaltered during the dynamics. In the last case (iii), the first 100 in the $\psi_{M}$ has to be removed due to the same reason mentioned in the first two cases. Following the above, the effective Hamiltonians for all such fragments reduce to a nearest-neighbor tight-binding model where a hole, 0, hops in the background of 1's in OBCs~\cite{East_Sreemayee}. The dimension of such fragments thus again becomes $D(L)=(N_{A}+1)$, where $N_{A}$ is the units of 1's in $\psi_{R}$ and furthermore, the dispersion can be readily shown to be $E_{p}=-2\cos(\pi\,p/(N_{A}+2))$, where $p=1,2,\cdots, (N_{A}+1)$.

 An identical root state construction representing similar integrable fragments can also be generalized to models with range-$q$ constraints after considering appropriate choices of $\psi_{L}$, $\psi_{M}$ and $\psi_{R}$.

\section{Description of the largest fragment}
\label{largest}
\subsection{range-1}
We will now show that this family of models demonstrates a strong-to-weak fragmentation transition as a function of filling fraction, which is called the freezing transition~\cite{Morningstar_2020,East_Sreemayee,wang_2023,brian_thermal}. We captured several features of the range-1 model in the previous paper~\cite{East_Sreemayee}, which we will again review in brevity for the sake of completeness and further to make a direct comparison with the longer-range variants discussed shortly after this. To begin with, one can check that the root state representing the largest fragment has the following form $1010..1011\cdots11$ at a filling $N_{f}=L/2+a$, where $a>0$ and $\psi_{M}$ and $\psi_{R}$ include $(L/2-a)$ and $2a$ numbers of $10$'s and $1$'s respectively. It can be further readily validated that the growth of such a fragment ($D_{frag}$) can be captured utilizing the well-studied Dyck sequence~\cite{catalan} with the help of off-diagonal elements of the Catalan triangle as 

\bea
    D_{max}(L)=C(n+k,n) = \frac{(2 a + 1)}{ (\frac{L}{2} + a + 1)} \frac{L!}{(\frac{L}{2} - a )! (\frac{L}{2} + a )!},\non\\
    \label{offCat}
\eea
which exhibits the following asymptotic growth in $L\to\infty$ limit
\bea
D_{max}(L)\,\simeq\, 2^{L+1} \frac{(2 a + 1)}{L} \sqrt{\frac{2}{\pi L}} e^{-\frac{2 a^2}{L}}.\label{largeLdim}
\eea
This asymptotic growth at $N_{f}=L/2+a$ after assuming $a=\rho L$ implies weakly fragmented Hilbert space above half-filling since $D_{max}(L)/D_{sum}(L)=4\rho/(1+2\rho)$, a constant in $L\to\infty$. Here, $D_{sum}$ refers to the full Hilbert space growth at the given filling fraction, which is $\left(\begin{array}{cc} L\\ L/2+\rho L\end{array}\right)$.  Subsequently, it can be proved that the filling at which this model manifests the largest fragment shifts as $L/2+a$, where $a\sim \sqrt{L}/2$ in the large-$L$ limit. This can be checked by extremizing Eq. \ref{largeLdim} w.r.t $a$~\cite{East_Sreemayee}. On the other hand, the largest fragment at $N_{f}=L/2$ is characterized by the root state $1010..10$ with $L/2$ numbers of 10's. This fragment follows the dimensional growth dictated by the diagonal Catalan number~\cite{catalan} as
\bea
D_{max}(L)=C(L/2,L/2)=\frac{1}{L/2+1}\left(\begin{array}{cc} L \\ \frac{L}{2}\end{array}\right).
\label{diag}
\eea
Eq. \eqref{diag} therefore suggests an asymptotic growth of $2^{L}/L^{3/2}$ in $L\to\infty$ and accordingly, $D_{max}/D_{sum}=1/L$ at this specific filling. The above decay points toward that $n_{c}=1/2$ is the critical filling fraction for the strong-to-weak fragmentation transition since $D_{max}/D_{sum}\to0$ polynomially~\cite{wang_2023,East_Sreemayee}, which is much slower than the exponential decay. Furthermore, the largest fragment below half-filling ($N_{f}=L/2-\alpha L$) is labeled by the root state $00..01010..10$ with $L/2-\alpha L$ number of 10's in $\psi_{M}$ and $2\alpha L$ number of 0's in $\psi_{L}$, respectively. The dimensional growth of such a fragment can again be apprehended employing the diagonal Catalan number~\cite{catalan}, i.e., $D_{max}(L)=C(L/2-\alpha L,L/2-\alpha L)$. This further implies an asymptotic growth of $\sim \left(\frac{2}{4^\alpha}\right)^{L}$ (only keeping the exponential part), thus suggesting the strong fragmentation below half-filling, i.e., $D_{max}(L)/D(L)\to 0$ exponentially in large-$L$ limit, for $0<\alpha\ll1$.

\subsection{range-2}
We will now proceed to the range-2 model and probe the nature of the freezing transition both analytically and numerically~\cite{Morningstar_2020,East_Sreemayee,wang_2023,brian_thermal}. Further, we will also capture the filling fraction where this model manifests the largest fragment, which is important to understand the ground state behaviors~\cite{East_Sreemayee,Deepak_HSF} that dominate the low-energy properties of this model.

As discussed earlier, the dimensional growth of a prominent subclass of root states labeling the fragments can be captured utilizing the Frey-Sellers sequence~\cite{East_Sreemayee} with $m=3$, whose details have been elucidated in Appendix \ref{combo}. At first, we will concentrate on the dimensional growth at $N_{f}=L/3$ for $L$ being multiples of three. The largest fragment in this case is specified by the following root state $00100..1001$, with $\psi_{L}$, $\psi_{M}$ and $\psi_{R}$ comprised of 2 $0$'s, $(L-3)/3$ 100's, and single 1, respectively. Following the rules discussed earlier in Sec. \ref{root} and thereafter utilizing Eqs. \ref{exactFS} and \ref{exactFS2}, it can be shown that the dimension of such a fragment is given by
\bea
D_{frag}(L)\,=\,\left\{\begin{array}{cc}n\\2n-2\end{array}\right\}_{2}=\frac{1}{2n+1}\left(\begin{array}{cc}3n \\n \end{array}\right),\label{onethird}
\eea
where $n$ is the number of 1's in the root state. Thereafter, utilizing the fact $n=L/3$ and subsequently using Stirling's formula, it can be shown that
$D_{frag}(L)\simeq 1.89^{L}/L^{3/2}$ in $L\to\infty$. On the other hand, the dimension of the Hilbert space at $N_{f}=L/3$ grows as
\bea
D_{sum}(L)=\left(\begin{array}{cc}L \\L/3 \end{array}\right),\label{onethirdfull}
\eea
which indicates the asymptotic growth of $1.89^{L}/\sqrt{L}$ in $L\to\infty$ limit. This thus implies that $D_{max}/D_{sum}(L)= \frac{1}{L}$, which tends to zero polynomially in $L\to\infty$~\cite{East_Sreemayee,wang_2023}, however, much slower than exponential decay. This further suggests that $n_{c}=1/3$ is the critical filling fraction for the freezing transition in the range-2 case. One can, therefore, infer that the increasing range of constraints pushes this dynamical phase transition to occur at a lower filling fraction in this family of models. The first few values of the dimension of such fragments are exhibited in Table-\ref{frag1},
\begin{table}[htb]
\begin{tabular}{|c|c|c|c|c|c|c|c|c|c|c|c|c|}
\hline
No. of $100$ in $\psi_{M}$&0&1&2&3&4&5&6&7 \\
\hline
$D(L)$&1&3&12&55&273&1428&7752&43263\\
\hline
\end{tabular}
\caption{ The dimension of fragments represented by the root states with $\psi_{L}=0$'s, $\psi_{M}=100$'s and $\psi_{R}=1$ obtained using the Frey-Sellers sequence, which has further been validated using numerical basis enumeration method.}
\label{frag1}

\end{table}
which we obtained using the FS formula and are in perfect agreement with the numerical results. 

One can also argue that the largest fragment below the critical filling, i.e., $N_{f}=L/3-a$ with $a>0$ can be labeled by the following root state, $0^{\otimes(3a+2)}(100)^{\otimes(L/3-a-1)}1$, where $\psi_{L}=0^{\otimes(3a+2)}$, $\psi_{M}=(100)^{\otimes(L/3-a-1)}$ and $\psi_{R}=1$, respectively. After imposing the rules elucidated in Sec. \ref{root} and utilizing the Frey-Sellers sequence, it can be checked the dimensional growth of such a fragment is as follows
\bea
D_{frag}(L)&=&\left\{\begin{array}{cc}L/3-a\\2L/3-2a-2\end{array}\right\}_{2},\non\\
&=&\frac{1}{2(L/3-a)+1}\left(\begin{array}{cc}(L-3a)\\(L/3-a)\end{array}\right),
\label{strongr2}
\eea
which we acquire utilizing Eqs. \eqref{exactFS} and \eqref{exactFS2}. Eq. \eqref{strongr2} can further be simplified using Stirling's formula in $L\to\infty$ limit as $D_{frag}(L)\simeq \left(3/2^{2/3}\right)^{L} \left(4/27\right)^{\alpha L}$, where we consider $a=\alpha L$ and $0<\alpha\ll 1$ (only keeping the exponential part). This indicates strong fragmentation since $D_{frag}/D_{sum}\simeq(4/27)^{\alpha L}\to0$ asymptotically (exponentially tending to zero) for $\alpha>0$.

Thereafter, we examine the filling fraction where this model manifests the largest fragment. Our numerical examination suggests that the largest fragment in this model (using numerical enumeration for system sizes up to $L=24$) lies at $N_{f}=L/2$ and $N_{f}=L/2+1$ for $L$ being even, and $N_{f}=(L-1)/2+1$ and $(L-1)/2+2$ for $L$ being odd. The root state (at $N_{f}=L/2$ and $N_{f}=(L-1)/2+1$ for even and odd $L$'s, respectively) and dimensional growth of such fragments for the first few $L$'s are presented in Table \ref{groundr2}, which can be easily procured using basis enumeration method.
\begin{table}[htb]
\begin{tabular}{|c|c|c|c|c|c|c|c|c|c|c|c|c|}
\hline
$L$& Root state& $D_{frag}(L)$\\
\hline
6&010011&6\\
\hline
7& 0100111&10\\
\hline
8&01001011&19\\
\hline 
9&010010111&34\\
\hline
10&0100100111&65\\
\hline
11&01001001111&120\\
\hline
12&010010010111&228\\
\hline
13&0100100101111&431\\
\hline
14&01001001001111&822\\
\hline
15&010010010011111&1575\\
\hline
16&0100100100101111&3006\\
\hline
17&01001001001011111&5820\\
\hline
18&010010010010011111&11139\\
\hline
19&0100100100100111111&21717\\
\hline
20&01001001001001011111&41643\\
\hline
\end{tabular}
\caption{Root states and dimensional growth of the largest fragments for the first few $L$'s in the case of range-2 constraint obtained using numerical basis enumeration method.}
\label{groundr2}
\end{table}
Subsequently, utilizing the rules discussed earlier, one can show that this growth follows the FS sequence~\cite{FS_article} with $m=3$ as
\bea
D_{max}(L)&\,=\,&\left\{\begin{array}{cc}n\\n-1\end{array}\right\}_{2}\non\\
&=&\left(\begin{array}{cc}2n-1\\n\end{array}\right)-\left\la\begin{array}{cc}n\\n-1\end{array}\right\ra_{2},\non\\
\left\la\begin{array}{cc} n\\n-1\end{array}\right\ra_{2}&=&\sum_{k=0}^{P}\frac{1}{2k+1}\left(\begin{array}{cc} 3k\\k\end{array}\right)\left(\begin{array}{cc}2n-3k-2\\n-k\end{array}\right),\non\\
&&\text{where}~~~P=\Large\lceil\frac{n-1}{2}\Large\rceil-1,\non\\
\label{Dfragr2}
\eea
where $n$ is the number of 1's (for even $L$). 
One should first note that the reason for the range-2 constraint displaying two fillings for embodying the largest fragments is two possible choices of $\psi_{L}$, which are $0$ or $1$ ($\psi_{M}\otimes \psi_{R}$ identical for both cases). Both of them cause identical dimensional growth and can be checked with the help of rules given in Sec. \ref{root}. %Additionally, the filling possibilities for the above increase with an increasing range of constraints due to various choices of $\psi_{L}$ giving rise to the same dimensional growth.
Interestingly, we also observe that the above fragment does not move away significantly from these two $N_{f}$'s with increasing system sizes (validated up to $L=26$). This is distinct from the range-1 case where the filling for the largest fragment shifts as $L/2+a$, where $a\sim\sqrt{L}/2$ in $L\to\infty$. However, it is hard to prove the absence of any shift analytically in the $L\to\infty$ limit due to the lack of a closed form of Eq. \eqref{Dfragr2}. Nevertheless, we numerically argue that there is, in fact, no significant shift in the thermodynamic limit by examining the variation of $D_{frag}$ vs $N_{f}$ for $L=150$ followed by the identification of the correct representative state (root state located at the minimum filling involving $\psi_{L}=0$, and remaining part of the root state is fixed by $L$ and filling) and afterward using the Frey-Sellers formula with $m=3$~\cite{FS_article}, as demonstrated in Fig. \ref{larger2r3}. Fig. \ref{larger2r3} thus facilitates the understanding that the minimum filling where $D_{max}$ first appears in the range-2 case is at $N_{f}=L/2+1$. This is almost comparable to that observed for finite $L$'s, thus corroborating no anomalous shift, unlike the range-1 case~\cite{East_Sreemayee}.
\begin{comment}
\begin{figure}[htb]
\subfigure[]{\includegraphics[width=0.8\hsize]{plots/exponential_fitting_r2_model_one_third_filling.png}}\\
{\includegraphics[width=0.8\hsize]{plots/exponential_growth_r2_model.png}}
\caption{(a) Plot showing the dimensional growth of the largest fragment with $L$ at $n_{f}=1/3$ obtained from a simple numerical enumeration. (b) Plot showing the dimensional growth of the largest fragment with $L$ obtained using simple numerical enumeration, which occurs at $n_{f}=L/2$. Numerical fitting shows that $D_{frag}$ in (a) grows as $1.88^{L}/L^{1.31}$ in the large $L$ limit, which indicates strong fragmentation at this particular filling. On the other hand, the largest fragment within the model grows as $1.99^{L}/L^{0.75}$ ($\sim 2^{L}$) in the large $L$ limit, as suggested by the numerical fitting. This, therefore, indicates a weak fragment at $N_{f}=L/2$. Further, both the numerical fittings are in complete agreement with our analytical findings. }\label{large_frag_r2}
\end{figure}
\end{comment}
\subsection{range-3}
We now extend a similar examination to the range-3 case where we find the critical filling fraction~\cite{Morningstar_2020,wang_2023,two_site_East_model} to be $n_{c}=1/4$ using both numerical enumeration and analytically using the Frey-Sellers sequence~\cite{FS_article}. To do so, one can argue that the root state specifying the largest fragment at $N_{f}=L/4$ is $0001000\cdots10001$, where $\psi_{L}$, $\psi_{M}$ and $\psi_{R}$ are 3 0's, $(L-4)/4$ 1000's and 1, respectively. Thereby, employing the rules given for the range-3 case exhibited in Sec. \ref{root}, it can be demonstrated that these fragments satisfy the growth dictated by the Frey-Sellers sequence~\cite{FS_article} with $m=4$ as
\bea
D_{frag}\,=\,\left\{\begin{array}{cc}n \\3n-3\end{array}\right\}_{3}\,=\,\frac{1}{3n+1}\left(\begin{array}{cc}4n\\ n\end{array}\right),
\eea
where $n=L/4$ is the number of 1's in the root state, and further, $D_{frag}\sim 1.75^{L}/L^{3/2}$ in $L\to\infty$, thus implying $D_{frag}/D_{sum}\sim1/L$ at $n=1/4$. This also suggests that this is the critical filling in the range-3 case due to the slow polynomial decay of the largest fragment compared to the full Hilbert space growth at this particular filling.

It can also be easily verified that the largest fragment at $N_{f}=L/4-\alpha L$ for $0<\alpha\ll1$ is described by the root state $0^{\otimes(4\alpha L+3)}(1000)^{\otimes(L/4-\alpha L-1)}1$ with $\psi_{L}$, $\psi_{M}$ and $\psi_{R}$ being $0^{\otimes(4\alpha L+3)}$, $(1000)^{\otimes(L/4-\alpha L-1)}$ and $1$, respectively. Thereafter, similar to the range-2 case, the dimension of the given fragment can be shown to be
\bea
D_{frag}(L)&=&\left\{\begin{array}{cc}(L/4-\alpha L)\\(3L/4-3\alpha L-3)\end{array}\right\}_{3},\non\\
&=&\frac{1}{3(\frac{L}{4}-\alpha L)+1}\left(\begin{array}{cc}L-4\alpha L\\(\frac{L}{4}-\alpha L)\end{array}\right),
\label{strongr3}
\eea
which we obtain using the rules given in Sec. \ref{root} and afterward using the properties of the Frey-Sellers sequence with $m=4$ given in Eqs. \eqref{exactFS} and \eqref{exactFS2}. Eq. \eqref{strongr3} further reduces to 
$D_{frag}\sim \left(4/3^{3/4}\right)^{L}(27/64)^{\alpha L}$ in $L\to\infty$ limit employing the Stirling's formula (only keeping the exponential growth). This therefore means $D_{frag}/D_{sum}=(27/256)^{\alpha L}\to 0$ in $L\to\infty$, hinting at strong HSF below $n_{c}=1/4$ in the range-3 case.
\begin{comment}
We then investigate the nature of fragmentation at $n=1/3$ in this model to get further insight into the problem. First, we identify the root state, which represents the largest fragment at $n=1/3$ for $L=3p$, which with their dimension are listed below
\begin{table}[htb]
\begin{tabular}{|c|c|c|c|c|c|c|c|c|c|c|c|c|}
\hline
$L$& Root state& $D_{frag}(L)$\\
\hline
3&001&1\\
\hline
6& 0010001&3\\
\hline
9&001000101&14\\
\hline 
12&001000100011&74\\
\hline
15&001000100010011&409\\
\hline
18&001000100010001011&2379\\
\hline
21&001000100010001000111&14223\\
\hline
24&001000100010001000100111&86496\\
\hline
27&01000100010001000100010111&534604\\
\hline

\end{tabular}
\label{freezr3}
\caption{Root states representing largest fragments at $n=1/3$ with the value of largest fragment in range-3 model.}
\end{table}
We further validate that the dimensional growth of such fragment can again be analytically found utilizing the Frey Sellers formula
\bea
D_{frag}^{{\rm max}}(L)&\,=\,&\left\{\begin{array}{cc}n\\r\end{array}\right\}_{3}\non\\
&=&\left(\begin{array}{cc}n+r\\n\end{array}\right)-\left\la\begin{array}{cc}n\\r\end{array}\right\ra_{3},
\label{Dfragr3}
\eea
where $n=L/3$ and $q=2L/3-2$. It can be further shown using Stiring's formula that the first term grows as $1.89^{L}$ (approximately). This exponential growth can not be altered by the presence of the second term, which again implies this model demonstrates strong fragmentation at $n=1/3$ with range-3 constraint as well.
\end{comment}

\begin{figure}
\includegraphics[width=0.82\hsize]{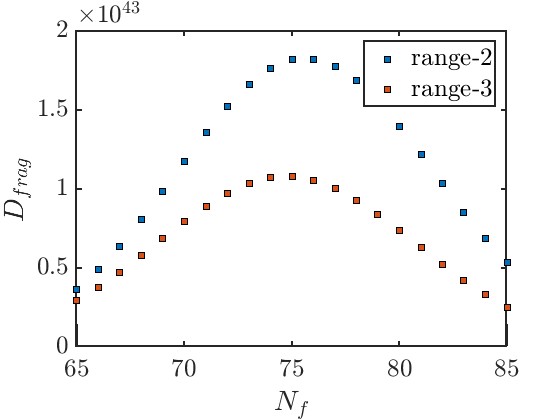}
\caption{Plot showing $D_{frag}$ vs $N_{f}$ in the range-2 and range-3 cases for $L=150$, obtained using the Frey-Sellers sequence. We observe that the largest fragment appears at $N_{f}=L/2+1$ (but $D_{frag}$ for $N_{f}=L/2$ is almost comparable) in $L\to\infty$ limit for the range-2 case, which does not significantly differ from the finite-size result. In the range-3 case, $D_{frag}$ turns out to be the largest at $N_{f}=L/2$ in $L\to\infty$ limit; this is identical to that observed for finite $L$'s.}
\label{larger2r3}
\end{figure}

We now explore the filling fraction for the largest fragment, which turns out to be at $N_{f}=(L/2$, $L/2+1$, and $L/2+2$) for even $L$ whereas the same lies at $N_{f}=(L-1)/2$, $(L-1)/2+1$, and $(L-1)/2+2$ for odd $L$ (for $L\geq 10)$, as seen from the numerical basis enumeration method. The root states and numerically obtained $D_{frag}$ at $N_{f}=L/2$ and $(L-1)/2$ (for $L\geq10$) for the first few even and odd $L$'s, respectively, are illustrated in Table-\ref{groundr3}.
\begin{table}[htb]
\begin{tabular}{|c|c|c|c|c|c|c|c|c|c|c|c|c|}
\hline
$L$& Root state& $D_{frag}(L)$\\
\hline
6&001011&3\\
\hline
7& 0010111&6\\
\hline
8&00100111&10\\
\hline 
9&001001111&20\\
\hline
10&0010001111&35\\
\hline
11&00100010111&69\\
\hline
12&001000101111&125\\
\hline
13&0010001001111&246\\
\hline
14&00100010011111&455\\
\hline
15&001000100011111&896\\
\hline
16&0010001000111111&1680\\
\hline
17&00100010001011111&3308\\
\hline
18&001000100010111111&6626\\
\hline
19&0010001000100111111&12343\\
\hline
20&00100010001001111111&23559\\
\hline
\end{tabular}
\caption{Root states and numerically obtained values of the largest fragment for the first few $L$'s in range-3 model.}
\label{groundr3}
\end{table}
Thereafter, utilizing our rules for the root states as discussed earlier and with the help of FS sequence with $m=4$, we show the largest fragments  within the full Hilbert space follow the dimensional growth as 
\bea
D^{{ max}}_{{frag}}(L)&=&\left\{\begin{array}{cc} n\\ n-2\end{array}\right\}_{3},\non\\
&=&\left(\begin{array}{cc}2n-2\\ n\end{array}\right)-\left\la\begin{array}{cc}n\\ n-2\end{array}\right\ra_{3},\non\\ \label{large_r3}
\eea
where $n$ is the number of 1's in the root state (for even $L$). It can also be argued, similar to the range-2 case, that the largest fragment lies at three fillings due to four alternate possibilities of $\psi_{L}$ (with identical $\psi_{M}\otimes\psi_{R}$), which are $00,10,01,11$; however, all such fragments display an identical dimensional growth. In this case, we again see that the largest fragment for the minimum filling (root state involving $\psi_{L}=00$) appears at $N_{f}\sim L/2$ even in $L\to\infty$ limit and does not show any shift with increasing $L$'s, unlike the range-1 case. It has been numerically studied using the Frey-Sellers sequence~\cite{FS_article} in Fig. \ref{larger2r3} for $L=150$, similar to the examination performed in the range-2 case.

\begin{comment}
\begin{figure}[htb]
\subfigure[]{\includegraphics[width=0.8\hsize]{plots/exponential_fitting_r3_model_one_third_filling.png}}\\
\subfigure[]{\includegraphics[width=0.8\hsize]{plots/exponential_fitting_r3_model_ground_state.png}}
\caption{(a) Plot showing the dimensional growth of the largest fragment with $L$ at $n_{f}=1/3$ obtained from a simple numerical enumeration. (b) Plot showing the dimensional growth of the largest fragment with $L$ obtained using simple numerical enumeration, which occurs at $n_{f}=L/2$ even for this case. Numerical fitting shows that $D_{frag}$ in (a) grows as $1.91^{L}/L^{1.01}$ in the large $L$ limit, which indicates strong fragmentation at this particular filling. On the other hand, the largest fragment within the model grows as $1.99^{L}/L^{0.44}$ ($\sim 2^{L}$) in the large $L$ limit, as suggested by the numerical fitting. This, therefore, indicates a weak fragment at $N_{f}=L/2$. Further, both the numerical fittings are in complete agreement with our analytical findings. }
\end{figure}
\end{comment}
\subsection{range-$q$}
\begin{figure}
\subfigure[]{\includegraphics[width=0.54\hsize,height=3.7cm]{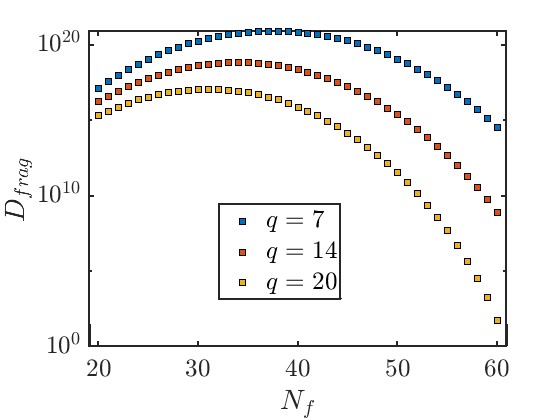}}%
\subfigure[]{\includegraphics[width=0.54\hsize,height=3.7cm]{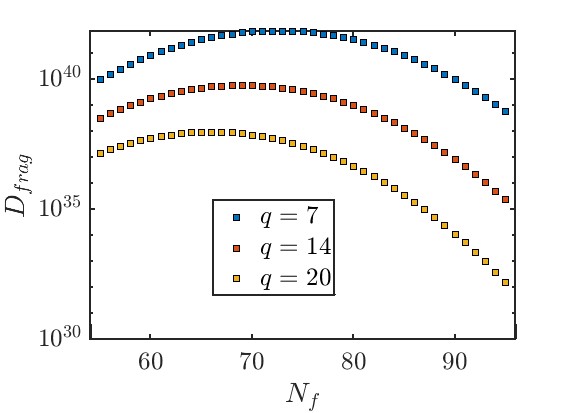}}
\caption{(a-b) Plot showing $D_{frag}$ vs $N_{f}$ for $L=80$ and $L=150$, respectively, for constraints with ranges, $q=7,14$ and $20$, obtained using root identification method and thereafter utilizing the Frey-Sellers formula with $m=(q+1)$. We see that the minimum fillings for the largest fragment appear at $N_{f}=L/2-2,L/2-6$, and $L/2-9$ for $q=7,10$ and $20$, respectively. This thus indicates no anomalous shift in the filling fraction with system sizes, unlike the range-1 constraint.}
\label{qrange}
\end{figure}
The above observations motivate us to make a general remark on the nature of fragmentation in the range-$q$ model, which can be understood with the help of the Frey-Sellers sequence with $m=(q+1)$~\cite{FS_article}. It can be readily inspected that the growth of the largest fragment at $n_{c}=\frac{1}{q+1}$ specified by the root state $0^{\otimes q}(10^{ q})^{\otimes\left((L-q-1)/(q+1)\right)}1$ is 
\bea
D_{n=\frac{1}{(q+1)}}=\left\{\begin{array}{cc}n \\q(n-1)\end{array}\right\}_{q}=\frac{1}{(qn+1)}\left(\begin{array}{cc}(q+1)n\\n \end{array}\right),\non\\
\eea
where $n=L/(1+q)$ is the number of 1's, which one can again obtain using the properties of the Frey-Sellers formula with $m=(q+1)$ given in Eqs. \eqref{exactFS}, \eqref{FS} and \eqref{exactFS2}. Hence, this fragment yields the following asymptotic growth as
\bea
D_{n=\frac{1}{(q+1)}}&\simeq&\frac{L!}{\left(L/(1+q)\right)!\left(qL/(1+q)\right)!}\non\\&\simeq& \frac{1}{L^{3/2}}\left(\frac{(1+q)}{q^{q/(1+q)}}\right)^{L},\non\\
\eea
which is the critical filling fraction for the freezing transition in the range-$q$ model as $D_{frag}/D_{sum}\sim 1/L$ in $L\to\infty$ at this specific filling. 

Our root identification also enables us to determine the number of fillings at which this family of models with range-$q$ constraint will embody the largest fragment (with the same dimensional growth) in OBCs. In doing so, we identify the several possibilities of $\psi_{L}$'s, which are $\psi_{L}=0^{ (q-1)},0^{k_{1}}1^{ k2}$, $1^{k_{1}}0^{k_{2}}$ and $1^{(q-1)}$, where $k_{1}+k_{2}=q-1$. These root states lead to $q$ in total fillings to incorporate the above. Next, we explore the minimum filling at which $D^{max}_{frag}$ appears for the very first time in the range-$q$ model, i.e., fragment labeled by the root state comprising $\psi_{L}=0^{q-1}$. After identifying the representative state, one can now utilize Eq. \ref{FS} with $m=(q+1)$ along with the constraint $n+r=L-(q-1)$, which yields
\bea
&&\left\{\begin{array}{cc} n\\ r\end{array}\right\}_{q}=\left(\begin{array}{cc}n+r\\n\end{array}\right)-\left\la\begin{array}{cc}n\\r \end{array}\right\ra_{q},\non\\
&&\left\la\begin{array}{cc} n\\r\end{array}\right\ra_{q}=\sum_{k=0}^{P}\frac{1}{qk+1}\left(\begin{array}{cc} (q+1)k\\k\end{array}\right)\non\\&&
\times\left(\begin{array}{cc}n+r-(q+1)k-1\\n-k\end{array}\right),~~~~\text{where}~~~P=\large\lceil\frac{r}{q}\large\rceil-1.\non\\
\label{FSq}
\eea

Afterward, it is required to extremize the above equation w.r.t $n$ in $L\to\infty$ limit to obtain the value of $n$ (or filling). Nevertheless, this extremization strategy is again tricky due to the lack of any closed form of Eq. \ref{FSq}. We, therefore, examine the minimum filling by examining the variation of $D_{frag}$ vs $N_{f}$ using the Eq. \ref{FS} after the identification of the appropriate root states (with $\psi_{L}=0^{q-1}$ and $\psi_{M}\otimes\psi_{R}$ fixed by $L$ and filling) for $q=7,14$ and $20$ and $L=80$ and $150$, respectively, as illustrated in Figs. \ref{qrange} (a-b). Our analysis indicates that the minimum values of $N_{f}$ for the occurrence of the largest fragment are $N_{f}=L/2-2,L/2-6,$ and $L/2-9$ for $q=7,14$ and $20$, respectively (for $L$ being even). Therefore, one can say that the minimum $N_{f}$ in the range-$q$ constraint are $(L-q)/2-1$ (for $L$ and $q$ being both even) and $(L-q-1)/2-1$ for (even $L$ and odd $q$).
We now provide a heuristic argument to capture this behavior by considering the first term in Eq. \eqref{FSq}, which is equal to $\left(\begin{array}{cc} (L-q)\\ n \end{array}\right)$ where we assume $L,q\gg1$. This term offers the maximum contribution when $n=(L-q)/2$, which is close to what we observe from our numerical study shown in Fig. \ref{qrange}. 

%Further, noting Eq. \ref{FS}, it can be clearly argued that the dimension of the largest fragment can only grow as $2^{L}$, when $n=r=L/2$, which further implies that this model, even for constraints with range-$q$ can only demonstrate the strong-to-weak fragmentation close to half-filling. The freezing transition in this class of models is quite universal in a sense, as the filling fraction for transition does not change with the change in the range of the constraints. This is rather opposite to what occurs in the spin-1 dipole conserving model, where the range of hopping can induce an alternation in the nature of the fragmentation. While strong fragmentation is favored by the strong-range model, the long-range constraints have inclinations toward weak fragmentation.
It has been demonstrated earlier in the context of dipole conserving models that increasing the range of kinetic constraints compels systems exhibiting strong HSF to go towards weak HSF~\cite{sala_ergo_2020,Morningstar_2020,abhisodh2024}. On the other hand, in the family of facilitated quantum East models, where the simplest short-range variant itself displays a freezing transition, increasing the range of constraint pushes the critical filling fraction towards a lower value, which we have concretely established through our analysis. Furthermore, increasing the range of constraints also allows multiple largest fragments at several fillings where the minimum value of $N_{f}$ for the manifestation of $D_{max}$ decreases with an increasing range of constraints, as we have argued earlier.
\begin{figure}[h!]
\subfigure[]{\includegraphics[width=\columnwidth,height=5.6cm]{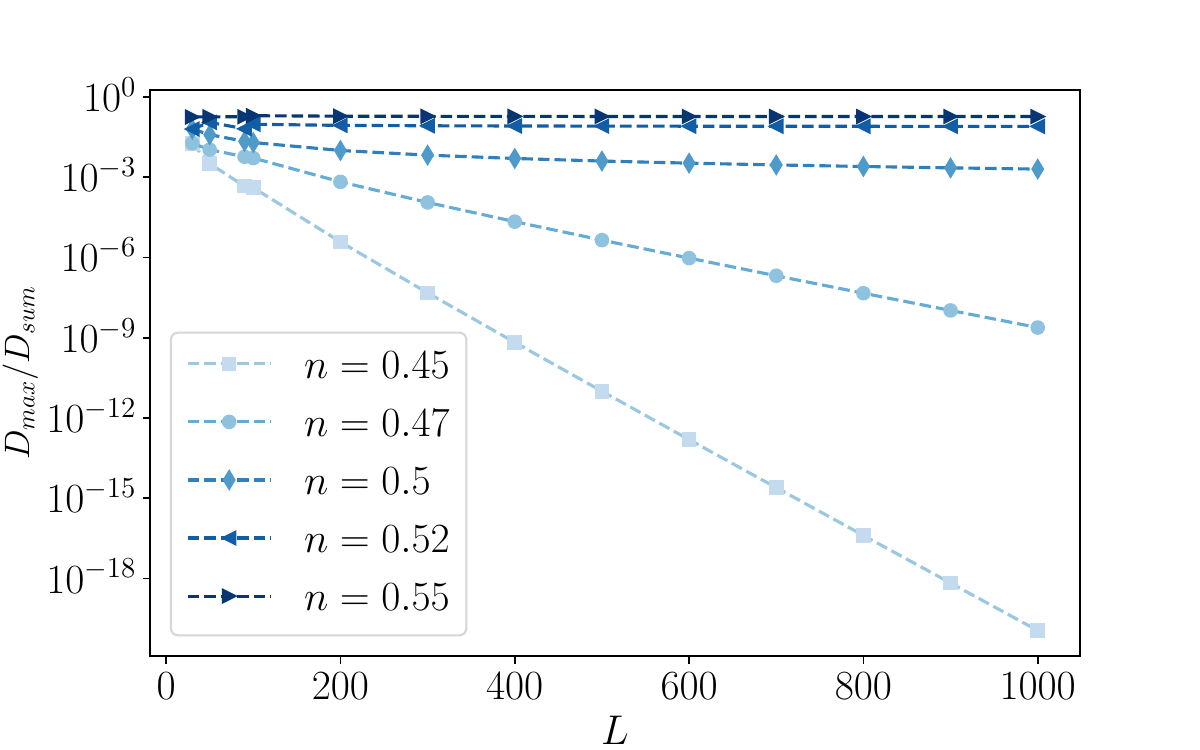}}\\
\subfigure[]{\includegraphics[width=\columnwidth,height=5.6cm]{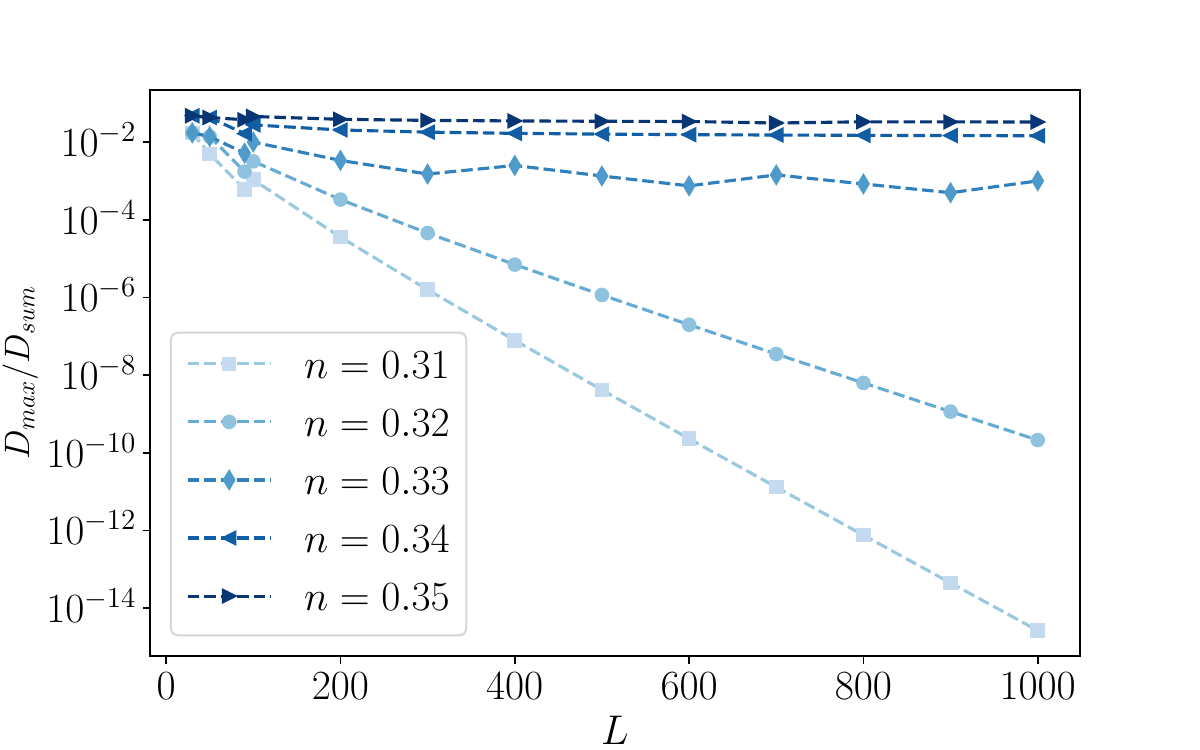}}\\
\subfigure[]{\includegraphics[width=\columnwidth,height=5.6cm]{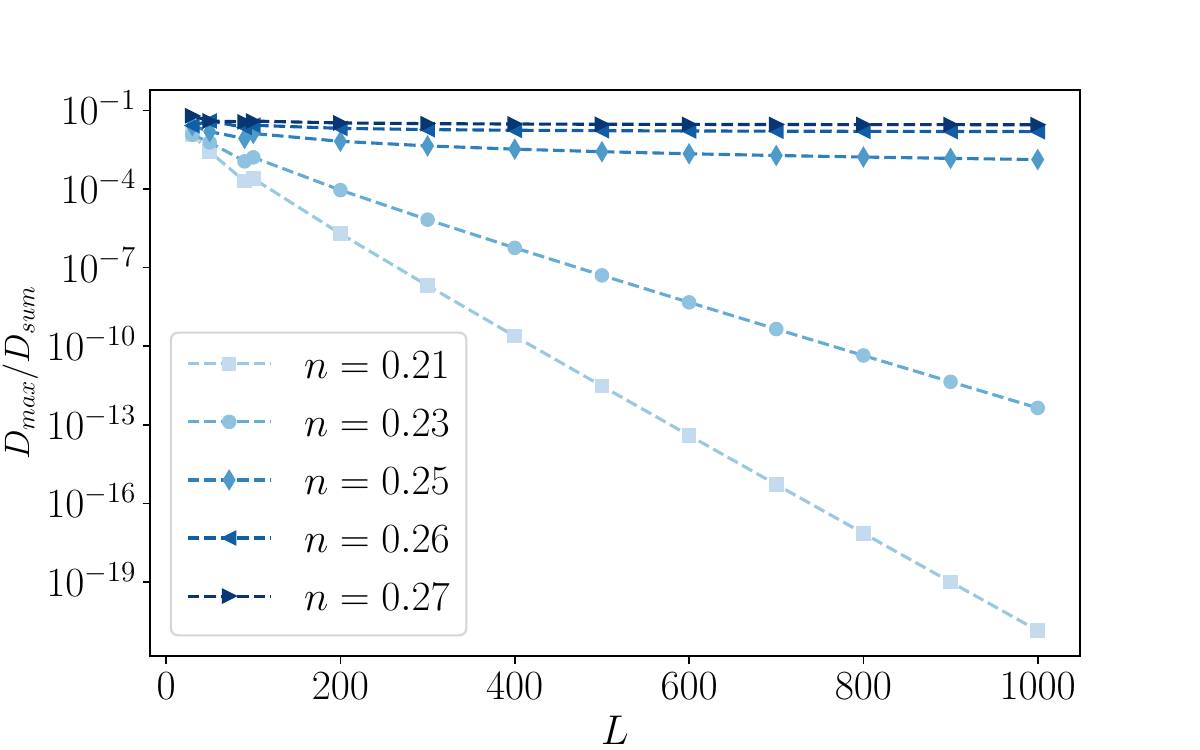}}
\caption{(a-c): The variation of $D_{max}/D_{sum}$ vs $L$ for several filling fractions for models with range-1, range-2, and range-3 constraints, respectively, obtained followed by the identification of appropriate root states and afterward using the Catalan triangle ($q=1$) and Frey-Seller sequence with $m=3$ and $4$ for $q=2$ and $q=3$ cases, respectively. It is quite evident from Figs. (a-c) that the behavior of $D_{max}/D_{sum}$ switches from the exponential decay to a constant at filling fractions $n_{c}=1/2$, $1/3$, and $1/4$, respectively.}
\label{dimfreez}
\end{figure}

\section{Freezing transition in this class of models and its sensitivity to boundary conditions}
\label{freezing}
We will now examine the sensitivity of this freezing transition~\cite{Morningstar_2020,wang_2023,East_Sreemayee,brian_thermal} due to the change in boundary conditions. To compare the nature of transitions in various boundary conditions, we again begin with OBCs, where the features are analytically tractable using the enumerative combinatorics method~\cite{generating1994,FS_article}, as discussed in Sec. \ref{largest}. In Figs. \ref{dimfreez} (a-c), we demonstrate the variation of $D_{max}/D_{sum}$ vs $L$ for various filling fractions in OBCs for the range-1, 2, and 3 constraints, respectively, using the Catalan and Frey-Sellers sequence. It is evident from Figs. \ref{dimfreez} (a-c) that $D_{max}/D_{sum}$ radically switches its behavior from exponential decay to a constant at a filling fraction close to $n_{c}=1/2,1/3$ and $1/4$ for range-1, 2, and 3 constraints, respectively, which are the critical filling for the freezing transition~\cite{wang_2023,East_Sreemayee}, as elucidated earlier.

\begin{figure*}
\begin{comment}
\subfigure[]
{\includegraphics[width=0.35\linewidth,height=5cm]{freezing_PBC_r1.pdf}}%
\subfigure[]{\includegraphics[width=0.35\linewidth,height=5cm]{freezing_PBC_r2.pdf}}%
\subfigure[]{\includegraphics[width=0.35\linewidth,height=5cm]{freezing_PBC_r3.pdf}}
\end{comment}
\includegraphics[width=0.95\hsize, height=5.5cm]{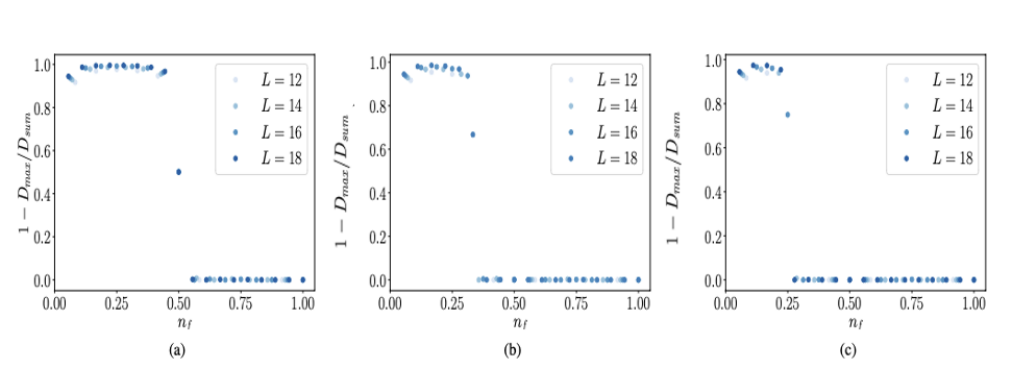}
\caption{(a-c): $D_{max}/D_{sum}$ vs $n_{f}$ in the models with range-1, range-2, and range-3 constraints in PBCs, respectively, for $L=12,14,16$ and $18$. In Figs. (a-c), it can be readily noticed that the behavior changes sharply at filling fractions $n_{c}\simeq1/2$, $1/3$, and $1/4$, respectively, which agrees well with the analytical behaviors obtained in OBCs.}
\label{freezPBC}
\end{figure*}

Next, we perform a similar analysis in periodic boundary conditions (PBCs) where this analytical understanding is tricky. Nevertheless, to proceed further, we probe the variation of $1-D_{max}/D_{sum}$ vs filling fractions $(n_{f})$ for finite $L$'s utilizing the numerical basis enumeration method. In Figs. \ref{freezPBC} (a-c), we inspect this in cases of range-1, range-2, and range-3 constraints, respectively, for $L=12,14,16$ and $18$. Accordingly, we witness $1-D_{max}/D_{sum}$ sharply change from $1$ (strong fragmentation~\cite{Moudgalya_review_2022,wang_2023,East_Sreemayee}) to $0$ (weak fragmentation~\cite{Moudgalya_review_2022,wang_2023,East_Sreemayee}) close to $n_{c}=1/2$, $1/3$ and $1/4$ in range-1, range-2 and range-3 models, respectively, thus indicating that the transition is robust even in PBCs.

\section{Ground state properties of this family of models}
\label{ground state}
We will now investigate how the distinct fragmentation structure affects the low-energy properties of this family of kinetically-constraint models~\cite{salberger2016fredkinspinchain,Garahhan_fredkin,East_Sreemayee,khagebdra2_fredkin,two_site_East_model,Lenart_hetero}. In doing so, we first compare the ground state fillings for these models with range-1, range-2, and range-3 constraints~\cite{brighi_2023,wang_2023,East_Sreemayee}, respectively. Further, we begin this investigation in OBCs since we comprehensively understand the fragmented Hilbert space in this specific scenario.

\begin{figure*}
\subfigure[]{\includegraphics[width=0.34\linewidth,height=4.5cm]{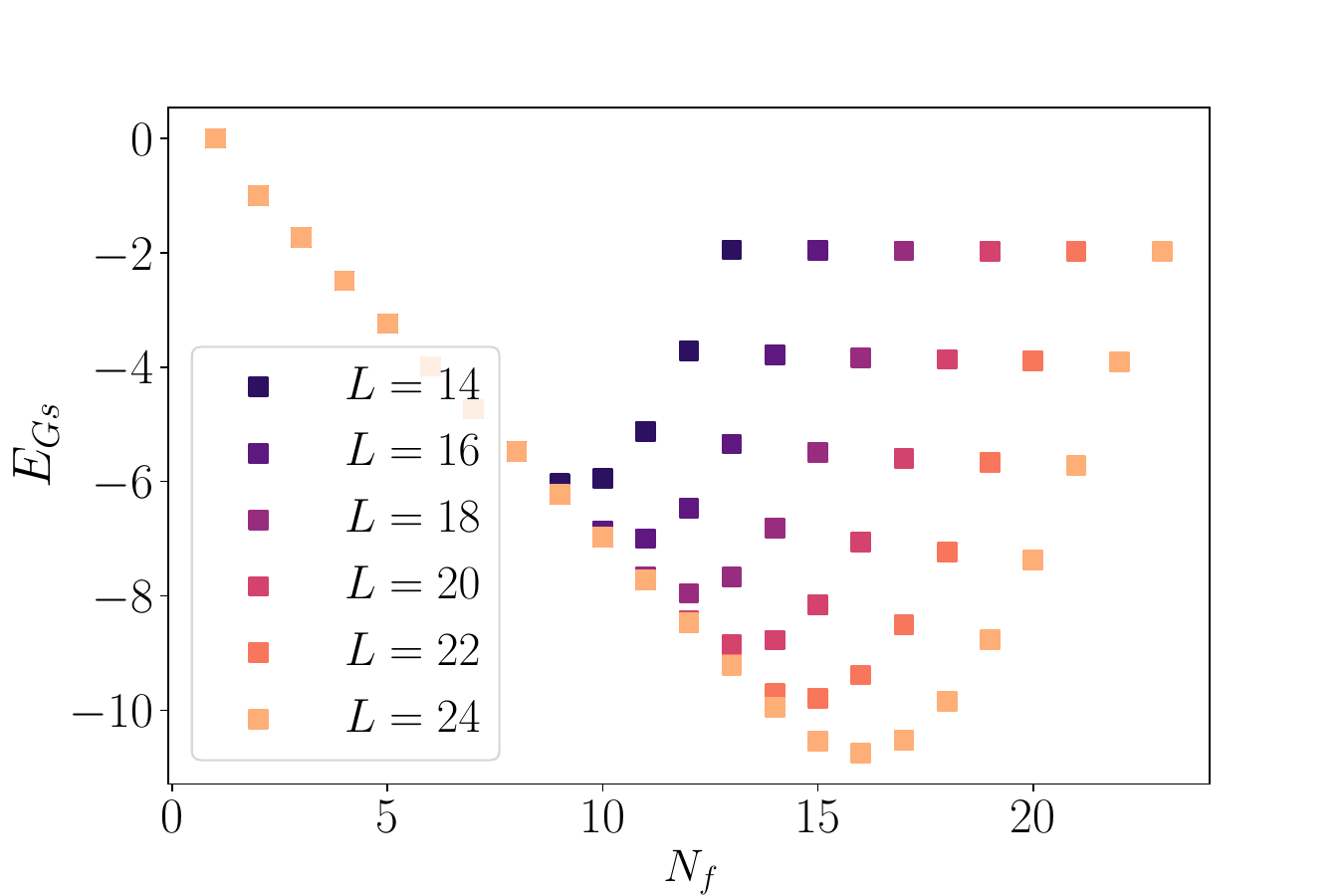}}%
\subfigure[]{\includegraphics[width=0.34\linewidth,height=4.5cm]{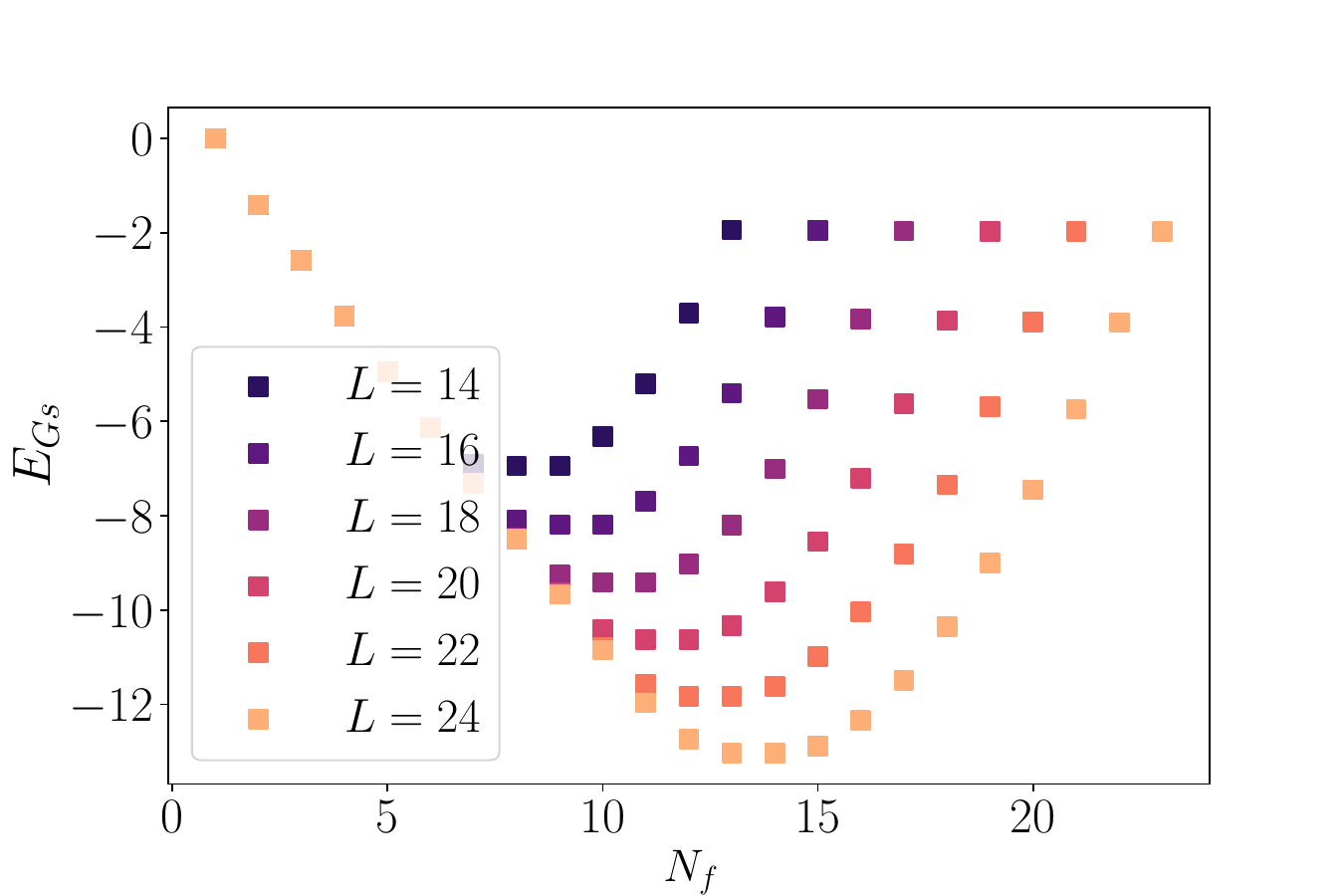}}%
\subfigure[]{\includegraphics[width=0.34\linewidth,height=4.5cm]{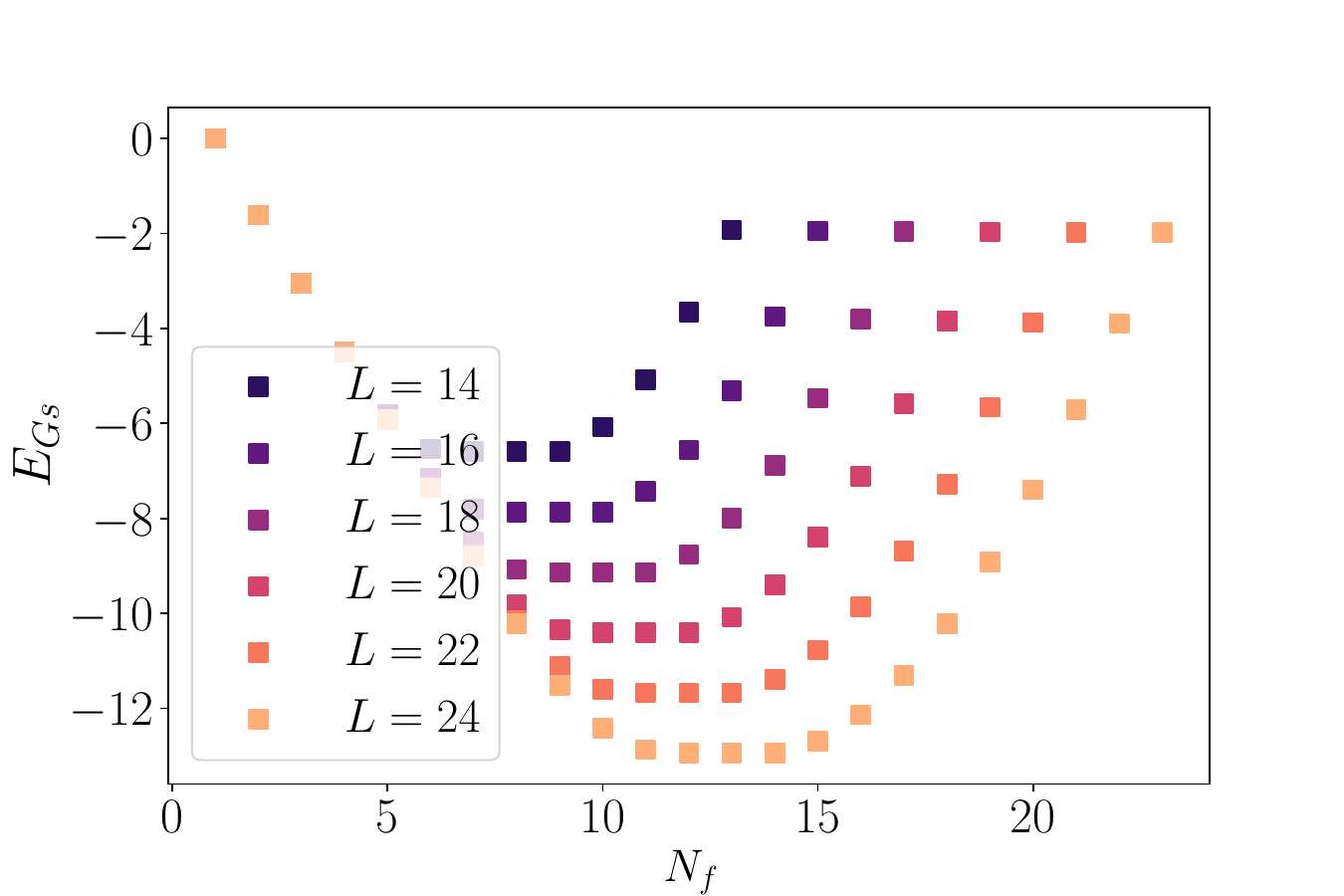}}
\caption{(a-c): The variation of ground state energy, $E_{GS}$ at different fillings, $N_{f}$ for several $L$'s  in OBCs for the models with range-1, range-2 and range-3 constraints, respectively. (a) In the case of range-1 constraint, $E_{Gs}$ lies at $N_{f}=L/2+2$, $L/2+3$ and $L/2+4$ for $L=14$, $L=(16,18,20$), and $L=(22,24$), respectively, revealing that the ground state filling shifts away from $N_f=L/2$ with increasing system sizes. (b) In the range-2 case, $E_{Gs}$ lies at two different fillings, $N_{f}=L/2+1$ and $L/2+2$, due to two possible root representatives located at two $N_{f}$'s with identical dimensional growths. Further, the filling fraction does not move away from these two specific fillings with increasing system sizes. (b) In the range-3 case, $E_{Gs}$ appears at three distinct fillings, $N_{f}=L/2,L/2+1$ (two of them) and $L/2+3$, respectively, which includes four ground states. In addition, the filling fraction for the ground state again does not move away from these specific $N_{f}$'s with growing $L$'s like the range-2 case.}
\label{EGsOBC}
\end{figure*}

In Figs. \ref{EGsOBC} (a-c), we compare the behaviors of $E_{Gs}$ vs $N_{f}$ involving range-1, range-2 and range-3 constraints, respectively~\cite{brighi_2023,wang_2023,East_Sreemayee}. In Fig. \ref{EGsOBC} (a), we witness that $E_{Gs}$ within the full Hilbert space lies at fillings, which constantly shift with increasing system sizes ($N_{f}=L/2+2$, $L/2+3$ and $L/2+4$ for $L=14$, $L=(16,18, 20)$ and $L=(22,24)$, respectively). This shift in the filling fraction can be comprehended by taking the distinct fragmentation structure of the range-1 case into account in terms of the conventional Dyck paths~\cite{catalan}, as we have disclosed in our previous paper. It can also be argued that the largest fragment (where lies the ground state) lies at a filling, which shifts as $L/2+a$, where $a=\sqrt{L}/2$ in the large-$L$ limit. Accordingly, the ground state filling also displays a shift even for finite $L$'s, as evident from Fig. \ref{EGsOBC} (a).

Contrary to this behavior seen in Fig. \ref{EGsOBC} (a), an identical investigation for the range-2 and range-3 constraints does not demonstrate this anomalous shift for finite system sizes, as can be seen from Figs. \ref{EGsOBC} (b-c). Also, the plot shown in Fig. \ref{larger2r3} confirms the absence of any shift in the filling fraction for the largest fragment in range-2 and range-3 cases in $L\to\infty$, which we have obtained using the root identification~\cite{East_Sreemayee,khagebdra2_fredkin,salberger2016fredkinspinchain} method and Frey-Seller sequence~\cite{FS_article}. This suggests that ground state filling (provided it is located at the largest fragment, which is in general true) will also not display any anomalous shift in $L\to\infty$ limit. In Fig. \ref{EGsOBC} (b), we also observe that the range-2 model manifests two ground states at $N_{f}=L/2+1$ and $L/2+2$, respectively. The existence of ground states at two fillings is the consequence of two possible choices of $\psi_{L}$ (either 0 or 1 and $\psi_{M}\otimes \psi_{R}$ are identical for both cases) with the same dimensional growth as explained for the largest fragment in Sec. \ref{largest}. In Fig. \ref{EGsOBC} (c), we perform an equivalent examination in the range-3 case, where the variation of $E_{Gs}$ vs $N_{f}$ reveals the presence of four ground states at $N_{f}=L/2$, $L/2+1$ (two of them lying in two distinct fragments) and $L/2+2$, respectively. One can again argue, like the range-2 case, that the appearance of four ground states at three fillings (two located at $N_{f}=L/2+1$) in this case is due to four potential alternatives of four distinct root representatives labeling four fragments with identical dimensional growth, which involve $\psi_{L}=00/01/10$ or $11$, and identical $\psi_{M}\otimes \psi_{R}$, as reasoned for the largest fragment in Sec. \ref{largest}.

Further, it can also be shown that the range-$q$ constraint facilitates an exponentially large number (in the range of constraints) of ground states appearing at $q$ number of fillings in OBCs. This is due to several alternate possibilities of root representatives (at different fillings) with the same dimensional growth of all such fragments (as clarified in Sec. \ref{largest}) incorporating the ground state. Furthermore, these potential $q$ number fillings comprise the total number of GS which is $N_{Gs}^{t}={q-1\choose 0}+{q-1\choose 1}+\cdots+{q-1\choose q-1}=2^{(q-1)}$. We further note that the lowest (root state with $\psi_{L}=0^{(q-1)}$) and highest fillings (root state with $\psi_{L}=1^{(q-1)}$) embodies precisely one ground state, while the other $(q-2)$ fillings include multiple ground states. This number can be calculated as $N_{k}={q-1 \choose k}$, where $k=1,2,\cdots,(q-2)$. In addition, since the largest fragment in the range-$q$ model emerges at a minimum filling of $N_{f}^{min}\sim (L-q)/2$, as discussed earlier, the longer range constraints assist the ground state to transpire at a comparatively low filling. All these features indicate that a single filling fraction is not adequate to capture the low-energy properties as the range of the constraints increases.

\begin{figure*}
\stackon{\includegraphics[width=0.345\linewidth,height=4.3cm]{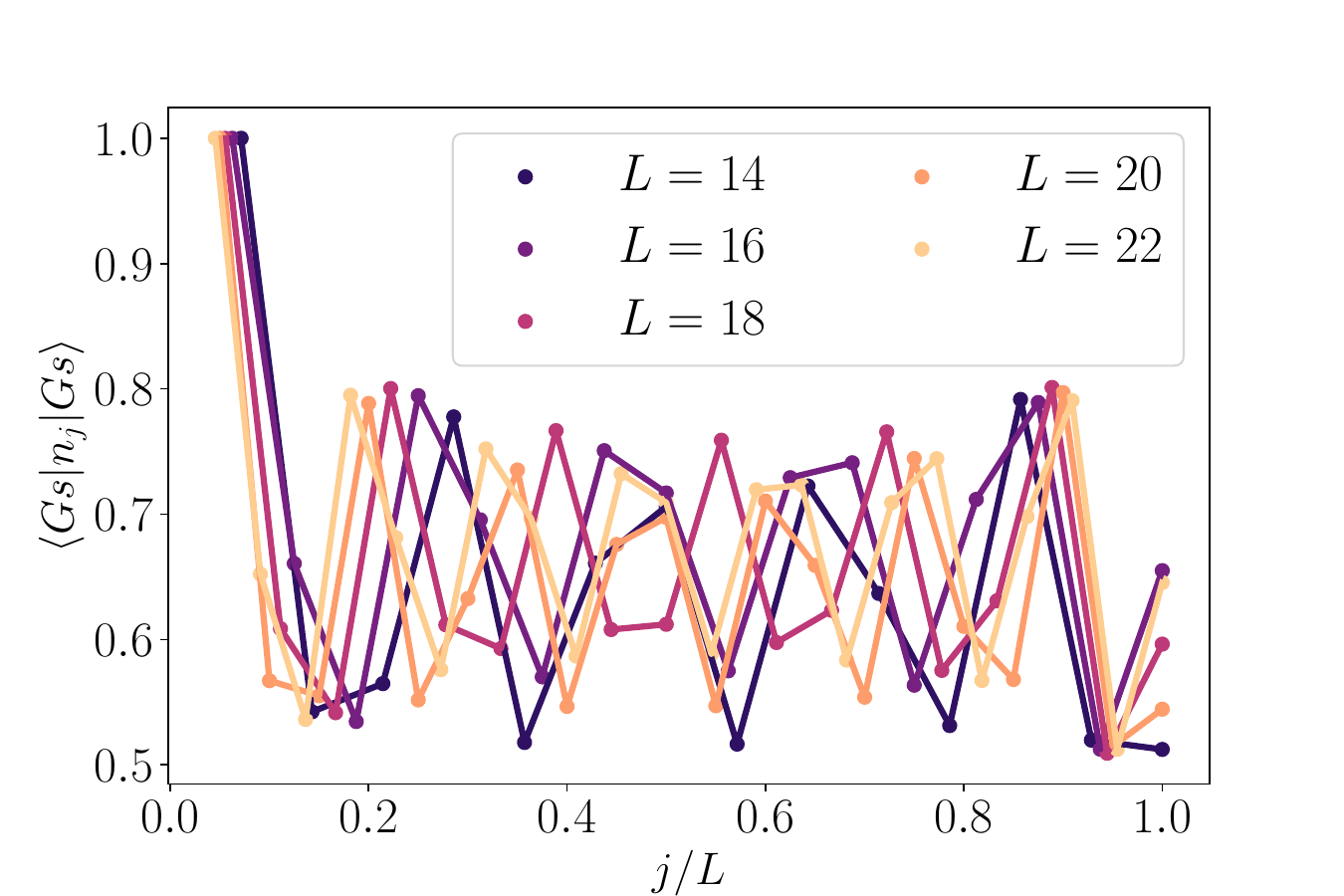}}{(a)}%
\stackon{\includegraphics[width=0.345\linewidth,height=4.3cm]{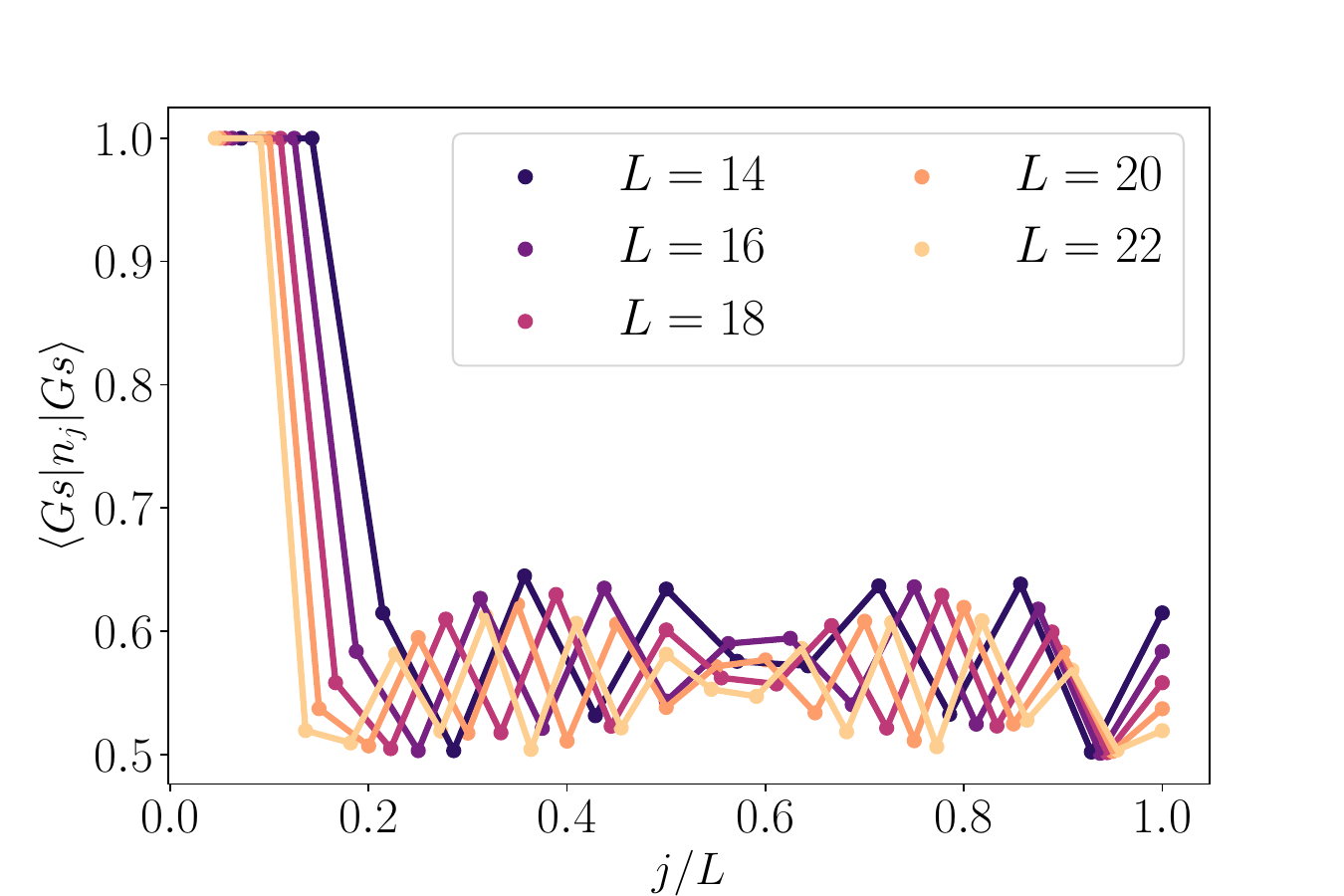}}{(b)}%
\stackon{\includegraphics[width=0.345\linewidth,height=4.3cm]{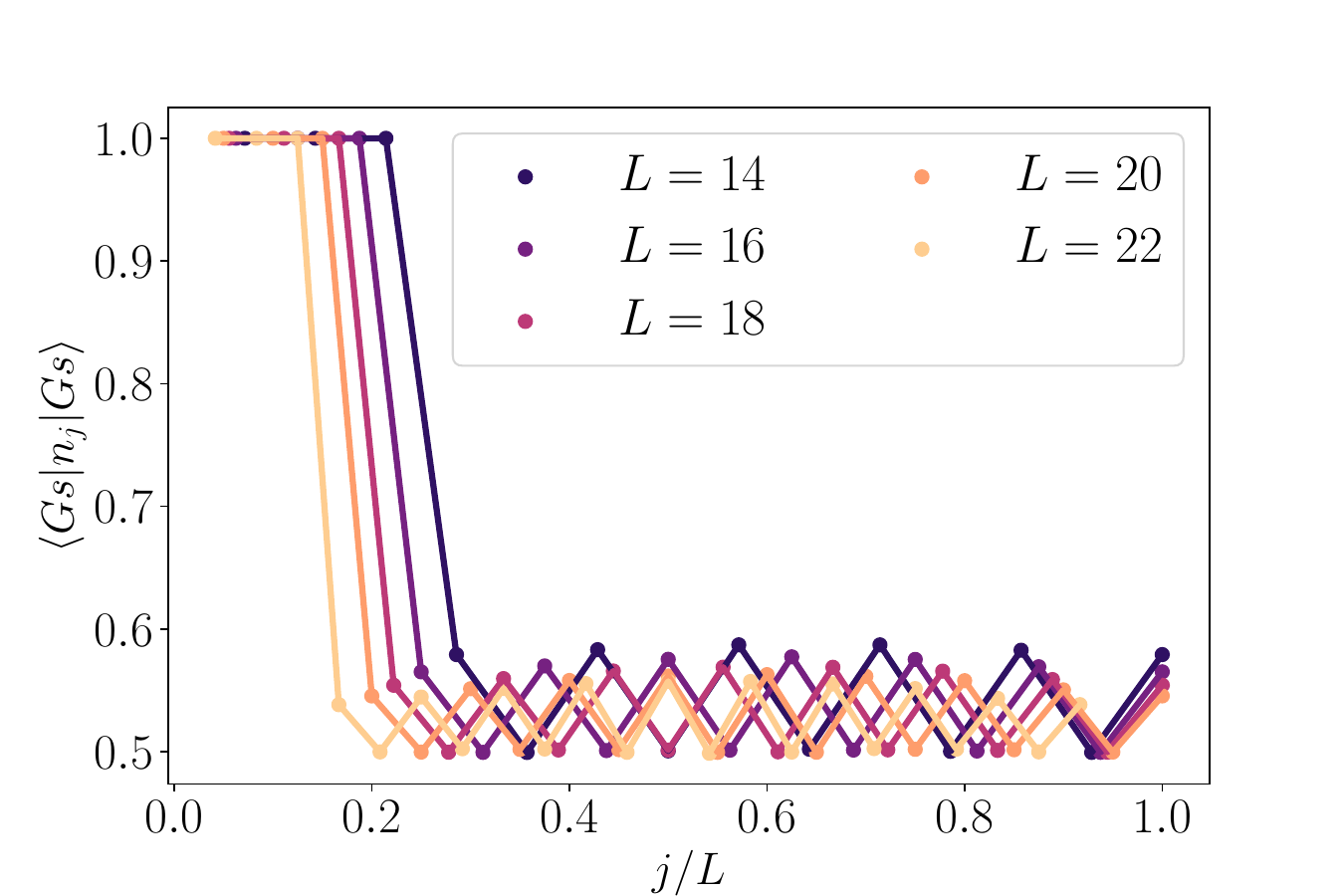}}{(c)}\\\stackon{\includegraphics[width=0.345\linewidth,height=4.3cm]{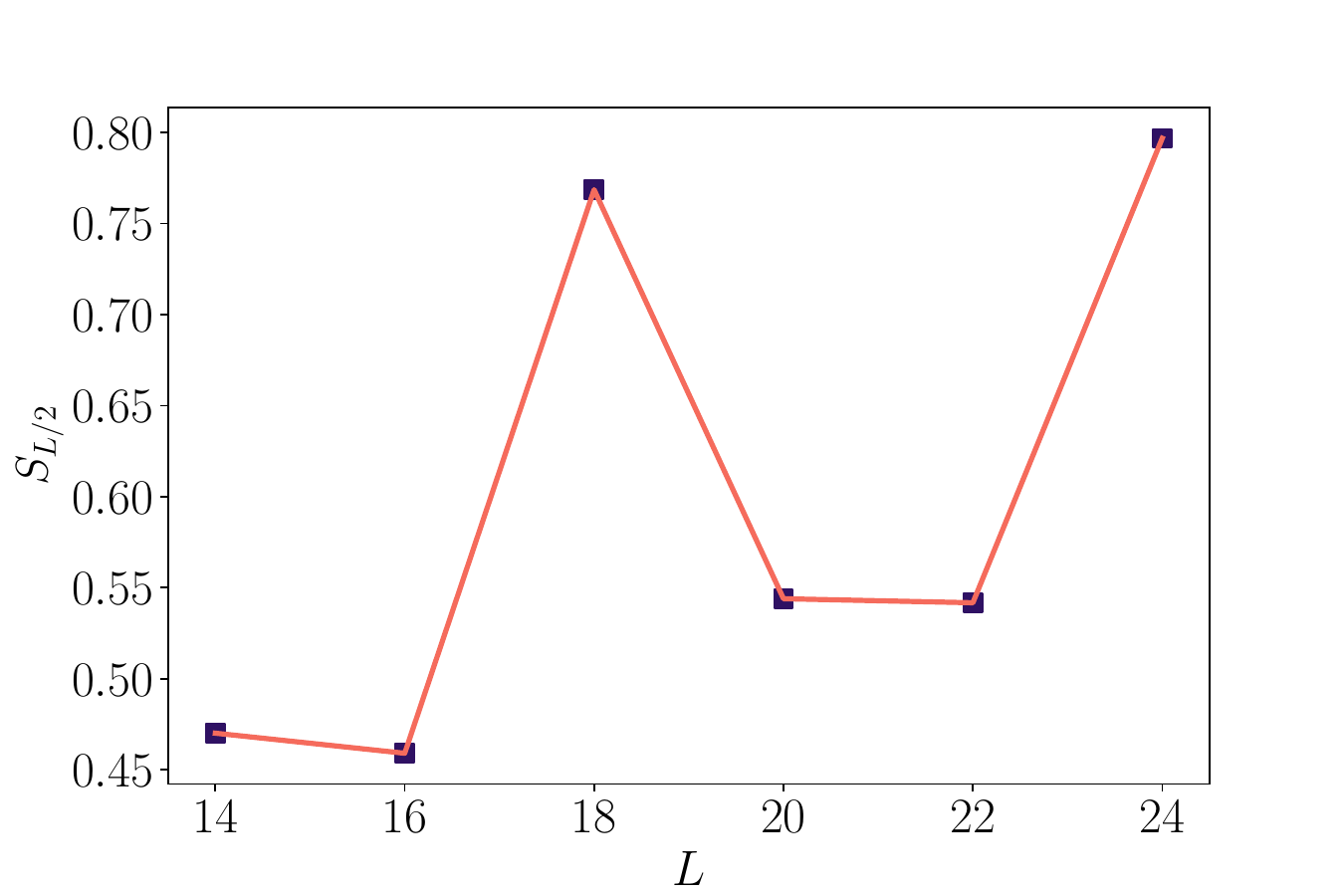}}{(d)}%
\stackon{\includegraphics[width=0.345\linewidth,height=4.3cm]{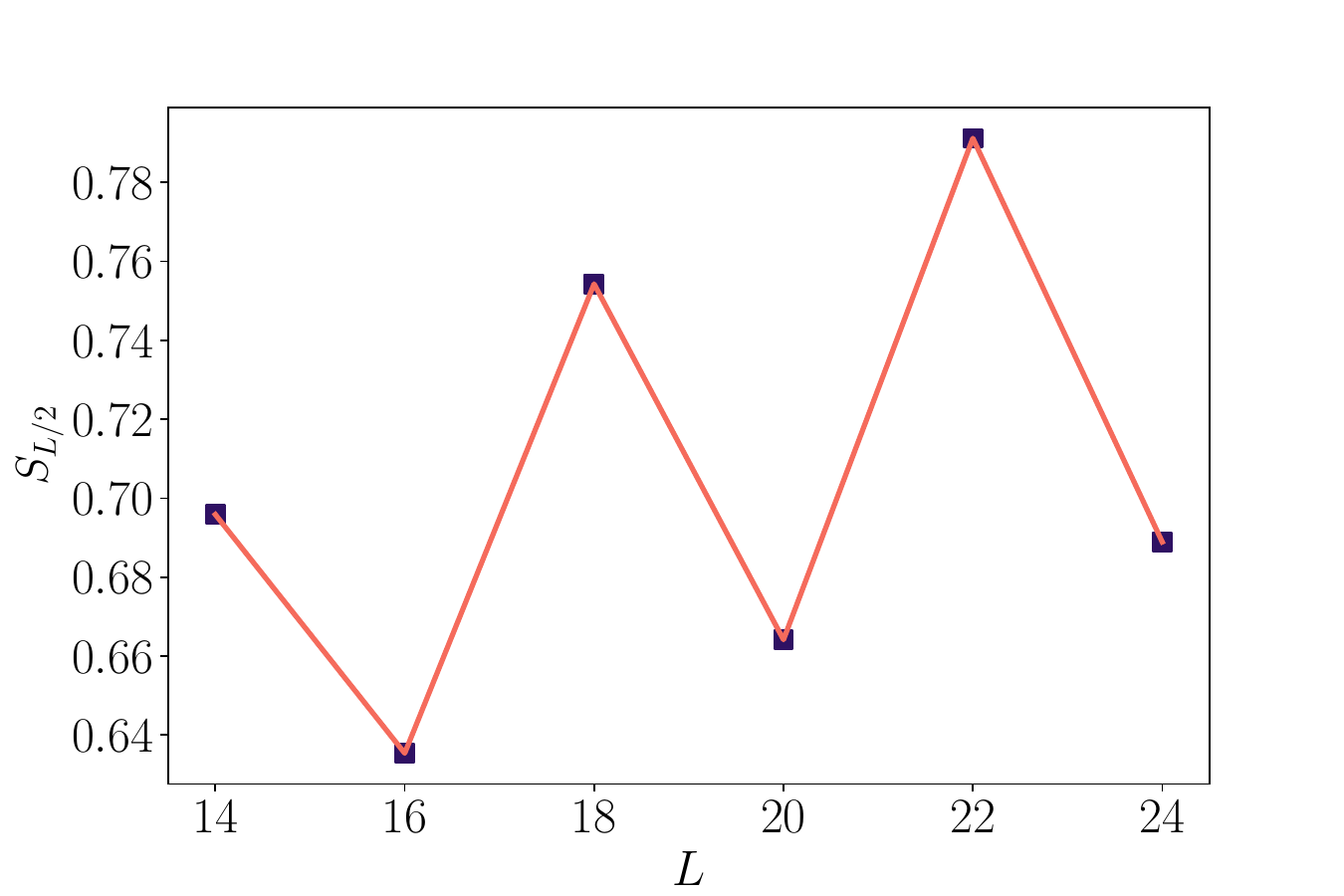}}{(e)}\stackon{\includegraphics[width=0.345\linewidth,height=4.3cm]{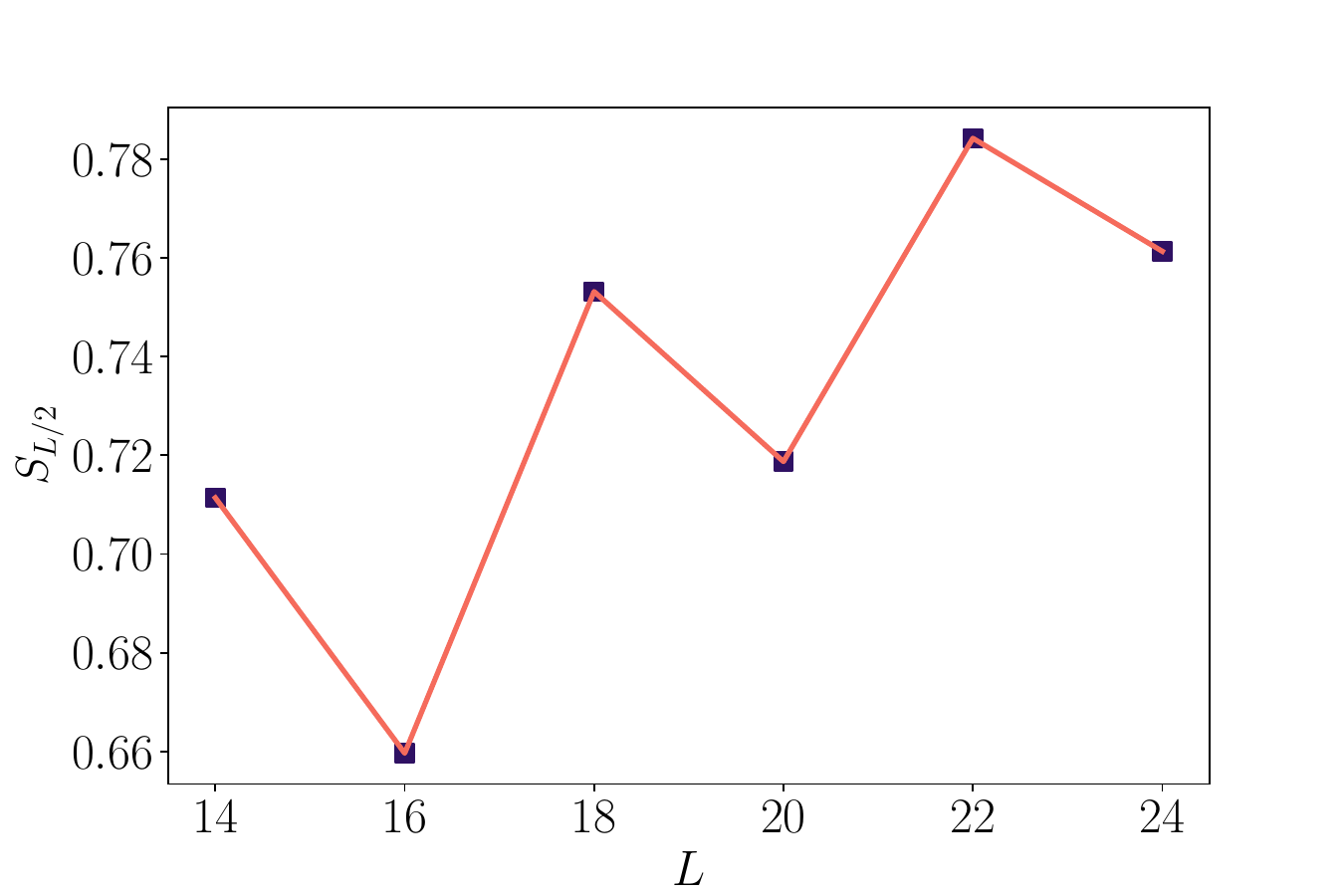}}{(f)}
\caption{(a-c) The ground state density profile for the range-1, range-2, and range-3 models, respectively, for several $L$'s in OBCs. (d-f) The variation of bipartite entanglement entropy across the central cut, $S_{L/2}$ with $L$ for the same three cases. (a) The density profile of the ground state in the range-1 case exhibits oscillations ($\la n_{j}\ra\in (0.5,0.8)$) with no regular pattern. (b) The same for the range-2 model demonstrates a more regular oscillation pattern ($\la n_{j}\ra\in (0.5,0.65)$) across the chain except for $j/L\in(0.55,0.65)$ where $\la n_{j}\ra$ abruptly becomes constant. (c) In the range-3 case, the same quantity displays the most regular oscillation pattern; nevertheless, the range of average density is significantly suppressed ($\in (0.5,0.6)$) than the former two cases. Additionally, one can also see the leftmost one, three, and four sites are frozen at $\la n_{j}\ra=1$ in Figs. (a-c), respectively, which can be explained by considering the root representatives of those fragments. (d) In the range-1 case, capturing any scaling for $S_{L/2}$ is quite difficult since the ground state filling constantly shifts with increasing $L$. (e-f) In the range-2 and range-3 cases, the scaling of $S_{L/2}$ vs $L$ seems to follow two distinct scaling behaviors depending on $L$ being $4n+2$ or $4n$. }
\label{density}
\end{figure*}

We then investigate the density profile of the ground state for range-1, range-2, and range-3 constraints, as displayed in Figs. \ref{density} (a-c), respectively, for several $L$'s in OBCs to make a comparison with the two-site East model~\cite{two_site_East_model} that has been vastly examined in the literature. In the two-site East model, the ground state exhibits a special kind of emergent super spin structure~\cite{two_site_East_model}, and it would thus be encouraging to see whether the ground state wavefunctions for this class of models showcase a similar structure. In Fig. \ref{density} (a), we note that the density profile spreads uniformly across the chain, demonstrating oscillations with no structured pattern with an average density between $0.5$ and $0.8$. Furthermore, the first site is strictly frozen at $\la n_{j}\ra=1$ due to the range-1 East constraint imposed in OBC. In Fig. \ref{density} (b), the density profile again spreads uniformly across the chain, exhibiting comparatively smooth oscillations across the chain for $j/L\in (0.2,0.55)$ and $\in (0.65,1)$, respectively, (approximately) while $\la n_{j}\ra$ almost remains constant at a value of $0.6$ at middle two sites in the bulk of the chain. Furthermore, the range of density oscillations lies between 0.5 and 0.65. This is less than the values in the range-1 case, along with the first three sites being frozen at $\la n_{j}\ra=1$. This can be easily understood by considering the root structure representing the fragment. In Fig. \ref{density} (c), the density profile again displays a uniform distribution with a perfect alternate density oscillation pattern lying in the range $\in( 0.5,0.6)$, which seems to be more suppressed than the other two cases. In this specific case, the first four sites are frozen as a consequence of the root structure representing the fragment where the ground state is located in the range-3 case. Nevertheless, we have not noticed any emergence of superspin structure in the ground state profile in none of these cases. 

\begin{figure*}
\subfigure[]{\includegraphics[width=0.335\hsize,height=4.4cm]{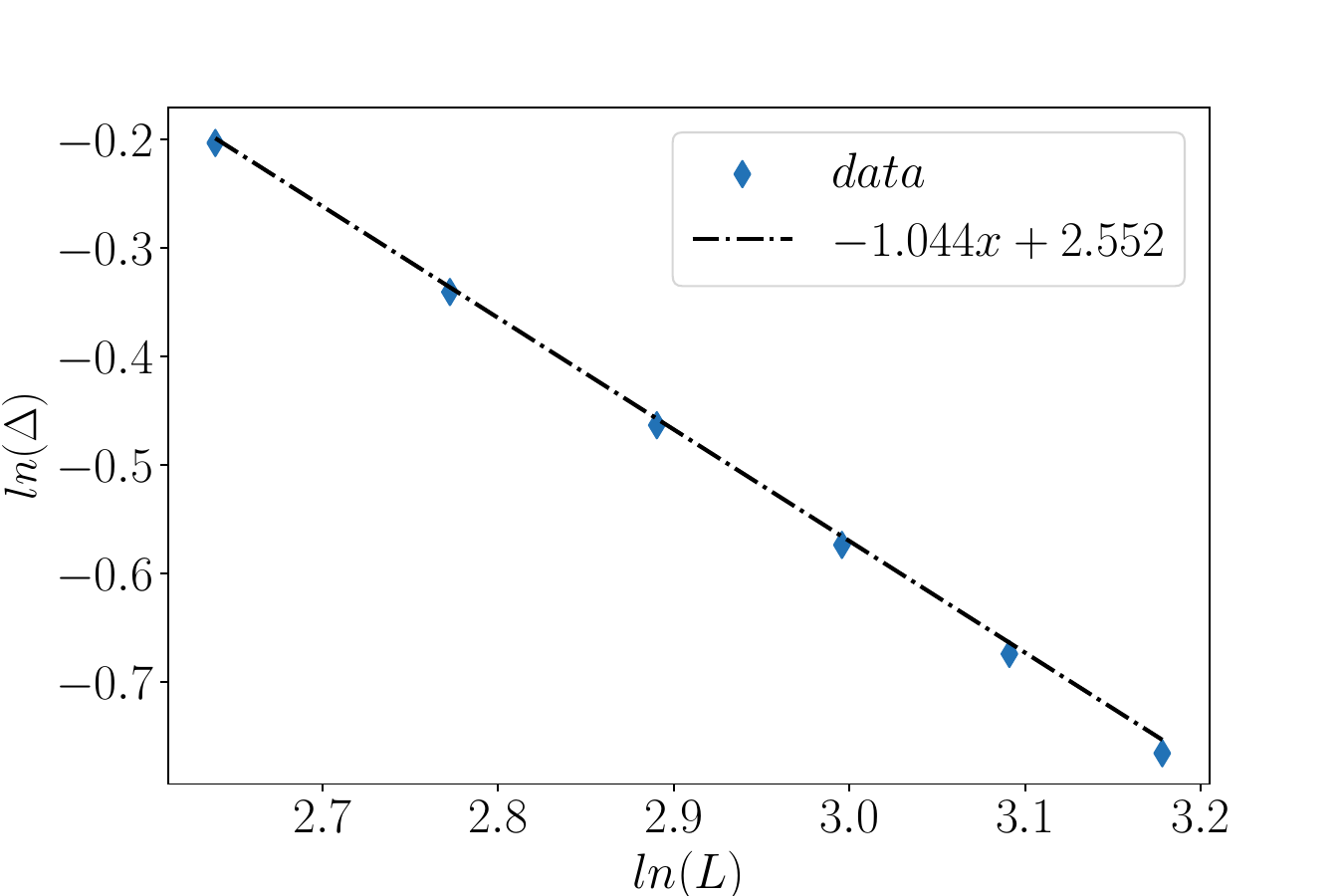}}%
\subfigure[]{\includegraphics[width=0.335\hsize,height=4.4cm]{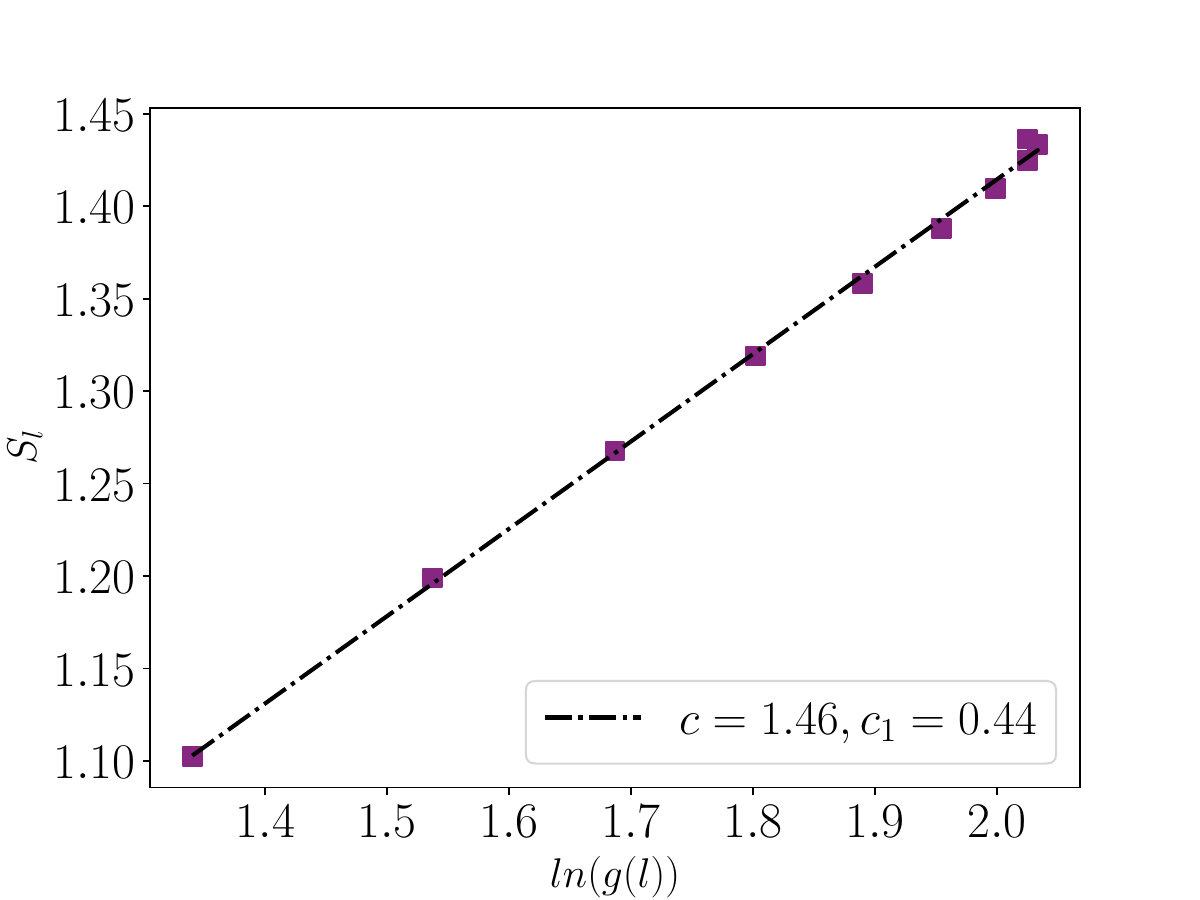}}%
\subfigure[]{\includegraphics[width=0.335\hsize,height=4.4cm]{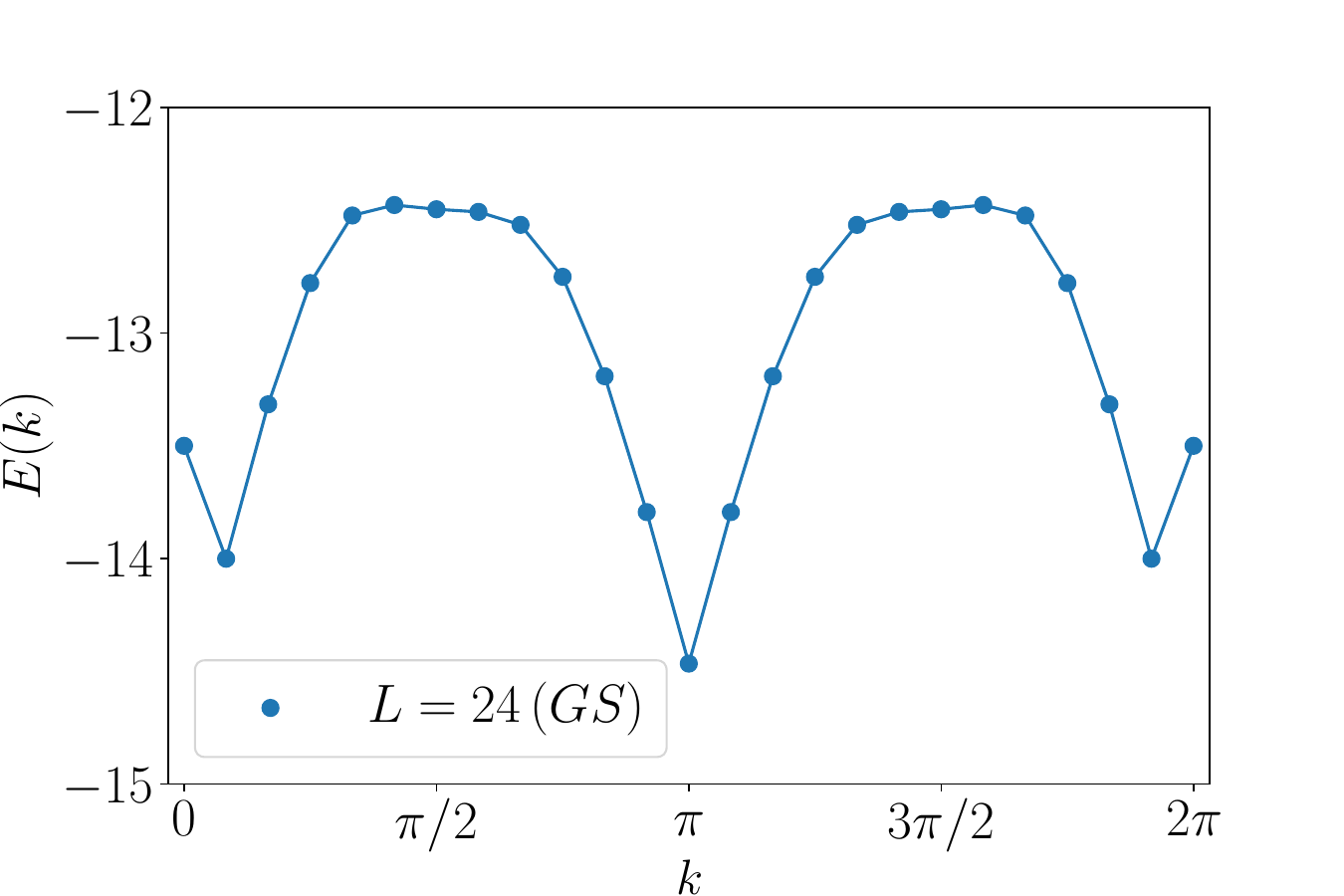}}
\caption{(a) The energy gap ($\Delta$) vs $L$ (both in log scale) for the range-2 East model at $N_{f}=L/2+1$ for $L=14-24$ in PBCs, revealing $\Delta$ scales as $\sim1/L$ in numerical fitting. This also implies that the dynamical exponent $z\sim1$. (b) $S_{l}$ for several subsystem cuts, $l$ as a function of $ln(g(l))$ is fitted linearly, disclosing the central charge of the CFT being $c=1.46$. (c) The dispersion $E(k)$ vs $k$ within the same particle-number symmetric sector ($N_{f}=L/2+1$) with $L=24$ in PBCs. The dispersion exhibits linear $k$ behavior near three $k$-points in the BZ (one linear $k$-mode at $k=\pi$ controlling the low-energy properties).}
\label{gsr2_PBC}
\end{figure*}

In Fig. \ref{density} (d-f), we plot the variation of $S_{L/2}$ vs $L$ for range-1, range-2, and range-3 models to understand the scaling of ground state entanglement entropy. It is hard to perform a scaling of $S_{L/2}$ for the range-1 case since the ground state filling continually shifts with $L$ as a consequence of Dyck combinatorics. Nevertheless, $S_{L/2}$ for the other two cases tries to display two distinct scaling behaviors depending on $L$ being $4n$ or $4n+2$, as can be seen in Fig. \ref{density} (e-f). In order to obtain further insight and smooth scaling behavior, we thereafter repeat a similar analysis for the range-2 case in PBCs (One can also repeat this analysis for all classes of models, which will not change the scaling properties as we have noted (plots not shown here)). At first, we observe that the range-2 constraint demonstrates a non-degenerate ground state in PBCs at $N_{f}=L/2+1$ for $L=14-24$ (plot not shown here). Next, our numerical analysis suggests that the next excited state in the full Hilbert space is the ground state located at $N_{f}=L/2+2$, thus implying that the low-energy properties of this model, even in PBCs, cannot be described while considering a single filling fraction. Furthermore, since this model conserves the total particle number, we, therefore, consider the ground state filling ($N_{f}=L/2+1$) to understand the low-energy dispersion and entanglement scaling of the ground state in this specific scenario (which also allows us to capture the transport properties at this particular filling as discussed later in the paper). Our investigation reveals that the energy gap ($\Delta$) between the $E_{Gs}$ and $E_{ex}$ within this sector scales as $1/L^{z}$ with dynamical exponent, $z=1.044$~\cite{Sachdev_2011}, as displayed in Fig. \ref{gsr2_PBC} (a), thus revealing the signature of a 1+1 dimensional conformal field theory. Furthermore, as the dynamical exponent $z\sim1$, one can also expect the von Neumann entanglement entropy for a subsystem size $l$ to obey the following finite-size scaling ansatz~\cite{PasqualeCalabrese_2004} as
\bea
S(l)=\frac{c}{3}ln(g(l))+c_{1}, 
\label{vonneu}
\eea
where $g(l)=\frac{L}{\pi}{\rm sin}(\frac{\pi l}{L})$, $c$ is the central charge of CFT and $c_{1}$ is a constant. In doing so, we consider $S(l)$ for various subsystem cuts, $l$ for $L=24$, plot it as a function of $ln(g(l))$, and then fit it linearly, as exhibited in Fig. \ref{gsr2_PBC} (b), which indicates $c=1.46\pm0.015$. Afterward, we also study the energy dispersion $E(k)$ vs. $k$ to understand the low-energy dispersion within the same sector as demonstrated in Fig. \ref{gsr2_PBC} (c). This reveals linear $k$ dispersion at three $k$-points in the Brillouin zone (low-lying excitation dominated by one linear-$k$ mode at $k=\pi$), hence supporting the $log$-scaling of $S_{l}$ shown in Fig. \ref{gsr2_PBC} (b).
 
\section{Bulk-boundary behavior of autocorrelation functions and freezing transition}
\label{bulk-boundary}
\begin{figure*}
\stackon{\includegraphics[width=0.33\linewidth,height=4.8cm]{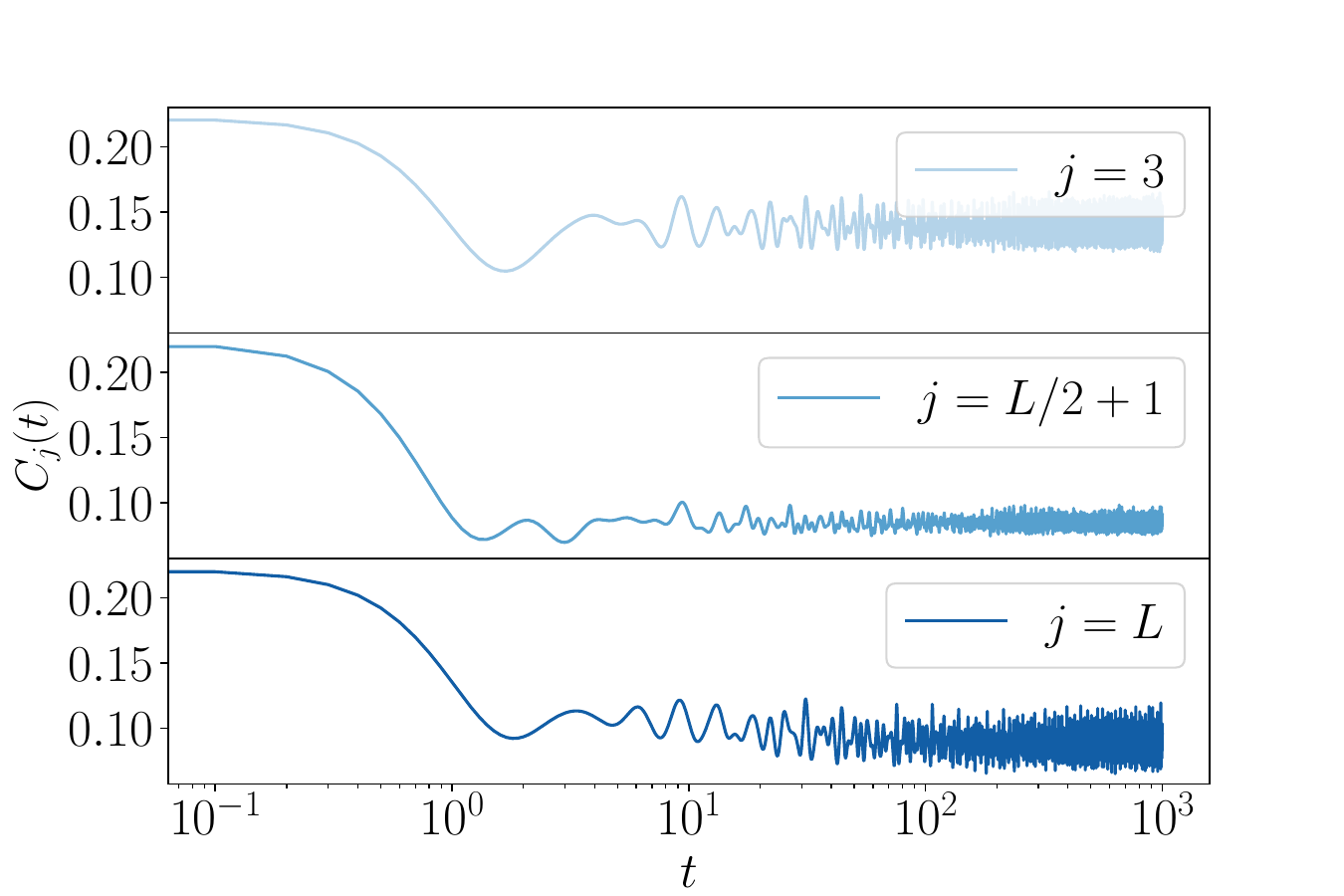}}{(a)}%
\stackon{\includegraphics[width=0.33\linewidth,height=4.8cm]{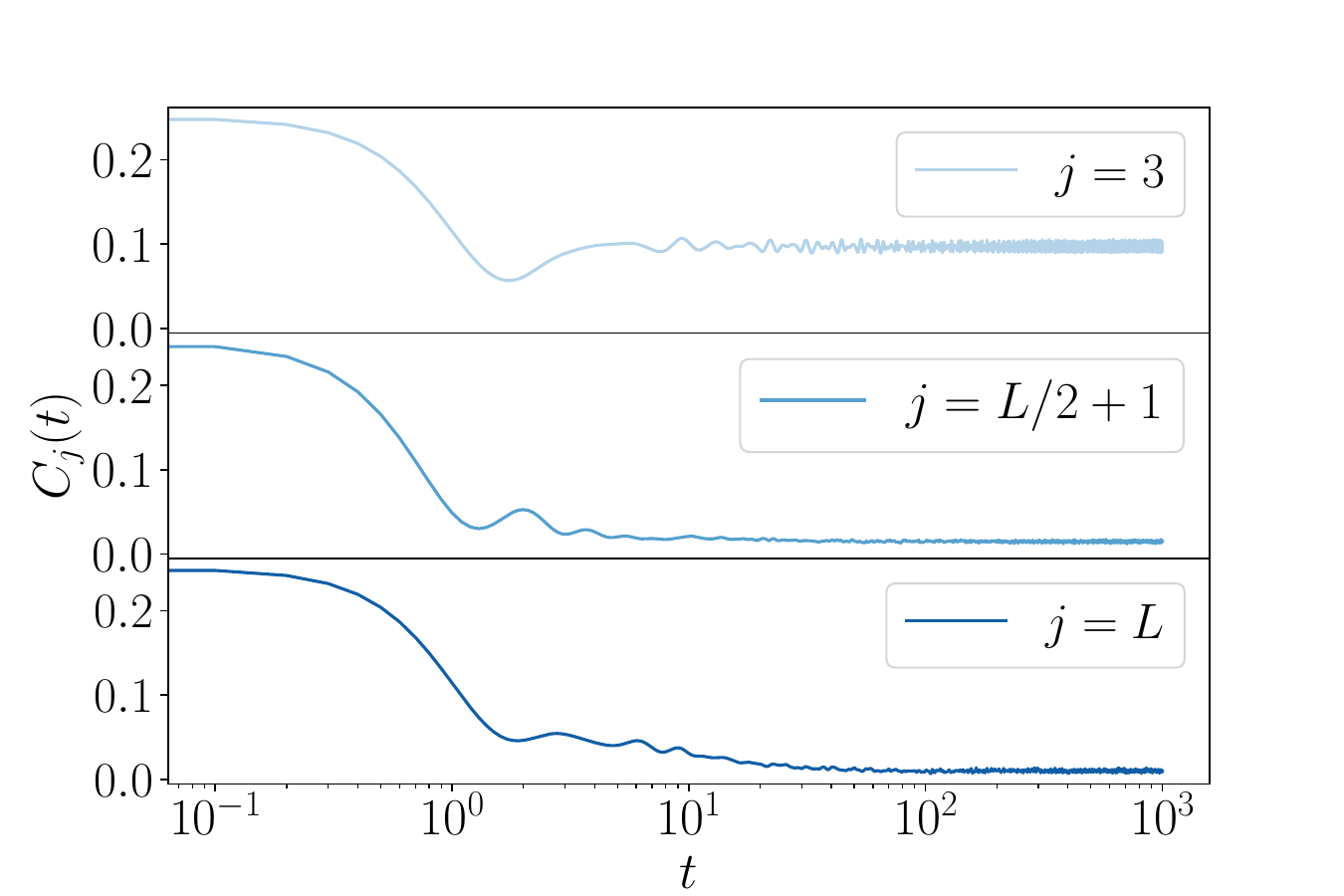}}{(b)}%
\stackon{\includegraphics[width=0.33\linewidth,height=4.8cm]{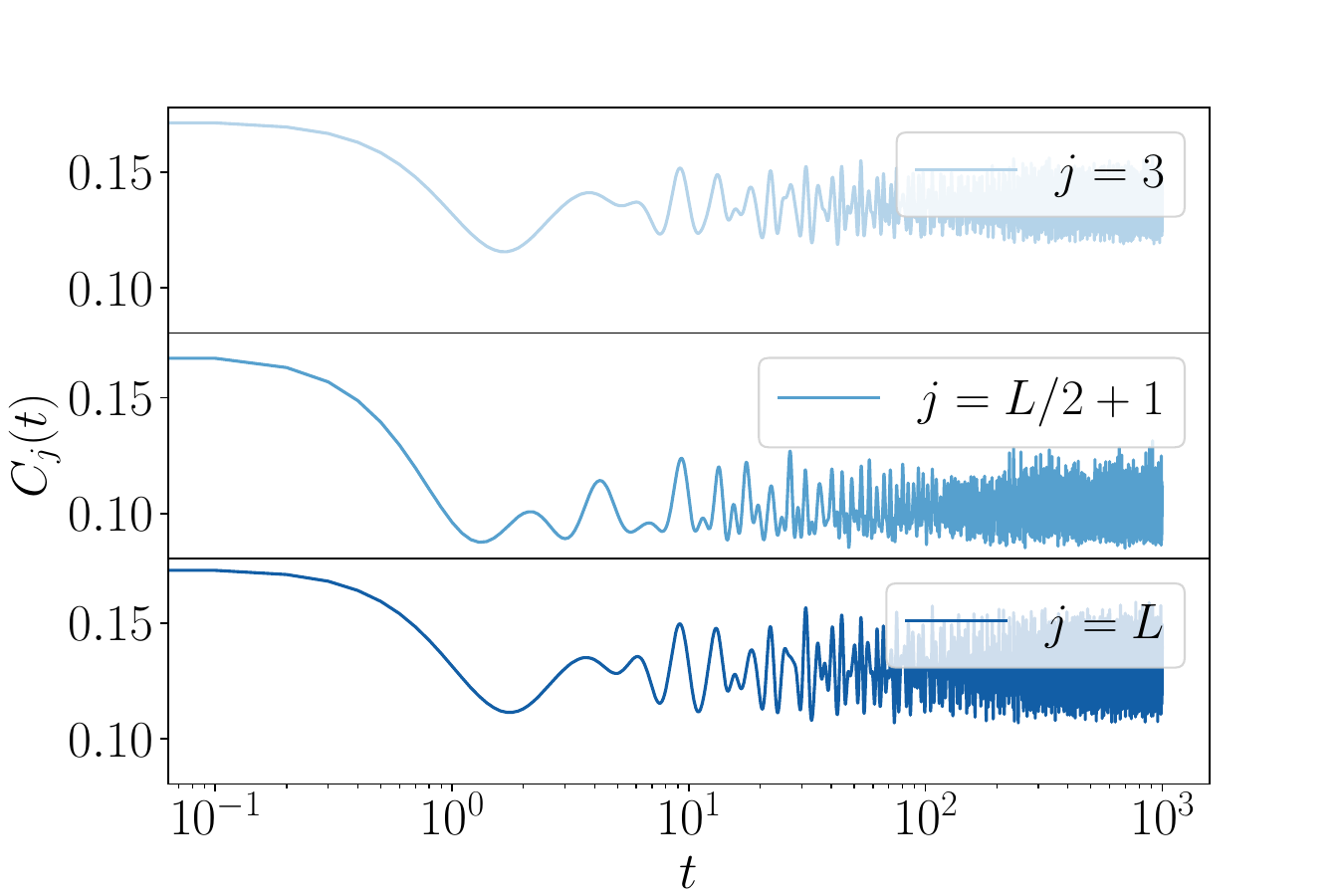}}{(c)}
\caption{(a-c) Plots showing $C_{j}(t)$ vs $t$ at the leftmost active site, $j=3$, in the bulk, i.e., $j=L/2+1$ and the rightmost boundary, $j=L$ for $L=18$ and $N_{f}=L/3$ (freezing transition), $N_{f}=L/2$ (weakly fragmented) and $N_{f}=L/3-2$ (strongly fragmented), respectively. (a): At the critical filling, $C_{j}$ at two boundaries and in the bulk saturate at finite values at long times, suggesting the non-thermal behavior of the autocorrelators. Also, we observe that the saturation value is the largest at the leftmost boundary, lower than the former one at the rightmost boundary, and finally, the lowest in the bulk of the chain. (b): In the weakly fragmented case, the saturation value is finite at the leftmost active site, while the bulk and rightmost boundary autocorrelators demonstrate nearly thermal behavior. (c): In the strongly fragmented case, all three autocorrelation functions exhibit finite saturation values similar to Fig. (a), revealing the non-thermal behaviors. In addition, the saturation value of the bulk autocorrelation is the lowest, while both the boundary saturates at values that are almost comparable to one another but larger than the bulk case.  }
\label{autor2}
\end{figure*}

We now focus on how the long-time bulk and boundary autocorrelators behave as we tune the filling fraction appropriately~\cite{hart2023exact,Moudgalya_review_2022,Deepak_HSF,East_Sreemayee,sala_ergo_2020}. It has been noted earlier that the long-time saturation value of boundary autocorrelators shows localized behavior in the presence of partially thermalizing bulk in strongly fragmented systems~\cite{hart2023exact,Moudgalya_review_2022,Deepak_HSF,East_Sreemayee,sala_ergo_2020}. Hence, an analogous investigation in a constrained system involving strong-to-weak fragmentation transition~\cite{Morningstar_2020,East_Sreemayee,wang_2023} with strongly broken inversion symmetry is capable of furnishing more fascinating characteristics. In doing so, we consider the range-2 model in OBCs and inspect the behaviors of the unequal time autocorrelation function starting from a typical random initial state, which is given below
\bea
C_{j}(t)=\la \psi|n_{j}(t)n_{j}(0)|\psi\ra-\la \psi|n_{j}(t)|\psi\ra\la \psi|n_{j}(0)|\psi\ra.
\eea
In Figs. \ref{autor2} (a-c), we scrutinize $C_{j}(t)$ at $N_{f}=L/3$ (at freezing transition), $N_{f}=L/2$ (weakly fragmented) and $N_{f}=L/3-2$ (strongly fragmented), respectively, for $L=18$. Furthermore, for each case, we investigate the same quantity at three sites: the leftmost active boundary ($j=3$), bulk ($j=L/2+1$), and rightmost boundary ($j=L$), as illustrated in Figs. \ref{autor2} (a-c). In Fig. \ref{autor2} (a) (at the freezing transition point), we note that $C_{3}(t)$ (leftmost boundary) saturates at a finite value in the long-time limit. This indicates non-thermal behavior at this boundary, which is the hallmark of fragmented systems~\cite{hart2023exact,Moudgalya_review_2022,Deepak_HSF,East_Sreemayee,sala_ergo_2020}. On the other hand, the saturation value of the autocorrelation function at the rightmost boundary, $C_{L}(\infty)$ appears to be also finite yet lower than that of $C_{3}(\infty)$, thus revealing the signature of lack of thermalization in the long-time limit. In addition, the bulk autocorrelation function ($C_{L/2+1}(t)$) also demonstrates a non-zero saturation value; nevertheless, it is much lower than the leftmost boundary and slightly less than the rightmost boundary. In Fig. \ref{autor2} (c), a similar investigation for $N_{f}=L/3-2$, i.e., strongly fragmented side of the Hilbert space, unveils similar characteristics, i.e., all three correlators saturate at a finite value in $t\to\infty$ limit. This again implies the absence of thermalization in the bulk and at the boundaries for the system size under consideration. We also note that the difference between saturation values between the active left and rightmost boundaries is rather small compared to that observed at critical filling. This seems to be an interesting trait, revealing the asymmetry between the two active boundaries in this inversion-symmetry broken model diminishes as we approach the strongly fragmented part of the system.
Contrary to these two behaviors, the same quantities disclose a radically distinct feature for the weakly fragmented part of the Hilbert space, as shown in Fig. \ref{autor2} (b). In this case, we see that the leftmost boundary autocorrelation function saturates at a finite value; this again reveals non-thermal behavior. However, the bulk and rightmost boundary almost approach close to the thermal value. This points towards the signature of violation of inversion symmetry becoming more prominent as one moves towards the weakly fragmented side of the Hilbert space. 

\begin{figure}
\subfigure[]{\includegraphics[width=0.8\hsize]{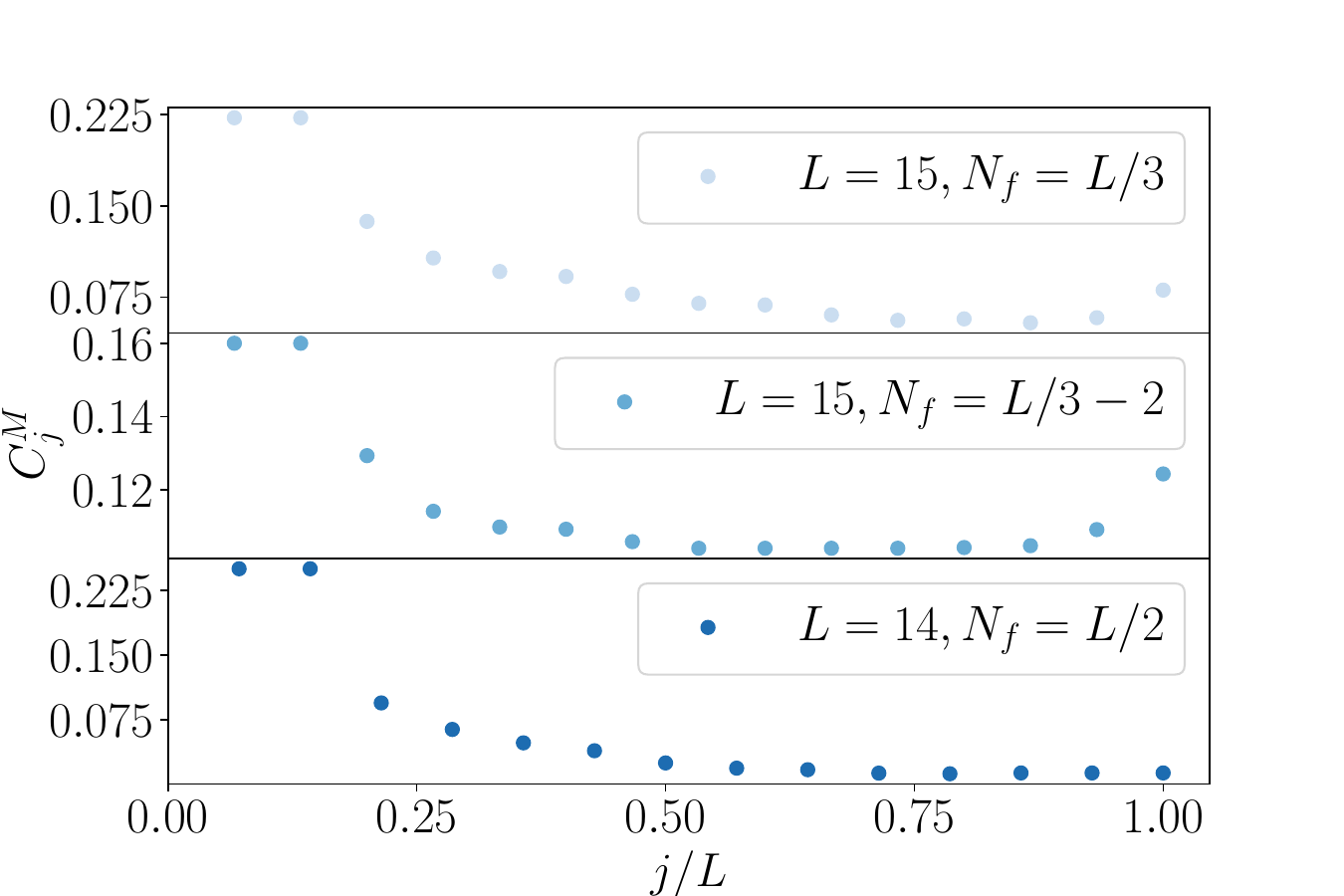}}\\
\subfigure[]{\includegraphics[width=\columnwidth,height=4.8cm]{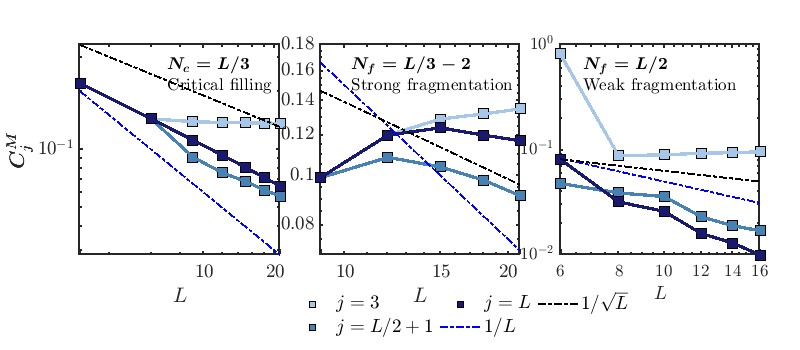}}
\caption{(a) The Mazur-bound predicted saturation value, $C_{j}^{M}$ vs $j/L$ for $(L,N_{f})=(15,L/3)$, $(15,L/3-2)$ and $(14,L/2)$. (b) The system size dependence of the lower-bound of the saturation value at the critical filling, $N_{f}=L/3$ (left panel), the strongly fragmented part of the Hilbert space,i.e.,  at $N_{f}=L/3-2$ (middle panel) and the weakly fragmented part of the Hilbert space at $N_{f}=L/2$ (right panel). (a): The behaviors of $C_{j}^{M}$ across the chain are in perfect agreement with those observed in Fig. \ref{autor2}. 
(b): In all cases, the saturation value indicates a completely non-uniform profile across the left edge of the chain, almost remaining constant for large enough $L$'s. Left panel: At the critical filling, the bulk and boundary saturation values diminish with $L$, following approximately the $1/\sqrt{L}$ guiding line. Middle panel: At the strongly fragmented side of the Hilbert space, the saturation value inside the bulk and rightmost boundary decays much more slowly with $L$, even much slower than $1/\sqrt{L}$. Right panel: In this case, the bulk and rightmost boundary decay almost follows the $1/L$ guiding line for large $L$, indicating typical thermal behavior. }
\label{mazurL}
\end{figure}

To get further insights into these anomalous properties of autocorrections in $L\to\infty$ limit, we probe the lower bound of the infinite-time saturation value utilizing the Mazur-Suzuki inequality~\cite{MAZUR1969533,SUZUKI1971277}at different filling fractions, as illustrated in Fig. \ref{mazurL} (a). The Mazur-Suzuki inequality in a fragmented system after incorporating the projectors of exponentially many disjoint sectors as non-local conserved quantities~\cite{sala_dipole} can be recast as
\bea
C^{M}_{j}\geq\frac{1}{D_{L}}\sum_{k}\frac{\left({\rm Tr}(P_{k}(n_{j}-\la n_{j}\ra)P_{k})\right)^2}{D_{k}},
\label{Mazur}
\eea
where $P_{k}$'s are the projectors of disjoint classical fragments, $D_{k}$'s and $D_{L}$ denote the dimensions of classical fragments and the dimension of particle-number resolved sector, respectively. Utilizing Eq. \eqref{Mazur}, we then compute $C_{j}^{M}$ for three different cases, $N_{f}=L/3$ (freezing transition), $N_{f}=L/3-2$ (strongly fragmented) and $N_{f}=L/2$ (weakly fragmented) for $L=15$, $L=15$ and $L=14$, respectively, as depicted in Fig. \ref{mazurL} (a). Precisely at the critical filling fraction \ref{mazurL} (a) (top panel), we notice that the Mazur-bound predicted saturation value exhibits a completely inhomogeneous profile, where $C_{j}^{M}$ gradually decreases as one moves away from the leftmost site to the bulk and afterward slightly increases near the rightmost edge of the chain (all of them still exhibiting non-thermal saturation value). However, the $C_{j}^{M}$ on the rightmost boundary is still lower than that of the leftmost boundary, as we have already seen for the case of exact time-evolution of the autocorrelation function illustrated in Fig. \ref{autor2} (a). Next, an identical analysis for the strongly fragmented part ($N_{f}=L/3-2$) (middle panel) showcases a behavior quite similar to the first case. However, the saturation value on the rightmost boundary is much larger compared to the first case, further being quite close to that of the active site on the left edge of the chain. This also validates our observation discussed in Fig. \ref{autor2} (c). At last, an equivalent investigation for the weakly fragmented Hilbert space ($N_{f}=L/2$, bottom panel) displays that $C_{j}^{M}$ slowly decreases from the leftmost active edge of the chain and saturates close to a thermal value, which is again well agreement with the behavior observed in Fig. \ref{autor2} (b). These rich features of inhomogeneous bulk and boundary saturation profile as a combined consequence of freezing transition and lack of inversion symmetry have been hardly reported in any other models exhibiting HSF until now, to the best of our knowledge, which opens up more possibilities for future analytical exploration.

Next, we turn to investigate the system size dependence of the saturation value of these correlators predicted by Mazur-Suzuki inequality~\cite{MAZUR1969533,SUZUKI1971277} in order to understand the thermalization properties in $L\to\infty$ limit, as depicted in Fig. \ref{mazurL} (b). In doing so, we investigate three cases: the critical filling at $N_{f}=L/3$ (left panel), the strongly fragmented part of the Hilbert space at $N_{f}=L/2-3$ (middle panel), and weakly fragmented part of the Hilbert space at $N_{f}=L/2$ (right panel) in Fig. \ref{mazurL} (b). At the critical filling, we observe that the leftmost active boundary ($j=3$) does not decrease with $L$, indicating a completely non-uniform profile near the left edge of the chain. However, $C_{j}^{M}$ on the rightmost boundary ($j=L$) and inside the bulk ($j=L/2+1$) decrease with increasing $L$. Moreover, for both cases, the dependence is almost parallel to $1/\sqrt{L}$ line~\cite{sala_ergo_2020,hart2023exact}; this implies anomalous thermalization in this specific case, as can be witnessed from the left panel of Fig. \ref{mazurL} (b). In this context, it should be noted that the typical $L$ dependence of long-time saturation value of autocorrelators in $U(1)$-conserving thermalizing systems is $1/L$, yet this is not the behavior noticed for the bulk and rightmost boundary at the critical filling.
We subsequently perform a similar investigation for the strongly fragmented side of the Hilbert space at $n_{f}=1/3-2/L$, which again demonstrates a non-uniform profile at the left edge of the chain, as shown in the middle panel of Fig. \ref{mazurL} (b). In addition, the saturation value in the bulk and at the rightmost edge does not decrease significantly with $L$ in this specific case. This reveals a possible signature of bulk localization. Finally, we examine the same for the weakly fragmented case, as depicted in the right panel of the same plot. Here, the saturation value at the leftmost edge again saturates with $L$ like the previous two cases. Nevertheless, the dependence of bulk and rightmost boundary long-time autocorrelators closely follow the $1/L$ line, indicating typical thermalization properties in the weakly fragmented part of the Hilbert space~\cite{Moudgalya_review_2022}. These behaviors, therefore, demand a deeper analytical investigation as our analysis is limited to finite system sizes.

In addition, we also note that all the models falling in this class demonstrate identical filling-dependent intriguing inhomogeneous profiles across the chain, which has been further validated using Mazur inequality~\cite{MAZUR1969533,SUZUKI1971277}in the range-1 and range-3 cases (the behaviors in the range-1 case are also shown in Appendix-\ref{mazurmore} along with the Mazur-predicted saturation profile across the chain).

\section{Infinite-temperature transport}
\label{inftran}
We now probe the infinite-temperature transport~\cite{transport_review} properties in this class of models by examining the relaxation behavior of the correlation function as given below
\bea
\tilde{C}(r,t)=\frac{1}{2^{L}}{ \rm Tr} \left(S_{j+r}(t)S_{j}(0)\right),
\label{transp1}
\eea
where $\rm Tr$ refers to the sum over all eigenstates of the given Hamiltonian, $S_{j}=(n_{j}-1/2)$, $S_{j}(t)=e^{iHt}S_{j}e^{-iHt}$, and $r$ is the distance between two sites. In general, $\tilde{C}(r,t)$ follows the following scaling relation
\bea
\tilde{C}(r,t)\sim \frac{1}{t^{1/z}}f\left(r/t^{1/z}\right), \label{spincorr}
\eea
where $z$ is the dynamical exponent, and $f$ represents a single-parameter scaling function. While $z=2$~\cite{transport_review} marks normal diffusion, $z>2$ represents subdiffusive transport properties~\cite{tracer,singh_2021,Kadhikari,pal_hydro}, which have been reported in several kinetically-constraint models. In addition, $\tilde{C}(0,t)$ takes a hydrodynamic tail as $t^{-\alpha}=t^{-1/z}$ in $t\to\infty$~\cite{transport_review}. Here, the primary focus would be to examine the above by employing the concept of quantum dynamical typicality (QDT)~\cite{QDT1,QDT2}. This would allow us to investigate relaxation behaviors beyond the numerical limitations imposed by exact diagonalization (ED) to a certain extent. The QDT~\cite{QDT1,QDT2} facilitates us to approximate Eq. \ref{transp1} as an average within a random typical state as
\bea
\tilde{C}(r,t)\simeq\la\psi|S_{j+r}(t)S_{j}|\psi\ra=
\la \psi(t)|S_{j+r}|\psi'(t)\ra,
\label{finaltra}
\eea
where $\ket{\psi'}=S_{j}\ket{\psi}$, and $\ket{\psi}=\sum_{l}(a_{l}+ib_{l})\ket{l}$, where $\ket{l}$ denotes the Fock space basis states allowed within a particle-number resolved sector, $a_{l}$ and $b_{l}$ stand for random numbers chosen from a Gaussian distribution with mean zero along with the normalization condition $\sum_{l}(|a_{l}|^2+|b_{l}|^2)=1$. This approximation thus enables us to simulate large enough system sizes that go beyond the ED analysis.

\begin{figure*}
\begin{comment}
\stackon{\includegraphics[width=0.52\hsize,height=5.3cm]{corr_freezing_transition_r1_stretched_expo_fit1.pdf}}{(a)}\stackon{\includegraphics[width=0.52\hsize,height=5.3cm]{corr_largest_fragment_r1.pdf}}{(b)}\\
\vspace{-0.1cm}
\end{comment}
\includegraphics[width=0.99\hsize]{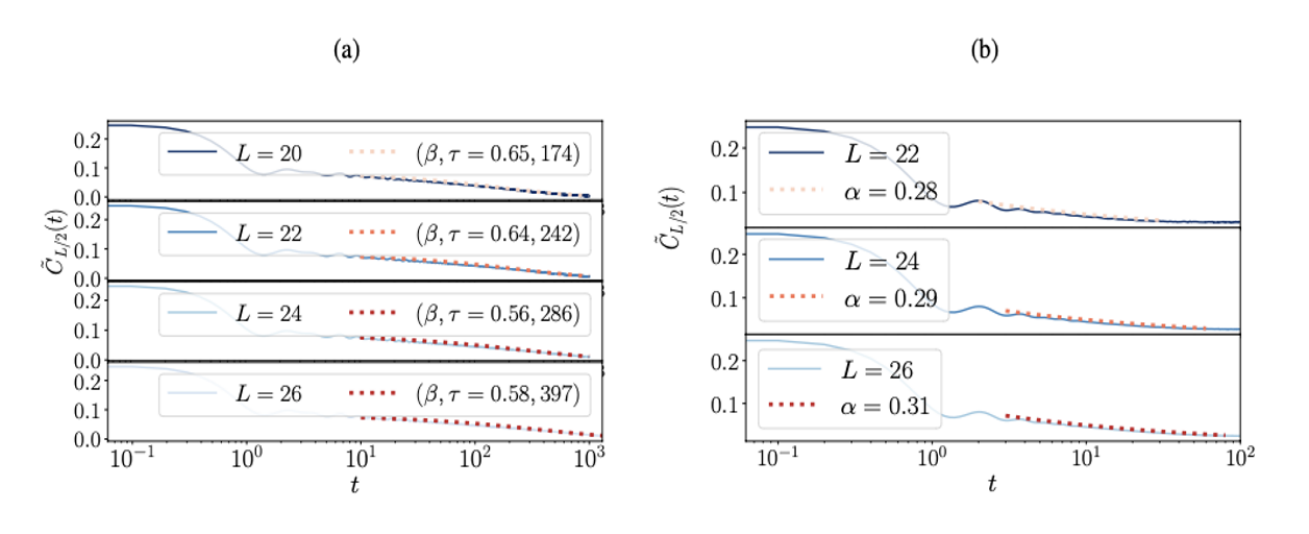}

\caption{(a-b) Plots showing $\tilde{C}_{L/2}(t)$ vs $t$ for the model with range-1 constraint for different systems sizes at critical filling fractions, $N_{f}=L/2$ and above the critical filling (weakly fragmented Hilbert space), where the system manifests the largest fragment, i.e., at $N_{f}=L/2+3$ for $L=22,24$ and $26$. (a): The infinite-temperature correlation function demonstrates size-stretched exponential relaxation (SSER) with $(\beta,\tau)=(0.65,174)$, $(0.64,242)$, $(0.56,286)$, and $(0.58,397$) for $L=20,22,24$ and $L=26$, respectively. (b): The same quantity above the critical filling fraction displays transient subdiffusive decay with $z=1/\alpha=3.57,3.45$ and $3.23$ for $L=22,24$ and $26$, respectively, thus pointing toward the profound impact of the freezing transition on the transport.}
\label{autocorr1}
\end{figure*}

\begin{figure*}
\begin{comment}
\stackon{\includegraphics[width=0.53\hsize,height=5.3cm]{corr_freezing_transition_fragment_range2.pdf}}{(a)}%
\stackon{\includegraphics[width=0.53\hsize,height=5.3cm]{corr_largest_fragment_range2.pdf}}{(b)}\\
\stackon{\includegraphics[width=0.53\linewidth,height=5.3cm]{corr_other_filling_fractions_range2.pdf}}{(c)}\stackon{\includegraphics[width=0.53\linewidth,height=5.3cm]{corr_largest_filling_range_fine_tuned.pdf}}{(d)}\\
\end{comment}
\includegraphics[width=0.98\hsize]{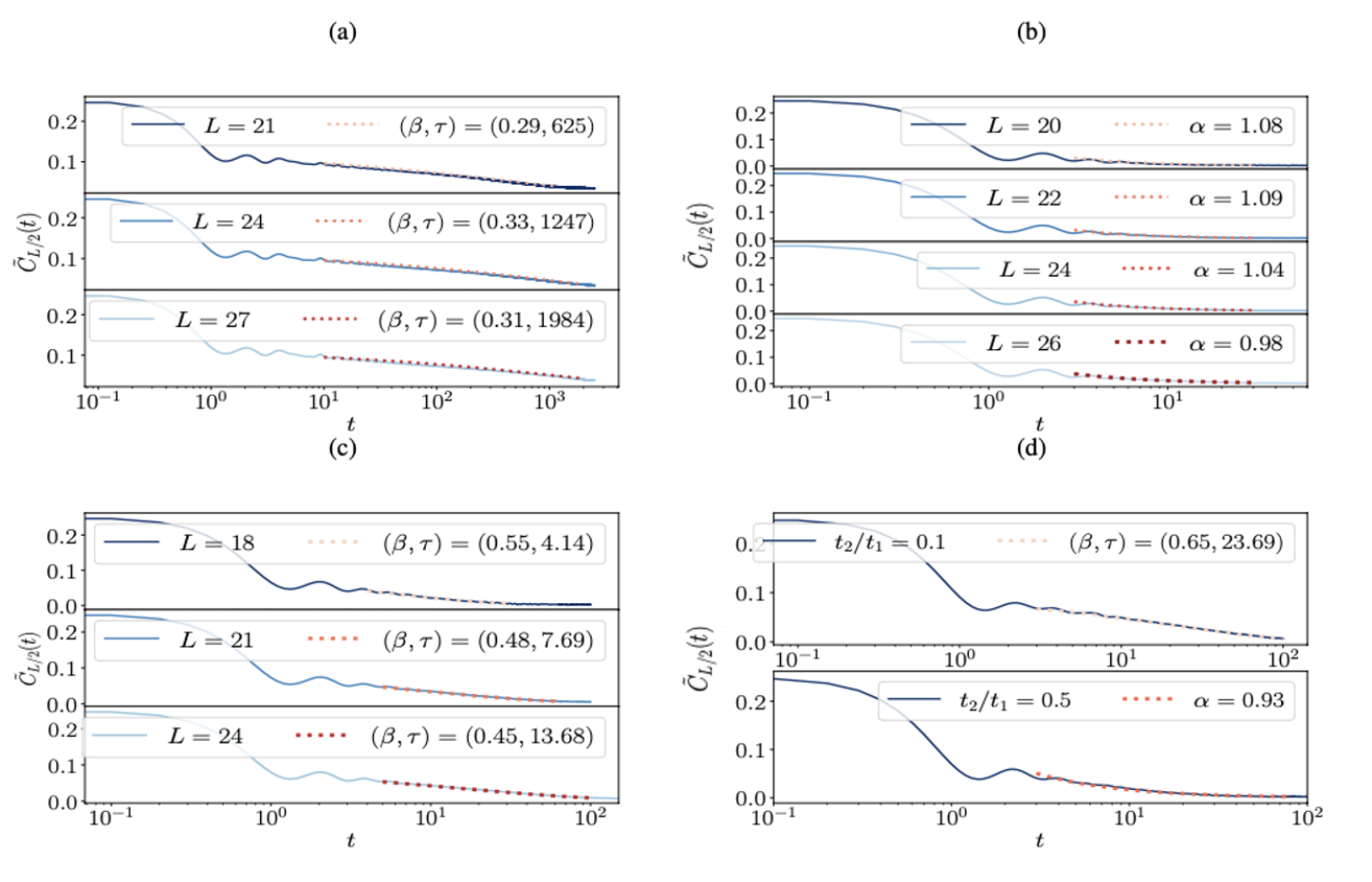}
\caption{(a-c) Plots showing the behavior of $\tilde{C}_{L/2}(t)$ vs $t$ at the freezing transitions, i.e., $N_{c}=L/3$, filling embodying the largest fragment, $N_{f}=L/2+1$ and several other fillings above the critical filling on the weakly fragmented side of the Hilbert space, i.e., $N_{c}<N_{f}<L/2+1$, respectively. (a): $\tilde{C}_{L/2}$ in this case displays SSER with $(\beta,\tau)=(0.29,625),(0.33,1247)$ and $(0.31,1984)$, respectively, for $L=21,24$ and $27$, respectively. (b): The same quantity in this specific case demonstrates close to transient ballistic behavior at intermediate times with $z=1/\alpha=0.93,0.92,0.96$ and $1.02$, respectively, for $L=20,22,24$ and $26$, respectively. (c) Here we consider $N_{f}=L/3+2$ for $L=18,21,24$, respectively, to examine the transport properties at intermediate times, which demonstrate the best numerical fitting with SSER involving $(\beta,\tau)=(0.55,4.14),(0.48,7.69), (0.45,13.68)$, respectively, for $L=18,21,24$, respectively. Fig. (d): The robustness of transient ballistic decay with changing values of $t_{2}/t_{1}$, one at 0.1 (top panel) and another at 0.5 (bottom panel), respectively, for $L=20$ and $N_{f}=L/2+1$. In the first case, we note the behavior agrees with SSER with $(\beta,\tau)=(0.65,23.69)$, whereas the bottom panel shows almost transient ballistic behavior at intermediate times, identical to that observed in Fig (b). }
\label{autocorr2}
\end{figure*}
Furthermore, we will consider PBCs for this analysis instead of OBCs to circumvent the edge effect~\cite{hart2023exact,sala_ergo_2020}, which appears to be quite prominent in systems with fragmented Hilbert space, as discussed earlier. Our objective is now to see how the freezing transition~\cite{East_Sreemayee,Morningstar_2020,abhisodh2024,wang_2023} impacts the transport behaviors in this class of models with an increasing range of constraints. In doing so, we further consider $\tilde{C}_{L/2}=\tilde{C}(0,t)$ putting $j=L/2$ (inside the bulk, but the reference position does not matter in PBCs) in Eq. \eqref{finaltra} to apprehend the relaxation properties for various filling fractions. First and foremost, we analyze the same for the range-1 constraint at two fillings, one at critical filling and another one where the model embodies the largest fragment, as illustrated in Figs. \ref{autocorr1} (a-b). In Fig. \ref{autocorr1} (a), we witness the relaxation behavior of $\tilde{C}_{L/2}$ exhibiting the best numerical fitting with a size-stretched exponential relaxation (SSER)~\cite{berma_system_stretched}, $\sim e^{-(t/\tau)^{\beta}}$, where $(\beta,\tau)=(0.65,174),(0.64,242),(0.56,286)$, and $(0.58,397)$, for $L=20,22,24$ and $26$, respectively, at large enough times. The size-stretched indicates that the relaxation time scale, $\tau$, increases with growing $L$'s. On the other hand, the same quantity demonstrates the signature of transient subdiffusive power-law decay~\cite{pal_2010,singh_2021,Morningstar_2020,tracer} in the second case where the model is weakly fragmented with $z=1/\alpha\simeq3.57,3.45$ and $3.23$ for $L=22,24$ and $26$, respectively, at intermediate times, as depicted in Fig. \ref{autocorr1} (b). One should note that it has been reported earlier that $U(1)$-conserving range-1 East model shows extremely slow transport behaviors at extremely long time scales with $z(t)\sim log(t)$~\cite{singh_2021}. Nevertheless, it was not known how freezing transition impacts the charge transport at intermediate times~\cite{wang_2023,East_Sreemayee} as one tunes the filling fraction. Hence, our primary purpose through this analysis is to concentrate on the intermediate time behavior accessible by numerics rather than focusing on $t\to\infty$ limit, which requires the large-scale numerical simulation using MPS. However, our above finite-size investigation evidently distinguishes the critical filling from the weakly fragmented side of the Hilbert space~\cite{Morningstar_2020}.

Next, we turn to the model with range-2 constraint where we again concentrate on three fillings: (i) critical filling ($N_{c}=L/3$), (ii) filling incorporating the largest fragment in PBCs ($N_{f}=L/2+1$), and finally (iii) at $N_{c}<N_{f}<L/2+1$ where the Hilbert space is again weakly fragmented, as portrayed in Figs. \ref{autocorr2} (a-c). At freezing transition in the range-2 case, we again observe the relaxation profile showing the best numerical fit with SSER~\cite{berma_system_stretched} with $(\beta,\tau)=(0.29,625),(0.33,1247)$, and $(0.31,1984)$ for $L=21,24$ and $27$, respectively, as shown in Fig. \ref{autocorr2} (a). This behavior switches to transient ballistic power-law decay~\cite{brighi_anomoulous} for case (ii) (where lies the largest fragment) with $z\simeq1$, as demonstrated in Fig. \ref{autocorr2} (b) for several $L$'s. Finally, a similar exploration for the third case (iii) unravels the best numerical fitting again with SSER~\cite{berma_system_stretched} with $\beta\simeq0.55,0.48$ and $0.45$ for $L=18,21$ and $24$, respectively, as shown in Fig. \ref{autocorr2} (c). Nevertheless, $\beta$ for this case is larger compared to that of $N_{c}$, revealing a faster relaxation at this filling (weakly fragmented) compared to the critical filling, as anticipated. This analysis thus implies three distinct dynamical relaxation behaviors based on the fragmentation structure of the Hilbert space~\cite{Morningstar_2020}.

Furthermore, the transient ballistic decay~\cite{brighi_anomoulous} seen in Fig. \ref{autocorr2} (b) is notably anomalous as the level spacing analysis of the spectrum within the largest fragment (in OBCs) shown in Fig. \ref{therma} (c) indicates the fragment being non-integrable~\cite{atas_2013,huse_2013,berry_1977,Wigner_1955} (which primarily dominates the dynamical behavior due to the weak fragmentation), while the ballistic transport is a characteristic feature of integrable quantum systems. This fact motivates us to scrutinize the robustness of this behavior against the change in $t_{2}/t_{1}$ (one should note that $t_{2}$ and $t_{1}$ are the parameters of the range-2 model as illustrated in Eq. \ref{Hamr123}), which has been illustrated in Fig. \ref{autocorr2} (d) for two parameter values, $t_{2}/t_{1}=0.1$ and 0.5, respectively. One should note that changing $t_{2}/t_{1}$ preserves the classical fragmentation structure of this model. This investigation allows us to comprehend whether this transient power-law decay is the sole outcome of the classical fragmentation structure or whether there are other mechanisms also involved in manifesting this behavior. In \ref{autocorr2} (d), we notice that the transport behavior for $t_{2}/t_{1}=0.5$ (bottom panel) exhibits nearly ballistic behavior with $z=1.08$, whereas it switches to a much slower SSER~\cite{berma_system_stretched} with $(\beta,\tau)=(0.65,23.69)$ (top panel). This observation thus facilitates this understanding that although it is robust for a broad range of $t_{2}/t_{1}$, it significantly changes when the ratio $t_{2}/t_{1}$ almost approaches the range-1 model.

\begin{figure*}
\subfigure[]{\includegraphics[width=0.55\hsize,height=4.5cm]{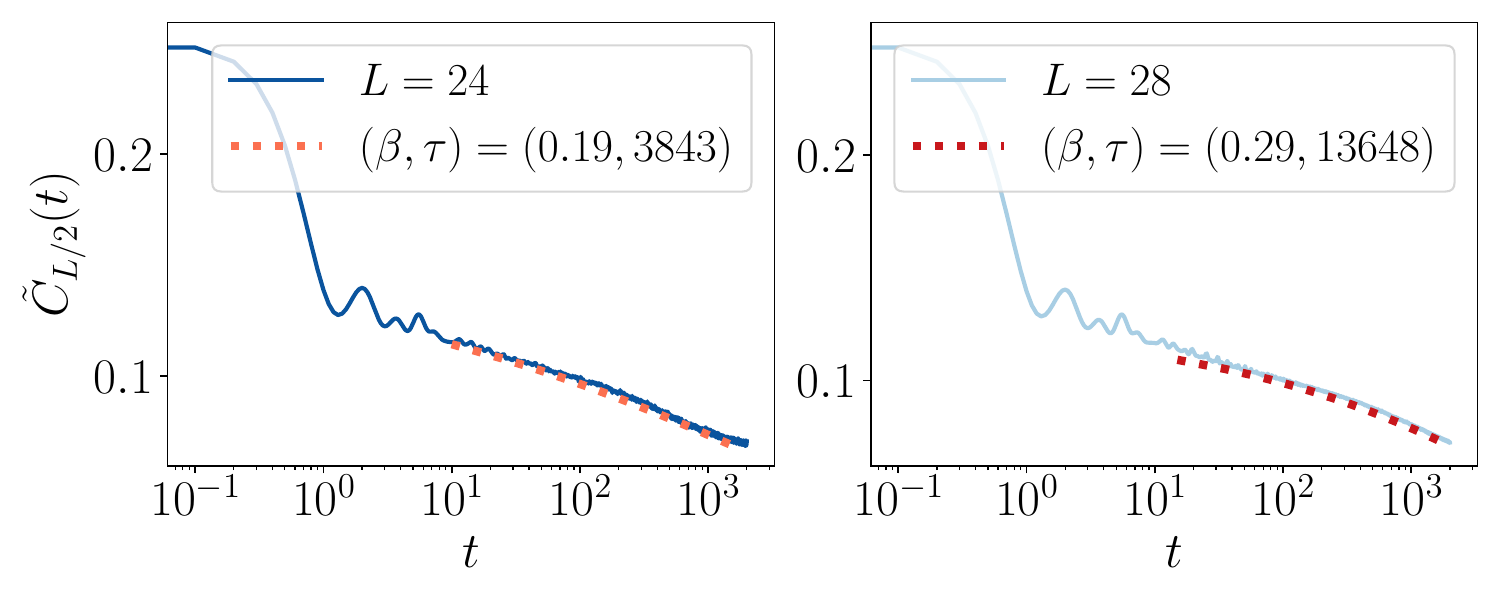}}%
\subfigure[]{\includegraphics[width=0.45\hsize]{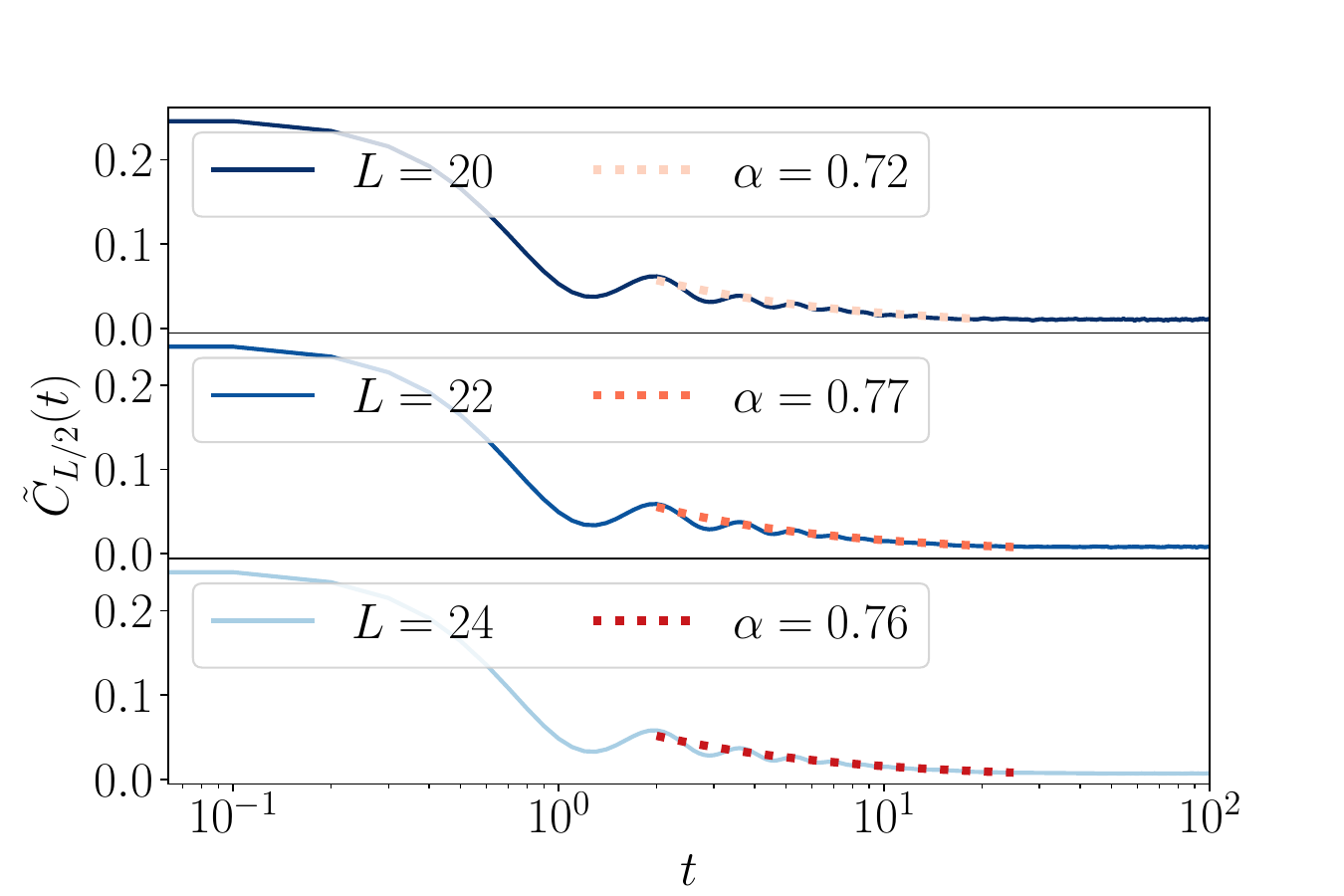}}\\

\caption{(a-b) Plots showing $\tilde{C}_{L/2}(t)$ vs $t$ at critical filling $N_{c}=L/4$, and $N_{f}>N_{c}$, respectively, for several system sizes. (a): $\tilde{C}_{L/2}$ in this case shows SSER relaxation with $(\beta, \tau)=(0.19,3843)$ and $(0.29,13643)$, respectively, for $L=24$ and $28$, respectively. However, the value of $\beta$ fails to converge for these two cases. (b): Here, the same quantity reveals transient superdiffusive behavior with $z=1/\alpha\simeq1.39,1.30$, and $1.32$, for $(L,N_{f})=(20,L/4+3),(22,(L-2)/4+4)$ and $(26,L/4+3)$, respectively, at intermediate times.}
\label{autocorr3}
\end{figure*}

Finally, we also perform a similar analysis for the range-3 constraint at $N_{c}=L/4$ (critical filling) and $N_{f}>N_{c}$ (weakly fragmented), as demonstrated in Figs. \ref{autocorr3} (a-b). Similar to the range-2 case, we note a slower SSER~\cite{berma_system_stretched}, as shown in Figs. \ref{autocorr3} (a). However, the stretched exponent $\beta$ fails to converge at a particular value like the range-2 and range-3 case and significantly changes with $L$'s under consideration. While $(\beta,\tau)=(0.19,3843)$ for $L=24$, the same thing appears to be $(0.29,13648)$ for $L=28$, as shown in Fig. \ref{autocorr3} (a). In a similar manner, the intermediate transport for several values of $N_{f}$ above $N{c}$ exhibit signatures of superdiffusive power-law decay~\cite{brighi_anomoulous} with $z=(1.39,1.30,1.32)$ for $L=20,22$ and $24$, respectively, at intermediate times, as presented in Fig. \ref{autocorr3} (b). This again confirms that the intermediate-time transport properties behave in a completely distinct manner due to the freezing transition~\cite{Morningstar_2020,East_Sreemayee,wang_2023,abhisodh2024} in this family of models and can further range from subdiffusive~\cite{tracer,singh_2021} through superdiffusive~\cite{brighi_2023} to ballistic~\cite{brighi_anomoulous} at the weakly fragmented side, depending on the range of the constraint. 

We will now furnish a heuristic argument for getting the SSER at the freezing transition point, which is vastly seen in the context of glassy dynamics~\cite{DAS1999245,barma_stretched_expo,Ritort01062003}. Typically, stretched exponential relaxation arises due to dynamic heterogeneity in glassy systems; this is further applicable to fragmented Hilbert spaces in the presence of blockades~\cite{moudgalya_memorial_2021}. At the critical filling, the dimension of the largest fragment decays very slowly (polynomially) to zero compared to the growth of the full Hilbert space, thus allowing the presence of blockades to play a significant role in charge transport. One can argue that the autocorrelation function at critical filling can thus take the following form
\bea
\tilde{C}(t)=\sum_{l,\rho} P(l,\rho)\,\tilde{C}(l,\rho,t),\label{sserform}
\eea
where $\tilde{C}(l,\rho,t)$ is the correlator in each active fragment with length $l$ and density $\rho$ and $P(l,\rho)$ is the distribution of such active fragments separated by blockades~\cite{barma_stretched_expo,DAS1999245}. Within each active fragment, one can further assume that the correlators relax exponentially with a relaxation time scale of $\tau_{l}$, as $e^{-t/\tau_{l}}$ at long times. In addition, one can also consider the particle spread in generic active fragments in a diffusive manner, which gives the relaxation time scale, $\tau_{l}\sim L^2$. Thereafter, one can get various stretched exponential relaxations depending on different distributions of $P(l,\rho)$ for different models (which typically take an exponential form as $e^{-\lambda l}$), and then the sum gets dominant contribution from an appropriate saddle point.
Therefore, it is plausible that various distributions of $P(l,\rho)$ may cause this SSER at critical filling, as demonstrated in Figs. \ref{autocorr1} (a), \ref{autocorr2} (a) and \ref{autocorr3} (a). 

On the other hand, the transient ballistic power-law relaxation in the range-2 model at the filling embodying the largest fragment, as demonstrated in Fig. \ref{autocorr2} (b), might be a consequence of gapless low-lying excitation in the largest fragment, as shown in Fig. \ref{gsr2_PBC}; this perhaps is the dominant hydrodynamic modes on the weakly fragmented side of the Hilbert space, thus causing ballistic transport at intermediate times. In addition, superdiffusive transport at intermediate times in the range-3 case at a filling fraction when the Hilbert space is weakly fragmented, as displayed in Fig. \ref{autocorr3} (b), also needs a deeper investigation by performing large-scale numerics and further using an effective hydrodynamic description. We anticipate the intricate interplay between ballistic hydrodynamic modes due to gapless low-energy excitations and fragmentation-induced blockades might be responsible for this superdiffusive dynamical behavior in the range-3 case on the weakly fragmented side of the Hilbert space. In this context, it should be noted that ballistic to superdiffusive transport~\cite{brighi_2023} has been reported earlier in the range-2 model while studying the domain wall dynamics, which we also see in the infinite-temperature transport in PBCs for this family of models.

\begin{comment}
\begin{figure}
\stackon{\includegraphics[width=\columnwidth]{corr_largest_filling_range2_J1_J0.5.pdf}}{(a)}\\
\stackon{\includegraphics[width=\columnwidth]{corr_largest_filling_range2_J1_J0.1.pdf}}{(b)}\\

\caption{(a-b) Plots showing the behavior of $\tilde{C}_{L/2}(t)$ vs $t$ at $N_{f}=L/2+1$ for $L=20$ with $t_{1}=1$ and $t_{2}=0.5$ and $0.1$, respectively. (a): In this case, we see the signature of superdiffusive transport with $z=1.08$. (b): On the other hand, here we see the behavior moves towards the stretched exponential decay, with $e^{-(t/\tau)^{\beta}}$, with $\beta=0.65$, which is close to that observed for the range-1 case exactly at the freezing transition. }
\label{autocorr2finetune}
\end{figure}
\end{comment}

\section{Summary and discussion}
\label{dis}

We will now summarize the main results obtained from our analysis elucidated above. In this paper, we furnish an exhaustive analytical understanding of the fragmentation structure of the one-dimensional facilitated East models with range-$q$ hoppings, albeit the analytical complexity introduced by the increasing range of constraints. To begin with, we first identify the correct root representation~\cite{East_Sreemayee,Deepak_HSF,khagebdra2_fredkin,salberger2016fredkinspinchain} describing each fragment in this class of models. This identification method allows us to determine the fragmentation structure of the Hilbert space employing the enumerative combinatorics and transfer matrix method~\cite{generating1994,Deepak_HSF,HariMenon_1995,menon_1997,Dhar_1993} , which includes capturing the growth of fragments and frozen eigenstates. Thereafter, we elaborate on determining the dimensional growth of the largest fragments at various filling fractions to get further insight into the nature of the fragmented Hilbert space~\cite{Moudgalya_review_2022,moudgalya_memorial_2021}. This analysis hints that a large class of fragments follows the growth specified by the generalization of the Catalan number introduced by Frey and Sellers~\cite{FS_article} for $q>1$ while all the fragments follow the combinatorics sequence of the widely-studied Dyck path~\cite{catalan} in the $q=1$ case. The investigation also enables us to understand the strong-to-weak fragmentation transition~\cite{East_Sreemayee,wang_2023,Morningstar_2020} in this family in terms of the above combinatorial sequence, further facilitating the extraction of the critical filling analytically, i.e., $n_{c}=1/(q+1)$. Hence, it infers that increasing the range of constraints allows the transition to occur at a lower filling.

Next, we turn to capture the dimensional growth of the largest fragment in the full Hilbert space to understand the nature of the ground state in this class of models. In the case of $q=1$ constraints, we see the existence of a non-degenerate ground state whose filling gradually shifts from half-filling as $L/2+a$, where $a\sim \sqrt{L}/2$ as an outcome of the Dyck combinatorics~\cite{East_Sreemayee}. While this behavior changes drastically in the presence of longer-range hoppings, where we observe the inclusion of multiple ground states in the full Hilbert space with a degeneracy of $2^{q-1}$ locating at $q$ numbers of filling fractions in total. This anomalous ground state behavior can further be readily apprehended utilizing the root identification method~\cite{East_Sreemayee,Deepak_HSF,khagebdra2_fredkin,salberger2016fredkinspinchain} and the dimensional growth determined by the Frey-Sellers sequence~\cite{FS_article}. In addition, it can argued that the minimum filling for the appearance of the ground state is around $(L-q)/2$
in OBCs. However, there is no shift in the above filling fraction in $L\to\infty$, unlike the range-1 model. Also, we note the ground state in this class of models appears to be critical with logarithmic scaling of entanglement with $L$.

Afterward, we examine the long-time behaviors of bulk and boundary autocorrelators in OBCs to comprehend the effect of freezing transition~\cite{East_Sreemayee,wang_2023,Morningstar_2020,brian_thermal} on the statistical edge localization~\cite{sala_ergo_2020,hart2023exact} in this family of models. Further, as we go from the strongly to the weakly fragmented side of the Hilbert space by tuning the fillings, the non-thermal bulk becomes thermal, while the leftmost active boundary continues to remain athermal. This distinct inhomogeneous profile~\cite{sala_ergo_2020,hart2023exact} of the long-time autocorrelation functions at various fillings is a combined outcome of the freezing transition~\cite{East_Sreemayee,wang_2023,Morningstar_2020,brian_thermal} and lack of inversion symmetry in this class of models. In addition, the lower bound of the long-time saturation value of the same has been further validated using the Mazur-Suzuki inequality~\cite{MAZUR1969533,SUZUKI1971277}.

Finally, we examine how the freezing transition impacts the infinite-temperature charge transport in this family of models using the concept of dynamical quantum typically (QDT)~\cite{QDT1,QDT2} in PBCs. Our investigation reveals the existence of rich transport behaviors at intermediate times depending on the filling fraction, which comprises the size-stretched exponential relaxation (SSER)~\cite{berma_system_stretched} at critical fillings, subdiffusive~\cite{pal_hydro,singh_2021} through superdiffusive~\cite{brighi_2023} to ballistic~\cite{brighi_anomoulous} in the weakly fragmented part of the Hilbert space (depending on the range of constraints), respectively.

We now outline some of the possible future avenues that might be worthwhile to look at in the future. A deeper analytical understanding of the transport properties~\cite{Dhar_1993,berma_system_stretched} would be an exciting avenue for the future. It is also worthwhile to understand the nature of ground state wavefunctions in terms of the Frey-Sellers sequence in the thermodynamic limit, which might provide more insight on this model like the examination done in cases of  Fredkin~\cite{salberger2016fredkinspinchain} and Motzkin spin chains~\cite{Movassagh_2017,motzkin2}. In addition, studying the effects of dissipation~\cite{Li_2023} and disorder~\cite{tomasi_2019} might also be promising in the future.

Recently, thermalization in some particular fragments of a model with HSF
has been observed in a Rydberg atom system in one dimension~\cite{dalta_2025}. Another observation of HSF has been noted in a superconducting processor in a system demonstrating Stark many-body localization~\cite{wang_2025}. We thus anticipate that our results can also be experimentally verified in cold-atom 
platforms~\cite{kohlert2021experimental,bloch_2008}.

\centerline{\bf Acknowledgments}
\vspace{0.4cm}
S.A. thanks Diptiman Sen, Chandan Dasgupta, and Subroto Mukerjee for many insightful discussions. S.A. thanks Diptiman Sen and Deepak Dhar for previous collaborations on this topic. S.A. thanks MHRD, India, and Alexander von Humboldt Foundation, Germany for financial support through a PMRF and a Humboldt research fellow.
\appendix
\section{Block identification for the range-2 model}
\label{rootr2}
Here, we will illustrate the procedure to find all possible choices of $\psi_{L}$, $\psi_{M}$, and $\psi_{R}$ in the context of the range-2 model, which is important to identify the general form of representative states describing the disjoint fragments. Nevertheless, a similar method, in principle, applies to other longer-range variants as well.

Before proceeding further, we again write the transitions entitled by the range-2 model as
\bea
1110\leftrightarrow 1101,~~~
1010\leftrightarrow 1001,~~~
0110\leftrightarrow0101.\label{trans}
\eea
We already argued in Sec. \ref{root} that unique root representations describing fragments can be achieved by successive substitutions of $1110$'s, $0110$'s, and $1010$'s starting from an arbitrary binary string configuration. This claim has been backed by the transfer matrix method as well as the numerical basis enumeration method. Now, to apprehend the potential $\psi_{L}$'s, $\psi_{M}$'s, and $\psi_{R}$'s, we employ a recursive trick, which is charted out below. First and foremost, we recursively increase the length of the binary strings highlighted in circles on the left side of the transitions in Eqs. \ref{p1}, \ref{p2} and \ref{p3} and afterward, continually substitute $1110$'s, $1010$'s, and $0110$'s by $1101$, $1001$ and $0101$'s until there is no $1110$, $1010$ and $0110$ left at the end. One should note that this recursive expansion of binary strings from the circled place is not special; it has been considered to systematically accomplish the operation. Following the above for multiple binary strings displayed in Eqs. \ref{p1}, \ref{p2}, and \ref{p3}, one can identify all the possible choices of $\psi_{R}$, $\psi_{M}$ and $\psi_{L}$ for the range-2 case as highlighted in rectangular boxes colored in blue, no color and red, respectively.
\begin{widetext}
\bea
1.&&11\,\circled{1}\,0\rightarrow \fbox{\red{1}}\fbox{\blue{101}},~~~~~~~~~ 11\,\circled{0}\,0\rightarrow \fbox{\red{1}}\fbox{\blue{100}}\non\\
2.&& 11\,\circled{11}\,0\rightarrow \fbox{\red{1}}\fbox{\blue{1011}},~~~ 11\,\circled{10}\,0\rightarrow \fbox{\red{1}}\fbox{100}\,\fbox{\blue1},~~~ 11\,\circled{01}\,0\rightarrow \fbox{\red1}\fbox{100}\,\fbox{\blue1},~~~ 11\,\circled{00}\,0\rightarrow \fbox{\red1}\fbox{100}\,\fbox{\blue0},\non\\
3.&&11\,\circled{111}\,0\rightarrow \fbox{\red1}\fbox{\blue{10111}},~~ 11\,\circled{110}\,0\rightarrow \fbox{\red1}\fbox{100}\fbox{\blue{11}},~~11\,\circled{101}\,0\rightarrow \fbox{\red1}\fbox{100}\fbox{\blue{11}},~~ 11\,\circled{011}\,0\rightarrow \fbox{\red{1}}\fbox{100}\fbox{\blue{11}},\non\\
&&11\,\circled{100}\,0\rightarrow \fbox{\red{1}}\fbox{100}\fbox{\blue{10}},~~ 11\,\circled{010}\,0\rightarrow \fbox{\red1}\fbox{100}\fbox{\blue{10}},~~11\,\circled{001}\,0\rightarrow \fbox{\red1}\fbox{100}\fbox{\blue{10}},~~ 11\,\circled{000}\,0\rightarrow \fbox{\red1}\fbox{100}\fbox{\blue{00}},\non\\
\label{p1}
\eea
%where the 1, 2, and $3$ in Eq. \eqref{first} indicate that the string in the circle on the left side of the transition is made of 1, 2, and 3 spinless fermions, respectively. The boxes on the right side of the transitions after removing all 1110's, 0110's, and 1010's signify the units of strings needed to represent the unique root state, which labels each fragment within this model. A similar procedure can also be repeated for the second transition given in Eq. \eqref{trans}, as follows
For the second case,
\bea
1.&&10\,\circled{1}\,0\rightarrow \fbox{100}\,\fbox{\blue1},~~~~~~~~~ 10\,\circled{0}\,0\rightarrow \fbox{100}\,\fbox{\blue 0}\non\\
2.&& 10\,\circled{11}\,0\rightarrow \fbox{100}\,\fbox{\blue{11}},~~~ 10\,\circled{10}\,0\rightarrow \fbox{100}\,\fbox{\blue{10}},~~~ 10\,\circled{01}\,0\rightarrow \fbox{100}\,\fbox{\blue{10}},~~~ 10\,\circled{00}\,0\rightarrow \fbox{100}\,\fbox{\blue{00}},\non\\
3.&&10\,\circled{111}\,0\rightarrow \fbox{100}\,\fbox{\blue{111}},~~ 10\,\circled{110}\,0\rightarrow \fbox{100}\,\,\fbox{\blue{101}},~~10\,\circled{101}\,0\rightarrow \fbox{100}\,\fbox{\blue{101}},~~ 10\,\circled{011}\,0\rightarrow \fbox{100}\,\,\fbox{\blue{101}},\non\\
&&10\,\circled{100}\,0\rightarrow \fbox{100}\,\fbox{\blue{100}},~~ 10\,\circled{010}\,0\rightarrow \fbox{100}\,\fbox{\blue{100}},~~10\,\circled{001}\,0\rightarrow \fbox{100}\,\fbox{\blue{010}},~~ 10\,\circled{000}\,0\rightarrow \fbox{100}\,\fbox{\blue{000}},\non\\
\label{p2}
\eea
and for the third process,
\bea
1.&&01\,\circled{1}\,0\rightarrow \fbox{\red0}\fbox{\blue{101}},~~~~~~~~~ 01\,\circled{0}\,0\rightarrow \fbox{\red0}\,\fbox{100}\non\\
2.&& 01\,\circled{11}\,0\rightarrow \fbox{\red0}\,\fbox{\blue{1011}},~~~ 01\,\circled{10}\,0\rightarrow \fbox{\red0}\,\fbox{100}\,\fbox{\blue{1}},~~~ 01\,\circled{01}\,0\rightarrow \fbox{\red0}\,\fbox{100}\,\fbox{\blue1},~~~ 01\,\circled{00}\,0\rightarrow \fbox{\red0}\,\fbox{100}\,\fbox{\blue0},\non\\
3.&&01\,\circled{111}\,0\rightarrow \fbox{\red0}\,\fbox{\blue{10111}},~~ 01\,\circled{110}\,0\rightarrow \fbox{\red0}\,\fbox{100}\,\fbox{\blue{11}},~~01\,\circled{101}\,0\rightarrow \fbox{\red0}\fbox{100}\,\fbox{\blue{11}},~~ 01\,\circled{011}\,0\rightarrow \fbox{\red0}\,\fbox{100}\,\fbox{\blue{11}},\non\\
&&01\,\circled{100}\,0\rightarrow \fbox{\red0}\,\fbox{100}\,\fbox{\blue{10}},~~ 01\,\circled{010}\,0\rightarrow \fbox{\red0}\,\fbox{100}\,\fbox{\blue{10}},~~01\,\circled{001}\,0\rightarrow \fbox{\red0}\,\fbox{100}\,\fbox{\blue{10}},~~ 01\,\circled{000}\,0\rightarrow \fbox{\red{0}}\,\fbox{100}\,\fbox{\blue{00}},\non\\
\label{p3}
\eea
\end{widetext}
These procedures facilitate an understanding of the root states where the block structure of these representative states is as follows:

Let us consider arbitrary strings, which by successive removal of $1100$, $0110$, and $1010$ reduce to the following separable form that identifies all the equivalence classes defining simple, fully-connected fragments with the following form $\psi_{L}\otimes\psi_{M}\otimes\psi_{R}$ where

\vspace{0.05cm}

\noi (i) $\psi_{L}$ can be either a null string ($\phi$), a string of all 0's, or a single 1.
\vspace{0.1cm}

\noi (ii) $\psi_{M}$ is units of 100's,

\vspace{0.1cm}

\noi (iii) $\psi_{R}$ can be either $\phi$, 0's, 0's followed by (1's/ a single 10), all 1's, or a single 10 followed by 1's. 
\section{Number of fragments in range-2 and range-3 model}
\label{frag}
\subsection{$N_{frag}$ in range-2 model}
To begin with, we now count the total number of fragments in the range-2 case employing the transfer matrix method~\cite{generating1994,Deepak_HSF,Dhar_1993,menon_1997,HariMenon_1995}. As this model allows transitions involving four consecutive sites, it thus requires us to construct a $8\times8$ transfer matrix, $T_{1}(c_{i},c_{j})$, where $\ket{c_{i}}$ denotes the following eight configurations where binary strings located at three consecutive sites, i.e., $\ket{111}$, $\ket{110}$, $\ket{101}$, $\ket{011}$, $\ket{100}$, $\ket{010}$, $\ket{001}$ and $\ket{000}$. In addition, as illustrated before, labeling each fragment requires the successive substitution of all $1110$, $1010$, and $0110$'s; the resultant transfer matrix thus must not include any $1110$, $0110$, and $1010$. The transfer matrix, therefore, becomes

\bea
T_{1}=\begin{pmatrix}
    1 & 0&  0&  0 & 0 & 0 & 0 & 0\\0 & 0 & 1 & 0 & 1& 0 & 0 & 0\\0 & 0 & 0& 1 & 0 & 0& 0 & 0\\1 & 0&  0&  0 & 0 & 0 & 0 & 0\\0& 0&  0&  0 & 0 & 0 & 1 & 1\\0 & 0 & 1 & 0 & 1& 0 & 0 & 0\\0 & 0 & 0 & 1 & 0& 1 & 0 & 0\\0 & 0 & 0 & 0 & 0& 0 & 1 & 1
\end{pmatrix},
\label{trans1}
\eea
whose eigenvalues are $1.466$, $-0.2328\pm0.7926i$, $1$, and four $0$'s. The number of fragments ($N_{frag}$) in this model thus grows as $1.466^{L}$ in $L\to\infty$ limit. In addition, $N_{frag}$ for the first few system sizes can be obtained using the formula $N_{frag}(L)=\sum_{ij=1}^{8}X_{ij}$, where $X=T_{1}^{L-3}$, as shown in Table-\ref{table1}, which is in perfect agreement with the results obtained using numerical basis enumeration method. Further, we consider $N_{frag}(0)=1$, $N_{frag}(1)=2$, $N_{frag}(2)=4$, and $N_{frag}(3)=8$.
\begin{table}[htb]
\begin{tabular}{|c|c|c|c|c|c|c|c|c|c|c|c|c|}
\hline
$L$&4&5&6&7&8&9&10&11&12&13&14&15 \\
\hline
$N_{frag}$&13&20&31&47&70&104&154&227&334&491&721&1058\\
\hline
\end{tabular}
\caption{ $N_{frag}$ vs $L$ for the first few system sizes obtained analytically in the range-2 case, which we validate using the numerical basis enumeration method.}
\label{table1}
\end{table}

It can be noted from Table-\ref{table1} that $N_{frag}(L)$ satisfies the following relation $N_{frag}(L)=2N_{frag}(L-1)+N_{frag}(L-3)-N_{frag}(L-2)-N_{frag}(L-4)$ for $L\geq 4$.

\subsection{$N_{frag}$ in range-3 model}
An identical method~\cite{generating1994,Deepak_HSF,Dhar_1993,menon_1997,HariMenon_1995} is also applicable to compute $N_{frag}$ in the case of range-3 constraint. As the transitions allowed in this case involve five consecutive sites, it is required to build a $16\times16$ transfer matrix, $T_{4}(c_{i},c_{j})$, where $c_{i}$ denotes the string configurations on four consecutive sites, $\ket{1111},\ket{1110},\ket{1101},\ket{1011},\ket{0111}$, $\ket{1100},\ket{0110},\ket{0011},\ket{1001},\ket{0101},\ket{1010},\ket{1000},\ket{0100}$, $\ket{0010},\ket{0001}$, and $\ket{0000}$. Moreover, the resultant transfer matrix should not contain any seven of these configurations, $11110$, $11010$, $10110$, $01110$, $10010$, $01010$, and $00110$, as the representation of unique root states requires the successive substitution of these binary strings. Following the rule, the transfer matrix thus becomes 
\bea
T_{4}=\begin{pmatrix}
    1 & 0 & 0& 0& 0& 0& 0 & 0 & 0 & 0 & 0 & 0 & 0 & 0 & 0 & 0 \\  0 & 0 & 1& 0& 0& 1& 0 & 0 & 0 & 0 & 0 & 0 & 0 & 0 & 0 & 0\\ 0 & 0 & 0& 1& 0& 0& 0 & 0 & 0 & 0 & 0 & 0 & 0 & 0 & 0 & 0\\ 0 & 0 & 0& 0& 1& 0& 0 & 0 & 0 & 0 & 0 & 0 & 0 & 0 & 0 & 0\\1 & 0 & 0& 0& 0& 0& 0 & 0 & 0 & 0 & 0 & 0 & 0 & 0 & 0 & 0\\ 0& 0 & 0& 0& 0& 0& 0 & 0 & 1 & 0 & 0 & 1 & 0 & 0 & 0 & 0\\ 0 & 0 & 1& 0& 0& 1& 0 & 0 & 0 & 0 & 0 & 0 & 0 & 0 & 0 & 0\\ 0& 0 & 0& 0& 1& 0& 0 & 0 & 0 & 0 & 0 & 0 & 0 & 0 & 0 & 0\\ 0 & 0 & 0& 0& 0& 0& 0 & 1 & 0 & 0 & 0 & 0 & 0 & 0 & 0 & 0\\ 0 & 0 & 0& 1& 0& 0& 0 & 0 & 0 & 0 & 0 & 0 & 0 & 0 & 0 & 0\\ 0 & 0 & 0& 0& 0& 0& 0 & 0 & 0 & 1 & 0 & 0 & 1 & 0 & 0 & 0\\ 0 & 0 & 0& 0& 0& 0& 0 & 0 & 0 & 0 & 0 & 0 & 0 & 0 & 1 & 1\\ 0 & 0 & 0& 0& 0& 0& 0 & 0 & 1 & 0 & 0 & 1 & 0 & 0 & 0 & 0\\ 0 & 0 & 0& 0& 0& 0& 0 & 0 & 0 & 1 & 0 & 0 & 1 & 0 & 0 & 0\\ 0 & 0 & 0& 0& 0& 0& 0 & 1& 0& 0 & 0 & 0 & 0 & 1 & 0 & 0\\ 0 & 0 & 0& 0& 0& 0& 0 & 0 & 0 & 0 & 0 & 0 & 0 & 0 & 1 & 1\\
\end{pmatrix}.
\label{trans3}
\eea
The eigenvalues of Eq. \ref{trans3} are given by 1.3803, $-0.8192$, $0.2194+\pm0.9145i$, 1, and eleven 0's. Hence, $N_{frag}$ in this case grows as $1.38^{L}$ in the asymptotic limit. $N_{frag}$ for the first few system sizes are shown in Table \ref{table45}, which can again be obtained using the formula $N_{frag}=\sum_{ij=1}^{16}X_{ij}$, where $X=T_{4}^{L-4}$, which are in perfect agreement with the numerically obtained values. Additionally, we also consider $N_{frag}(L)=0,1,4,8,16$ for $L=0,1,2,3$, and $4$, respectively.
\begin{table}[htb]
\begin{tabular}{|c|c|c|c|c|c|c|c|c|c|c|c|c|}
\hline
$L$&5&6&7&8&9&10&11&12&13&14&15 \\
\hline
$N_{frag}$&25&36&51&74&106&149&207&288&401&557&771\\
\hline
\end{tabular}
\caption{ $N_{frag}$ vs $L$ for the first few system sizes for the range-3 case in OBCs, which has been obtained analytically. This is also in perfect agreement with numerically acquired values.}
\label{table45}
\end{table}
\section{Counting the number of frozen states in range-2 and range-3 model}
\label{frozena}
\subsection{range-2}
Here, we discuss the counting of frozen fragments ($N_{froz}$) in the range-2 case by employing a similar transfer matrix method~\cite{generating1994,Deepak_HSF,Dhar_1993,menon_1997,HariMenon_1995} as discussed before. This counting again requires us to construct a $8\times 8$ transfer matrix, $T_{2}(c_{i},c_{j})$, where $\ket{c_{i}}$ denotes the following eight configurations on three consecutive sites, $\ket{111}$, $\ket{110}$, $\ket{101}$, $\ket{011}$, $\ket{100}$, $\ket{010}$, $\ket{001}$ and $\ket{000}$. However, the condition that has to be imposed in this case is different from the former case, i.e., no occurrences of $1110$, $1101$, $1010$, $1001$, $0110$, and $0101$ in the resultant transfer matrix. $T_{2}$ thus takes the following form
\bea
T_{2}=\begin{pmatrix}
    1 & 0&  0&  0 & 0 & 0 & 0 & 0\\0 & 0 & 0 & 0 & 1& 0 & 0 & 0\\0 & 0 & 0& 1 & 0 & 0& 0 & 0\\1 & 0&  0&  0 & 0 & 0 & 0 & 0\\0& 0&  0&  0 & 0 & 0 & 0 & 1\\0 & 0 & 0 & 0 & 1& 0 & 0 & 0\\0 & 0 & 0 & 1 & 0& 1& 0 & 0\\0 & 0 & 0 & 0 & 0& 0 & 1 & 1
\end{pmatrix},
\label{trans2}
\eea
which gives the following eigenvalues: $1.38$, $0.2194\pm0.9145i$, $1$, and 4 0's.  Hence, $N_{froz}$ grows as $1.38^{L}$ in the asymptotic limit. Furthermore, $N_{froz}$ for the first few system sizes can be obtained in OBCs with the help of $N_{froz}(L)=\sum_{ij}Y$, where $Y=T_{2}^{L-3}$, as shown in Table-\ref{table2}. In addition, one can also choose $N_{froz}(L)=1,2,4$, and $8$ for $L=0,1,2$, and $3$.
\begin{table}[htb]
\begin{tabular}{|c|c|c|c|c|c|c|c|c|c|c|c|c|}
\hline
$L$&4&5&6&7&8&9&10&11&12&13&14 \\
\hline
$N_{froz}$&10&13&19&28&39&53&73&102&142&196&270\\
\hline
\end{tabular}
\caption{$N_{froz}(L)$ vs $L$ for the first few system sizes for the range-2 case in OBCs obtained analytically, which agree well with our numerical results.}
\label{table2}
\end{table}

One can also perform a similar calculation in PBCs, which demands six consecutive sites $(L-2,L-1,L,1,2,3)$ not containing any $(1110,1101,0110,0101,1010,1001)$. $N_{froz}$ for first few $L$'s in PBCs is thus given by 
\bea
N_{froz}^{PBC}&=&Y(1,1)+Y(1,3)+Y(1,4)+Y(1,7)+Y(1,8)\non\\&&+Y(4,5)+Y(4,8)+Y(5,8)+Y(6,5)+Y(6,8)\non\\&&+Y(7,2)+Y(7,5)+Y(7,6)+Y(7,8)\non\\&&+Y(8,7)+Y(8,5)+Y(8,6)+Y(8,8),
\eea
as also shown in Table \ref{table3}.
\begin{table}[htb]
\begin{tabular}{|c|c|c|c|c|c|c|c|c|c|c|c|c|}
\hline
$L$&4&5&6&7&8&9&10&11&12&13&14 \\
\hline
$N_{froz}$&6&7&8&9&14&20&27&35&48&67&93\\
\hline
\end{tabular}
\caption{ $N_{froz}$ vs $L$ for the first few system sizes for the range-2 constraint in PBCs, which has been acquired analytically. Furthermore, these agree well with numerically obtained values.}
\label{table3}
\end{table}

Further, we validate the analytical results numerically utilizing the numerical basis enumeration method.
\subsection{range-3}
In a similar manner~\cite{generating1994,Deepak_HSF,Dhar_1993,menon_1997,HariMenon_1995}, one can also count the number of frozen states in the range-3 case, which requires the construction of a $16\times 16$ transfer matrix, $T_{5}$, as examined in the previous section. The constraint on the resultant transfer matrix would be to exclude all 16 resultant configurations allowed by the transitions enabled the range-3 constraints, which are $11110$, $11010$, $10110$, $01110$, $10010$, $01010$, $00110$, $11101$, $11001$, $10101$, $01101$, $10001$, $01001$, $00101$. The resultant transfer matrix for this case thus takes the following form
\bea
T_{5}=\begin{pmatrix}
    1 & 0 & 0& 0& 0& 0& 0 & 0 & 0 & 0 & 0 & 0 & 0 & 0 & 0 & 0 \\  0 & 0 & 0& 0& 0& 1& 0 & 0 & 0 & 0 & 0 & 0 & 0 & 0 & 0 & 0\\ 0 & 0 & 0& 1& 0& 0& 0 & 0 & 0 & 0 & 0 & 0 & 0 & 0 & 0 & 0\\ 0 & 0 & 0& 0& 1& 0& 0 & 0 & 0 & 0 & 0 & 0 & 0 & 0 & 0 & 0\\1 & 0 & 0& 0& 0& 0& 0 & 0 & 0 & 0 & 0 & 0 & 0 & 0 & 0 & 0\\ 0& 0 & 0& 0& 0& 0& 0 & 0 & 0 & 0 & 0 & 1 & 0 & 0 & 0 & 0\\ 0 & 0 & 0& 0& 0& 1& 0 & 0 & 0 & 0 & 0 & 0 & 0 & 0 & 0 & 0\\ 0& 0 & 0& 0& 1& 0& 0 & 0 & 0 & 0 & 0 & 0 & 0 & 0 & 0 & 0\\ 0 & 0 & 0& 0& 0& 0& 0 & 1 & 0 & 0 & 0 & 0 & 0 & 0 & 0 & 0\\ 0 & 0 & 0& 1& 0& 0& 0 & 0 & 0 & 0 & 0 & 0 & 0 & 0 & 0 & 0\\ 0 & 0 & 0& 0& 0& 0& 0 & 0 & 0 & 0& 0 & 0 & 1 & 0 & 0 & 0\\ 0 & 0 & 0& 0& 0& 0& 0 & 0 & 0 & 0 & 0 & 0 & 0 & 0 & 0 & 1\\ 0 & 0 & 0& 0& 0& 0& 0 & 0 & 0 & 0 & 0 & 1 & 0 & 0 & 0 & 0\\ 0 & 0 & 0& 0& 0& 0& 0 & 0 & 0 & 0 & 0 & 0 & 1 & 0 & 0 & 0\\ 0 & 0 & 0& 0& 0& 0& 0 & 1 & 0 & 0 & 0 & 0 & 0 & 1 & 0 & 0\\ 0 & 0 & 0& 0& 0& 0& 0 & 0 & 0 & 0 & 0 & 0 & 0 & 0 & 1 & 1\\

\end{pmatrix},
\label{trans4}
\eea
where eigenvalues of Eq. \eqref{trans4} are given by
1.3247,1, $-0.6624\pm0.5623i$, $0.5\pm0.8667i$, and 10 0's. Therefore, $N_{froz}$ in this case grows as $1.3247^{L}$. $N_{froz}$ for the first few $L$'s in OBCs are shown in Table \ref{table46}, using the formula $N_{froz}=\sum_{ij}Y_{ij}$, where $Y=T_{5}^{L-4}$.
\begin{table}[h]
\begin{tabular}{|c|c|c|c|c|c|c|c|c|c|c|c|c|}
\hline
$L$&5&6&7&8&9&10&11&12&13&14&15 \\
\hline
$N_{froz}$&18&21&27&39&56&75&97&125&165&222&298\\
\hline
\end{tabular}
\caption{ Analytically obtained $N_{froz}(L)$ vs $L$ for the first few system sizes in the range-3 model with OBCs, which are in perfect agreement with numerically obtained values.}
\label{table46}
\end{table}
A similar calculation can be extended to PBCs as well, like the range-2 case. However, this would be an extremely tedious task. Moreover, we validate our analytically obtained values using the numerical basis enumeration method, which is in again perfect agreement with one another.
\section{Combinatorics method: Frey Sellers sequence}
\label{combo}

Here, we will review the combinatorics method primarily employed to compute the dimension of the fragments in this family of models with range-$q$ constraint.
Bailey first derived a formula to compute the number of sequences with the form $a_{1}a_{2}a_{3}\cdots a_{n+k}$ comprising $n$ 1's and $k$ -1's~\cite{B1} such that
$a_{1}+a_{2}+...+a_{j}\geq0$, for all $1\leq j \leq n+k$. Alternatively, this can be equivalent to computing the number of strings made of $n$ $X$'s and $k$ $Y$'s such that no initial configuration has more $Y$'s than $X$'s. Bailey showed the number of such sequences $C_{1}(n,k)$ satisfies the following recursion relation
\bea C_{1}(n,0)&=&1~~~\text{for}~~n\geq0,\non\\
C_{1}(n,1)&=&1~~~\text{for}~~n\geq1,\non\\
C_{1}(n+1,k)&=&C_{1}(n+1,k-1)\,+\,C_{1}(n,k)\non\\&&~~~\text{for}~~1\leq n\leq n+1,\non\\
C_{1}(n+1,n+1)&=&C_{1}(n+1,n)~~~\text{for}~~n\geq1.
\eea
Further, the general formula for such a sequence with $n$ 1's and $k$ -1's is given by
\bea
C_{1}(n,k)\,=\,\frac{n-k+1}{n+1}\left(\begin{array}{cc}  n+k \\ k \end{array}\right),\label{cata}
\eea
which are the members of the Catalan triangle~\cite{catalan} of order $l=1$. In addition, Eq. \ref{cata} reduces to the Catalan number for $n$ 1's and $n$ 0's. We have already argued that the dimensional growth of all the fragments in the range-1 model can be computed with the help of this combinatorics sequence.

A generalization of the Catalan triangle~\cite{catalan} has been introduced later by Frey and Sellers~\cite{FS_article}, which counts the number of sequences $a_{1}a_{2}\cdots a_{n+r}$ with $n$ $(m-1)$'s and $r$ -1's for $m$ larger than 1. The number of such sequences again satisfies a constraint similar to the previous case as 
\bea
a_{1}+a_{2}+\cdots+a_{j}\geq 0~~{\rm for}~~ j=1,2,\cdots, (n+r). \non\\
\label{Fsconstraint}
\eea
They showed that such recurrences satisfy the following relations
\bea
\left\{\begin{array}{cc} n\\ r\end{array}\right\}_{m-1}&=&\left\{\begin{array}{cc} n-1\\ r\end{array}\right\}_{m-1}+\left\{\begin{array}{cc} n\\ r-1\end{array}\right\}_{m-1},\non\\
&&~~~~~~~\text{for}~~~~1<r<n,\non\\
\left\{\begin{array}{cc} n\\ (m-1)n\end{array}\right\}_{m-1}&=&\left\{\begin{array}{cc} n\\ (m-1)n-1\end{array}\right\}_{m-1}\non\\&&=
\left\{\begin{array}{cc} n\\ (m-1)n-2\end{array}\right\}_{m-1}\non\\
&&\cdots\non\\
&&=\left\{\begin{array}{cc} n\\ (m-1)n-(m-1)\end{array}\right\}_{m-1}.\label{exactFS}
\eea
In that paper~\cite{FS_article}, the authors also derived the form of $\left\{\begin{array}{cc}n\\r\end{array}\right\}_{m-1}$ for all $r$ with $1\leq r\leq (m-1)n$, which reads as
\bea
&&\left\{\begin{array}{cc} n\\ r\end{array}\right\}_{m-1}=\left(\begin{array}{cc}n+r\\n\end{array}\right)-\left\la\begin{array}{cc}n\\r \end{array}\right\ra_{m-1},\non\\
&&\left\la\begin{array}{cc} n\\r\end{array}\right\ra_{m-1}=\sum_{k=0}^{P}\frac{1}{(m-1)k+1}\left(\begin{array}{cc} mk\\k\end{array}\right)\non\\&&
\times\left(\begin{array}{cc}n+r-mk-1\\n-k\end{array}\right),\non\\
&&\text{where}~~~P=\large\lceil\frac{r}{m-1}\large\rceil-1.\non\\
\label{FS}
\eea
They also managed to derive a closed form~\cite{FS_article} of the Eq. \ref{FS} when $r=(m-1)n$, which is given below 
\bea
\left\{\begin{array}{cc}n\\(m-1)n\end{array}\right\}_{m-1}=\frac{1}{(m-1)n+1}\left(\begin{array}{cc}mn\\ n\end{array}\right).\label{exactFS2}
\eea
The expression shown in Eq. \ref{exactFS} is widely known as the Fuss-Catalan sequence~\cite{Fuss}. 

We now elucidate how this Frey-Sellers sequence is related to the growth of fragments in the facilitated East model with a varying range of constraints. In particular, one can show that the members of a dominant class of fragments labeled by the root state (we call it FS root states) for the range-$q$ model with $\psi_{M}\otimes\psi_{R}$ comprising $n$ 1's and $r$ 0's follow the same constraint imposed by Frey and Sellers with $m=(q+1)$. This correspondence becomes evident after mapping each 1 and 0 in the binary strings to the integers $q$ and $-1$ in the sequence of numbers following FS combinatorics, respectively.

To illustrate this mapping, consider an explicit example of dimensional counting for the range-2 model. We consider the root state composed of $\psi_L = 0$, $\psi_M = 100$, and $\psi_R = 1$. Using the local transitions allowed in the range-2 model (as described by Eq. \eqref{Hamr123}), one can verify that this root state leads to three accessible binary string configurations: 01001, 01010, and 01100.

To relate these to the FS sequence, we first exclude the leftmost bit $\psi_L$ from the counting, as it remains dynamically frozen. Then, we map 1's and 0's in the remaining strings to 2's and -1's, respectively. Applying this transformation to the configurations 1010, 1001, and 1100 yields the corresponding sequences in terms of 2's and -1's.  It can now be noted that all of them satisfy the FS constraint described in Eq. \eqref{Fsconstraint}, under the transformation. This thus implies that this fragment belongs to FS combinatorics. In this context, we further note that the value of $m$ for the FS series is $(q+1)$ for the range-$q$ model ($m=3$ for the range-2 case), which provides the correct dimensional counting for the FS class of root states.  In the particular case considered earlier, the mapped FS sequences contain $n = 2$ instances of the number $2$ (from binary 1’s) and $r = 2$ instances of $-1$ (from binary 0’s). Furthermore, this pair $(n, r)$ satisfies the FS condition $r=(m-1)n-(m-1)$. Due to this relation, using Eqs. \eqref{exactFS} and \eqref{exactFS2}, we analytically show the number of such FS sequences is exactly three
—precisely matching the number of allowed binary configurations obtained from the local dynamics of the range-2 East model.  
\section{Distribution of fragments in the range-2 model}
\label{distribution}
\begin{figure*}
\includegraphics[width=0.955\hsize]{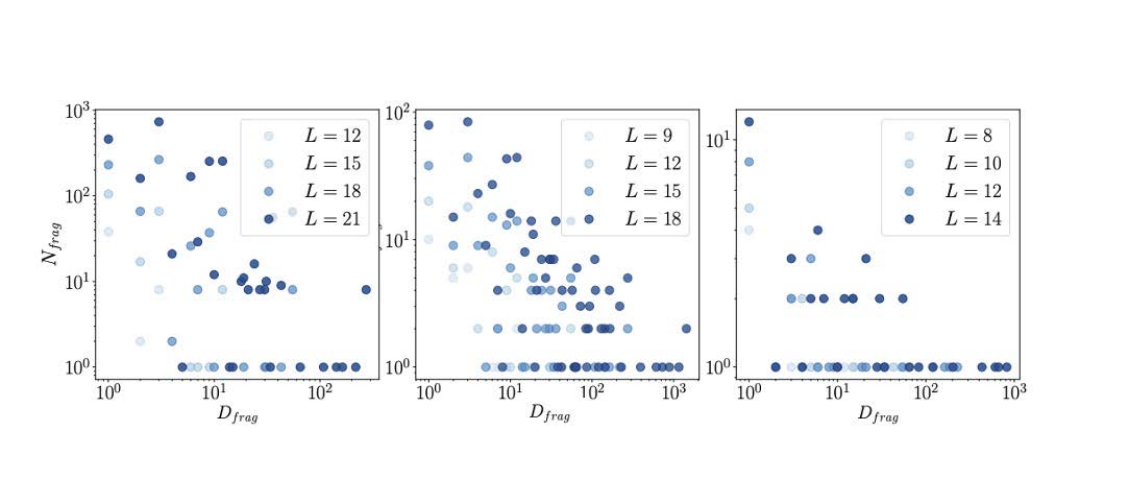}
\caption{Number of fragments ($N_{frag}$) vs. size of the fragments ($D_{frag}$) for different $L$'s in the case of range-2 constraint at freezing transition (middle panel), strongly fragmented (left panel) and weakly fragmented side of the Hilbert space (right panel), respectively. In all three cases, we note a common feature, i.e., $N_{frag}$ is the largest for the frozen fragments comprising one state and integrable fragments including few numbers of states, whereas $N_{frag}$ for large enough fragments with exponential asymptotic growth in $L$ are comparatively much smaller. It is evident that $N_{frag}$ for frozen and simple integrable fragments gradually diminishes as one moves from the strongly to weakly fragmented part of the Hilbert space (left-to-right panel). Apart from that, the number of the largest fragments at the critical filling ($N_{f}=L/3$), below the critical filling ($N_{f}=L/3-2$), and above the critical filling ($N_{f}=L/2$) appear to be two, eight, and one, regardless of system sizes.
%Left panel: Exactly at the critical filling, we observe that $N_{frag}$ is the largest for the frozen fragments and integrable fragments, including very few states. While the other extreme, i.e., $N_{frag}$ for the largest fragment, is exactly equal to two, irrespective of the system size. Middle panel: On the strongly fragmented part of the Hilbert space, we again see a similar feature; however, the number of largest fragments in this case turns out to be five, which is distinct from the earlier case. Right panel: In this case, $N_{frag}$ for the frozen states and integrable fragments appears to be much smaller than the earlier two cases, although their numbers are still the largest among all other fragments at this given filling fraction. Moreover, $N_{frag}$ for the largest fragment is exactly equal to one with no degeneracy, unlike the earlier two cases
}
\label{dis_fragments}
\end{figure*}

We now study the distribution of fragments~\cite{sala_ergo_2020,Deepak_HSF} in the range-2 constraint in OBCs for three specific cases: below the critical filling (strong fragmentation), at the critical filling, and above the critical filling (weak fragmentation), as depicted in Fig. \ref{dis_fragments} (left to right panel, respectively). In all three cases, we observe $N_{frag}$ is the largest for the frozen and integrable fragments with a single and a few states, respectively, while the same gradually decreases for typical non-integrable fragments comprising an exponentially large number of states. Furthermore, the number of the former generally reduces as one goes from the strong-to-weak fragmented side of the Hilbert space, as can be seen in Fig. \ref{dis_fragments}. Simultaneously, we also note another interesting feature that $N_{frag}$ of the largest fragment turns out to be distinct depending on the filling fraction, where $N_{frag}$ are two, eight, and one at the critical filling, strongly fragmented and weakly fragmented part of the Hilbert space, respectively, irrespective of system sizes. Such multiplicities of $D_{max}$ can be easily understood by taking the root structure representing the fragments into account. Precisely at the freezing transition, i.e., $N_{f}=L/3$, the largest fragments are represented by two root states, one with $(\psi_{L}, \psi_{M},\psi_{R})=(00, 100\cdots 100, 1$), and the other with $(0,100\cdots 100, 10$), respectively. Both of these fragments follow the same asymptotic growth dictated by the FS sequence with $m=3$~\cite{FS_article}, and hence $N_{frag}$ two as shown in the middle panel in Fig. \ref{dis_fragments}.
While the largest fragment at $N_{f}=L/3-2$ is labeled by the following eight root states with $(\psi_{L},\psi_{M},\psi_{R})=(0^{k_{1}},(100)^{(L/3-3)},10^{k_{2}})$, where $k_{1}+k_{2}=8$, and $k_{1}=1,\cdots,8$, and consequently, $N_{frag}$ reduces to eight as illustrated in the left panel of Fig. \ref{dis_fragments}. One can also show that the increase in the largest fragments occurs linearly with the decrease in the filling fraction below the freezing transition, and the number can be shown to be $(3a+2)$ at $N_{f}=L/3-a$ with $a>0$. Similarly, it can be argued that the largest fragment on the weakly fragmented side of the Hilbert space is described by the root state involving $\psi_{L}=0$ and $\psi_{M}\otimes\psi_{R}$ is fixed by $L$ and $N_{f}$. Consequently, there is only one such fragment in this case, as shown in the right panel of Fig. \ref{dis_fragments}. 

We also note that these features remain intact for all the models lying in this class of constraints. However, the multiplicities will be different depending on the root structure describing the largest fragment for those specific cases.
\section{The dimensional growth of non-FS root states in range-2 and range-3 models}
\label{nonFS}
We will now discuss some examples of fragments that do not belong to the FS class in the context of range-2 and range-3 constraints.

\subsection{range-2}
As emphasized earlier in Sec. \ref{root}, the primary distinct feature of these non-FS root states in the range-2 model occurs when $\psi_{L}$ is a null set ($\phi$). We will now elucidate a particular example whose dimensional growth can be captured utilizing another generalization of Catalan sequence, i.e., the concept of $k_{t}$-Dyck path. We first assume two sets of root states, $100...1001$ (root-1) and $0100100..1001$ (root-2), which differ from one another by a single 0 in $\psi_{L}$. The latter one falls under the FS class, while the earlier one does not. A simple numerical basis enumeration provides the following dimensional growth for these two root states with increasing lengths of $\psi_{M}$ as shown in Table-\ref{nonFS2}.
\begin{table}[htb]
\begin{tabular}{|c|c|c|c|c|c|c|c|c|c|c|c|c|}
\hline
No. of $100$ in $\psi_{M}$&0&1&2&3&4&5&6\\
\hline
$D^{root_{1}}(L)$&1&2&7&30&143&728&3876\\
\hline
$D^{root_{2}}(L)$&1&3&12&55&273&1428&7752\\
\hline
\end{tabular}
\caption{ Numerically obtained dimension of the fragments labeled by two root states with $\psi_{M}=100$'s, $\psi_{R}=1$ and $\psi_{L}=\phi$ and $0$, respectively, in the range-2 case. The second root state falls under the FS class with $m=3$, while the first one does not.}
\label{nonFS2}
\end{table}
 It can be readily shown that all the members lying in the fragment represented by the $root_{2}$ satisfy the generalization of the Catalan constraint introduced by Frey and Sellers. On the other hand, $\psi_{L}=\phi$ fails to generate all the transitions allowed by the FS sequence~\cite{FS_article}. Before proceeding further, we first introduce the concept of $k_t$-Dyck path~\cite{ALET2018498} with $n$ up-steps, which is a special lattice path in the two-dimensional $x-y$ plane, where each up and down step is represented by (1, $k$) and (1,\,-1), steps, respectively, in the coordinate space. Additionally, such paths start from the origin, terminate at $(n+1)k,0)$, and must remain weakly above the $y=-t$ line. It can also be shown that the number of transitions allowed by $k_{t}$-Dyck path with $n$ upsteps is 
\bea
C_{n}=\frac{(t+1)}{(t+1)+(k+1)n}\frac{\left((k+1)n+t+1\right)!}{n! (t+1+kn)!}.\label{kteq}
\eea
We will now argue that the growth of such fragments generated by the first representative root state, $root_{1}$, can be captured utilizing the combinatorics of $2_{1}$-Dyck path. To show that, we perform a mapping of the allowed transitions to the configuration enabled by the $2_{1}$-Dyck path~\cite{Asinowski_2022}, which can be done as follows. At first, one has to remove the first 10, which facilitates transitions yet remains unaltered during transitions. Thereafter, we map the binary bits 1 and 0 to up-steps and down-steps, respectively. In addition, it can also be checked that such Dyck paths always end with a downstep (due to the constraint given by $t=-1$), and we, therefore, insert a virtual downstep ($0$) at the end of all the allowed transitions since this is a trivial operation (the last $0$ remains dynamically frozen). The mapping is displayed in Table-\ref{root2} for the first few $L$'s, which reveals that the allowed binary strings are the exact paths enabled by the $2_{1}$-Dyck paths. Therefore, inserting $k=2$ and $t=1$ in Eq. \ref{kteq}, one can easily obtain the dimension of such fragments as
\bea
D_{root_{1}}(L)=\left(\frac{2}{2+3n}\right)\frac{(3n+2)!}{n! (2n+2)!},
\eea
where $n=(L-1)/3$ are the numbers of 1's in the root state (excluding the first 1), which perfectly agrees with the numerically obtained values. In the thermodynamic limit, this fragment again demonstrates the growth of $\left(3/2^{2/3}\right)^{L}\simeq 1.89^{L}$, which is the same asymptotic growth given by the root-1 state, as discussed earlier utilizing the FS sequence. Nevertheless, the non-FS root states with $\psi_{L}=\phi$ always generate less number of transitions compared to their FS counterparts involving $\psi_{L}=1$ or $0$ (while $\psi_{M}\otimes \psi_{R}$ for both cases are identical) at finite $L$'s. Hence, the information on FS root states is sufficient to understand the detailed fragmentation structure of this family of models. 
\begin{table*}[htb]
\begin{center}
\begin{tabular}{|c|c|c|c|c|c|c|c|c|c|c|c|c|c|}

\hline
Number of  & Root state& $D_{frag}$ &The allowed processes& Dyck path notations of allowed processes\\
$100$'s in $\psi_{M}$&&&&\\
\hline

1& 1001& 2 & 1001, 1010 & \begin{tikzpicture}[scale=0.33]
\draw[step=1cm,gray,very thin] (0,-1) grid (3,1);
\draw[thick,->] (0,0) -- (1,-1);
\draw[thick,->] (1,-1) -- (2,1);
\draw[thick,->][dotted] (2,1) -- (3,0);
\end{tikzpicture}
\begin{tikzpicture}[scale=0.33]
\draw[step=1cm,gray,very thin] (0,0) grid (3,2);
\draw[thick,->] (0,0) -- (1,2);
\draw[thick,->] (1,2) -- (2,1);
\draw[thick,->] [dotted](2,1) -- (3,0);
\end{tikzpicture}\\
\hline
2& 1001001&7&1001001, 1001010, 1001100, &\begin{tikzpicture}[scale=0.3]
\draw[step=1cm,gray,very thin] (0,-1) grid (6,1);
\draw[thick,->] (0,0) -- (1,-1);
\draw[thick,->] (1,-1) -- (2,1);
\draw[thick,->] (2,1) -- (3,0);
\draw[thick,->] (3,0) -- (4,-1);
\draw[thick,->] (4,-1) -- (5,1);
\draw[thick,->][dotted] (5,1) -- (6,0);
\end{tikzpicture}
\begin{tikzpicture}[scale=0.3]
\draw[step=1cm,gray,very thin] (0,-1) grid (6,2);
\draw[thick,->] (0,0) -- (1,-1);
\draw[thick,->] (1,-1) -- (2,1);
\draw[thick,->] (2,1) -- (3,0);
\draw[thick,->] (3,0) -- (4,2);
\draw[thick,->] (4,2) -- (5,1);
\draw[thick,->][dotted] (5,1) -- (6,0);
\end{tikzpicture}
\begin{tikzpicture}[scale=0.3]
\draw[step=1cm,gray,very thin] (0,-1) grid (6,3);
\draw[thick,->] (0,0) -- (1,-1);
\draw[thick,->] (1,-1) -- (2,1);
\draw[thick,->] (2,1) -- (3,3);
\draw[thick,->] (3,3) -- (4,2);
\draw[thick,->] (4,2) -- (5,1);
\draw[thick,->][dotted] (5,1) -- (6,0);
\end{tikzpicture}\\
&&&1010100, 1011000, 1010001, &\begin{tikzpicture}[scale=0.3]
\draw[step=1cm,gray,very thin] (0,0) grid (6,3);
\draw[thick,->] (0,0) -- (1,2);
\draw[thick,->] (1,2) -- (2,1);
\draw[thick,->] (2,1) -- (3,3);
\draw[thick,->] (3,3) -- (4,2);
\draw[thick,->] (4,2) -- (5,1);
\draw[thick,->][dotted] (5,1) -- (6,0);
\end{tikzpicture}
\begin{tikzpicture}[scale=0.3]
\draw[step=1cm,gray,very thin] (0,0) grid (6,4);
\draw[thick,->] (0,0) -- (1,2);
\draw[thick,->] (1,2) -- (2,4);
\draw[thick,->] (2,4) -- (3,3);
\draw[thick,->] (3,3) -- (4,2);
\draw[thick,->] (4,2) -- (5,1);
\draw[thick,->][dotted] (5,1) -- (6,0);
\end{tikzpicture}
\begin{tikzpicture}[scale=0.3]
\draw[step=1cm,gray,very thin] (0,-2) grid (6,2);
\draw[thick,->] (0,0) -- (1,2);
\draw[thick,->] (1,2) -- (2,1);
\draw[thick,->] (2,1) -- (3,0);
\draw[thick,->] (3,0) -- (4,-1);
\draw[thick,->] (4,-1) -- (5,1);
\draw[thick,->][dotted] (5,1) -- (6,0);
\end{tikzpicture}\\
&&&1010010 &
\begin{tikzpicture}[scale=0.3]
\draw[step=1cm,gray,very thin] (0,-1) grid (6,2);
\draw[thick,->] (0,0) -- (1,2);
\draw[thick,->] (1,2) -- (2,1);
\draw[thick,->] (2,1) -- (3,0);
\draw[thick,->] (3,0) -- (4,2);
\draw[thick,->] (4,2) -- (5,1);
\draw[thick,->][dotted] (5,1) -- (6,0);
\end{tikzpicture}\\

\hline
\end{tabular}
\end{center}
\caption{A specific example of non-FS root state in the range-2 case with $\psi_{L}=\phi$, $\psi_{M}=100$'s and $\psi_{R}=1$, whose growth can be captured using the $k_{t}$-Dyck sequence with $k=2$, and $t=1$ after mapping appropriately. The mapping between binary strings and the Dyck path is displayed above and can be achieved after removing the first 10 since it remains unaltered during transitions due to the range-2 constraint in OBCs. The last down step in the dotted line can be trivially added since all the $2_{1}$-Dyck paths must end with a down step (or in the binary bits language, the last $0$ always remains dynamically inactive).}
\label{root2}
\end{table*}
\subsection{range-3}
In a similar manner, one can also argue that there are specific non-FS root states that include $\psi_{L}=\phi$ or $0/1$, and their dimensional growth can be captured utilizing the combinatorics sequence of $3_{t}$-Dyck paths~\cite{Asinowski_2022}. Also, like the range-2 case, although these non-FS root states follow the same asymptotic growth as their FS counterpart (which means $\psi_{M}$ and $\psi_{R}$ for all cases are identical), such fragments always include comparatively fewer numbers of binary strings at finite $L$'s. To prove this claim, we assume three root states from three different combinatorics families with identical $L$ and $N_{f}$, which are $1000\cdots10001000$, $01000\cdots1000100$ and $001000\cdots 100010$, respectively, with $\psi_{L}$ being $\phi$, $0$ and $00$, respectively. $\psi_{M}$ is units of $1000$'s and $\psi_{R}$ are 1000, 100, and 10, respectively. The dimensions of fragments generated from these root states using a simple numerical enumeration are illustrated in Table-\ref{growth_range3} for the first few $L$'s. One can readily check the transitions enabled by $root_{3}$ follow the FS sequence with $m=4$, while the other two do not. We will now show that the same for the first two root states can be captured using the $3_{t}$-Dyck path. Further, it should be noted that the 0's followed by 1 in $\psi_{R}$ are dynamically frozen binary strings. Nevertheless, we will not deduct them immediately in order to take advantage during binary string to Dyck path mapping~\cite{Asinowski_2022}.
\begin{table}[htb]
\begin{tabular}{|c|c|c|c|c|c|c|c|c|c|c|c|c|}
\hline
No. of $100$ in $\psi_{M}$&0&1&2&3&4&5\\
\hline
$D^{root_{1}}(L)$&1&2&9&52&340&2394\\
\hline
$D^{root_{2}}(L)$&1&3&15&91&612&4389\\
\hline
$D^{root_3}(L)$&1&4&22&140&969&7084\\
\hline
\end{tabular}
\caption{ Numerically obtained dimensional growth of fragments for three different families of root states with $\psi_{M}=1000$'s, $\psi_{R}=1000$, $100$, $10$, respectively, and $\psi_{L}=\phi, 0$, and $00$, respectively, in the range-3 case. While the last root lies in the FS class with $m=4$, the other two root states do not follow the FS sequence.}
\label{growth_range3}
\end{table}
Also, inserting $k=3$ and $t=1$ and $t=2$ in Eq. \ref{kteq}, we note that the numerically acquired dimensional growth for these two cases agrees with the number of sequences allowed by $3_{1}$ and $3_{2}$-Dyck paths, respectively, which are
\bea
D^{root_1}(L)=\frac{2}{3n+2}\left(\begin{array}{cc}4n+1 \\n\end{array}\right),\non\\
D^{root_2}(L)=\frac{1}{n+1}\left(\begin{array}{cc}4n+2\\ n\end{array}\right),
\eea
where $n$ is the number of diagonal up-steps in these two Dyck paths, i.e., the number of 1's in the root state after excluding the first three binary bits. The members of the fragment labeled by the first root state and string-to-$3_{1}$-Dyck path mapping are shown in  Table-\ref{r3root}. This is achieved after removing the first three binary bits that remain unaltered during the transitions but facilitate correlated hoppings due to the range-3 constraint. A similar mapping to $3_2$-Dyck path can also be achievable for the second root state. Therefore, it can be concluded that among all three root states with identical $L$ and filling, FS root states cause the highest dimensional growth for finite $L$'s, thus again indicating the information of FS-root states is sufficient to apprehend key features of the nature of fragmentation.

\begin{table*}
\begin{center}
\begin{tabular}{|c|c|c|c|c|c|c|c|c|c|c|c|c|}
\hline
$1000$'s in $\psi_{M}$& Root state& $D_{frag}$ &The allowed processes in qubit language& Dyck path notations of allowed processes\\
\hline
&&&&\\
1& 1000100& 2&1000100, 1001000 &\begin{tikzpicture}[scale=0.33]
\draw[step=1cm,gray,very thin] (0,-1) grid (4,2);
\draw[thick,->] (0,0) -- (1,-1);
\draw[thick,->] (1,-1) -- (2,2);
\draw[thick,->] (2,2) -- (3,1);
\draw[thick,->] (3,1) -- (4,0);
\end{tikzpicture}
\begin{tikzpicture}[scale=0.33]
\draw[step=1cm,gray,very thin] (0,0) grid (4,3);
\draw[thick,->] (0,0) -- (1,3);
\draw[thick,->] (1,3) -- (2,2);
\draw[thick,->] (2,2) -- (3,1);
\draw[thick,->] (3,1) -- (4,0);
\end{tikzpicture}

\\
&&&&\\
\hline
&&&&\\
2& 10001000100&9&10001000100, 10001001000,
 10001010000,  &\begin{tikzpicture}[scale=0.25]
\draw[step=1cm,gray,very thin] (0,-1) grid (8,2);
\draw[thick,->] (0,0) -- (1,-1);
\draw[thick,->] (1,-1) -- (2,2);
\draw[thick,->] (2,2) -- (3,1);
\draw[thick,->] (3,1) -- (4,0);
\draw[thick,->] (4,0) -- (5,-1);
\draw[thick,->] (5,-1) -- (6,2);
\draw[thick,->] (6,2) -- (7,1);
\draw[thick,->] (7,1) -- (8,-0);
\end{tikzpicture}
\begin{tikzpicture}[scale=0.25]
\draw[step=1cm,gray,very thin] (0,-1) grid (8,3);
\draw[thick,->] (0,0) -- (1,-1);
\draw[thick,->] (1,-1) -- (2,2);
\draw[thick,->] (2,2) -- (3,1);
\draw[thick,->] (3,1) -- (4,0);
\draw[thick,->] (4,0) -- (5,3);
\draw[thick,->] (5,3) -- (6,2);
\draw[thick,->] (6,2) -- (7,1);
\draw[thick,->] (7,1) -- (8,-0);
\end{tikzpicture}
\begin{tikzpicture}[scale=0.25]
\draw[step=1cm,gray,very thin] (0,-1) grid (8,4);
\draw[thick,->] (0,0) -- (1,-1);
\draw[thick,->] (1,-1) -- (2,2);
\draw[thick,->] (2,2) -- (3,1);
\draw[thick,->] (3,1) -- (4,4);
\draw[thick,->] (4,4) -- (5,3);
\draw[thick,->] (5,3) -- (6,2);
\draw[thick,->] (6,2) -- (7,1);
\draw[thick,->] (7,1) -- (8,0);
\end{tikzpicture}\\
&&&10010000100, 10001100000, 10010100000, &\begin{tikzpicture}[scale=0.25]
\draw[step=1cm,gray,very thin] (0,-1) grid (8,4);
\draw[thick,->] (0,0) -- (1,3);
\draw[thick,->] (1,3) -- (2,2);
\draw[thick,->] (2,2) -- (3,1);
\draw[thick,->] (3,1) -- (4,0);
\draw[thick,->] (4,0) -- (5,-1);
\draw[thick,->] (5,-1) -- (6,2);
\draw[thick,->] (6,2) -- (7,1);
\draw[thick,->] (7,1) -- (8,0);
\end{tikzpicture}
\begin{tikzpicture}[scale=0.25]
\draw[step=1cm,gray,very thin] (0,-1) grid (8,5);
\draw[thick,->] (0,0) -- (1,-1);
\draw[thick,->] (1,-1) -- (2,2);
\draw[thick,->] (2,2) -- (3,5);
\draw[thick,->] (3,5) -- (4,4);
\draw[thick,->] (4,4) -- (5,3);
\draw[thick,->] (5,3) -- (6,2);
\draw[thick,->] (6,2) -- (7,1);
\draw[thick,->] (7,1) -- (8,-0);
\end{tikzpicture}
\begin{tikzpicture}[scale=0.25]
\draw[step=1cm,gray,very thin] (0,0) grid (8,5);
\draw[thick,->] (0,0) -- (1,3);
\draw[thick,->] (1,3) -- (2,2);
\draw[thick,->] (2,2) -- (3,5);
\draw[thick,->] (3,5) -- (4,4);
\draw[thick,->] (4,4) -- (5,3);
\draw[thick,->] (5,3) -- (6,2);
\draw[thick,->] (6,2) -- (7,1);
\draw[thick,->] (7,1) -- (8,-0);
\end{tikzpicture}\\
&&&10011000000, 10010010000, 10010001000&
\begin{tikzpicture}[scale=0.25]
\draw[step=1cm,gray,very thin] (0,0) grid (8,6);
\draw[thick,->] (0,0) -- (1,3);
\draw[thick,->] (1,3) -- (2,6);
\draw[thick,->] (2,6) -- (3,5);
\draw[thick,->] (3,5) -- (4,4);
\draw[thick,->] (4,4) -- (5,3);
\draw[thick,->] (5,3) -- (6,2);
\draw[thick,->] (6,2) -- (7,1);
\draw[thick,->] (7,1) -- (8,0);
\end{tikzpicture}
\begin{tikzpicture}[scale=0.25]
\draw[step=1cm,gray,very thin] (0,0) grid (8,4);
\draw[thick,->] (0,0) -- (1,3);
\draw[thick,->] (1,3) -- (2,2);
\draw[thick,->] (2,2) -- (3,1);
\draw[thick,->] (3,1) -- (4,4);
\draw[thick,->] (4,4) -- (5,3);
\draw[thick,->] (5,3) -- (6,2);
\draw[thick,->] (6,2) -- (7,1);
\draw[thick,->] (7,1) -- (8,-0);
\end{tikzpicture}
\begin{tikzpicture}[scale=0.25]
\draw[step=1cm,gray,very thin] (0,0) grid (8,3);
\draw[thick,->] (0,0) -- (1,3);
\draw[thick,->] (1,3) -- (2,2);
\draw[thick,->] (2,2) -- (3,1);
\draw[thick,->] (3,1) -- (4,0);
\draw[thick,->] (4,0) -- (5,3);
\draw[thick,->] (5,3) -- (6,2);
\draw[thick,->] (6,2) -- (7,1);
\draw[thick,->] (7,1) -- (8,-0);
\end{tikzpicture}\\
\hline
\end{tabular}
\caption{A particular example of non-FS root state in the range-3 case with $\psi_{L}=\phi$, $\psi_{M}=1000$'s and $\psi_{R}=100$, whose dimensional growth can be captured using $k_{t}$-Dyck sequence with $k=3$, and $t=1$. The mapping between binary strings and the Dyck path is shown in the above table and can be accomplished after removing the first 100 (since dynamically remains unaltered).}
\label{r3root}
\end{center}
\end{table*}
\section{Low-energy excitation spectrum}
\begin{figure*}
\includegraphics[width=1.01\hsize,height=6.2cm]{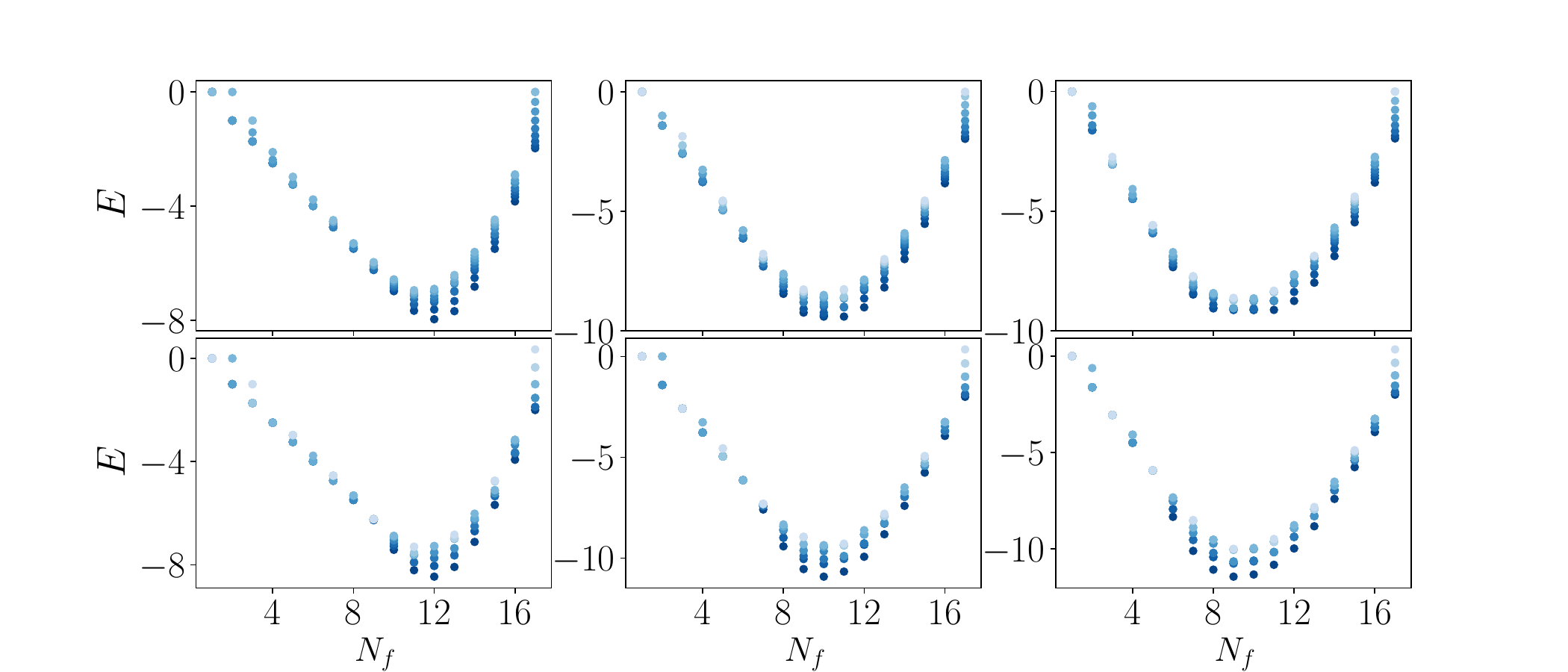}
\caption{Low-energy excitation spectrum for various $N_{f}$'s in the range-1, range-2, and range-3 models with OBCs (top three panels) and PBCs (bottom three panels), respectively, for $L=18$. We note that the ground state fillings in the range-1 case in OBCs (top left panel) and PBCs (bottom left panel) both appear at $N_{f}=L/2+3$. The same investigation in the range-2 case demonstrates two degenerate ($N_{f}=L/2+1,L/2+2$) and a non-degenerate ground state ($N_{f}=L/2+1$) in OBCs (top middle panel) and PBCs (bottom middle panel), respectively. Furthermore, the next low-energy excitation appears at a different filling than the ground state filling in PBCs ($N_{f}=L/2+2$). The features remain qualitatively the same even in the range-3 case, where we witness four degenerate ground states ($N_{f}=L/2,L/2+1, L/2+2$) in OBCs (top right) and a non-degenerate ground state with subsequent low-energy excitation emerging at different fillings in PBCs (bottom right), respectively. } 
\label{lowlying}
\end{figure*}

We now discuss the low-lying excitations in the facilitated East models for the range-1, range-2, and range-3 constraints, respectively, as demonstrated in Fig. \ref{lowlying} both in OBCs (top panel) and PBCs (bottom panel). As shown in Fig. \ref{lowlying}, the ground states in both OBCs (top left panel) and PBCs (bottom left panel) lie at $N_{f}=L/2+3$,  where the groundstate fillings shift with increasing $L$'s due to the fragmentation structure as argued in our previous study~\cite{East_Sreemayee}.
While in the range-2 case, the ground state appears at two distinct fillings in OBCs, $N_{f}=L/2+1$ and $L/2+2$ (as explained before due to the distinct fragmentation structure), as shown in the top middle panel. The same in PBCs turns out to be non-degenerate and is located at $N_{f}=L/2+1$.  The next few low-lying excitations sometimes come from different sectors (shown in the bottom middle panel). One can thus infer that a single filling is not sufficient to capture low-energy properties in this family of models with the increasing range of constraints, even in PBCs. An identical feature has been witnessed in the range-3 case, where four degenerate ground states appear at three distinct fillings due to the fragmentation structure in OBCs (top right panel). This again becomes non-degenerate in PBCs with the next low-lying excitations located at different fillings (bottom right panel). This again indicates that a single filling is insufficient to capture low-energy properties.

\begin{figure*}
\subfigure[]{\includegraphics[width=0.261\hsize,height=3.7cm]{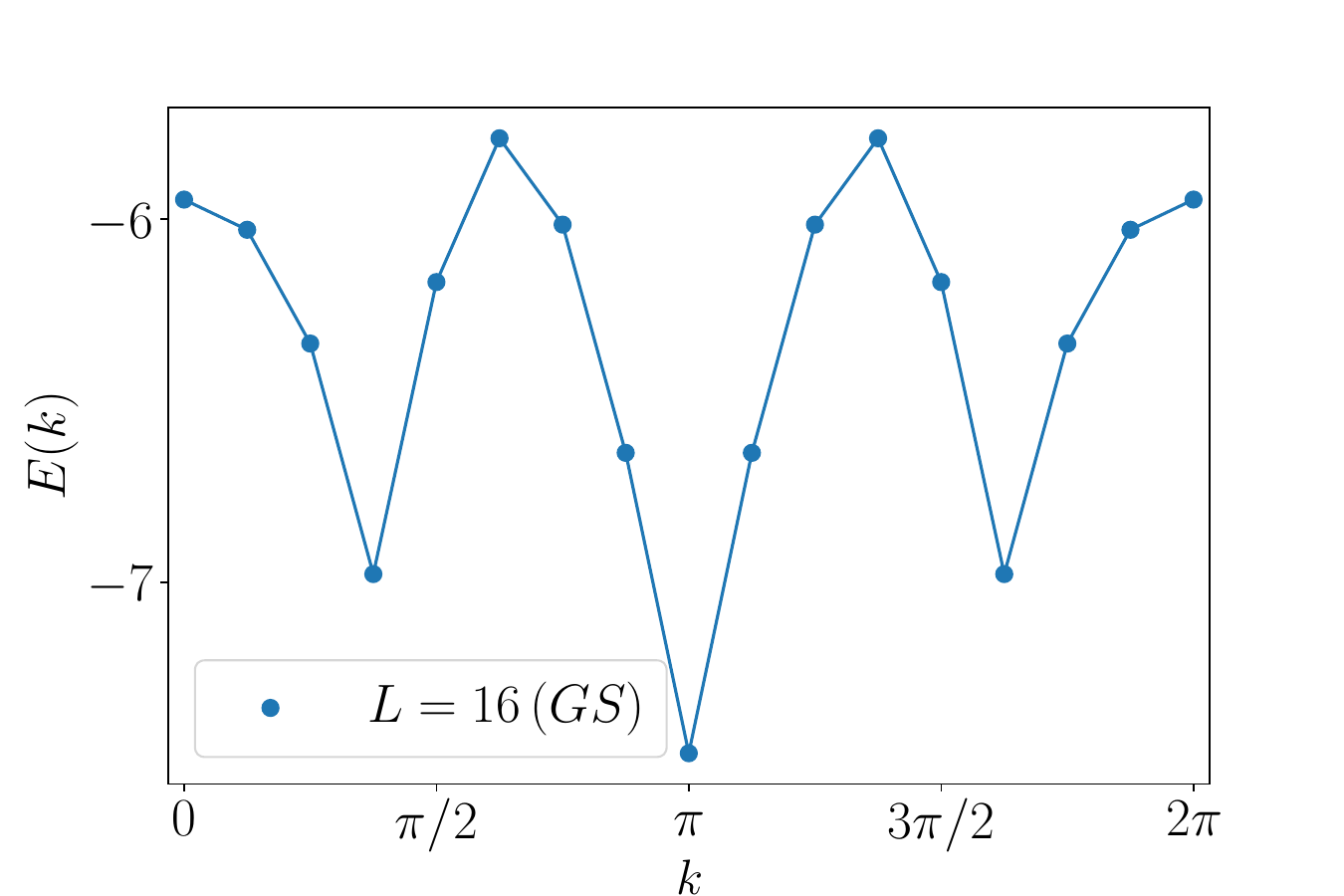}}%
\subfigure[]{\includegraphics[width=0.261\hsize,height=3.7cm]{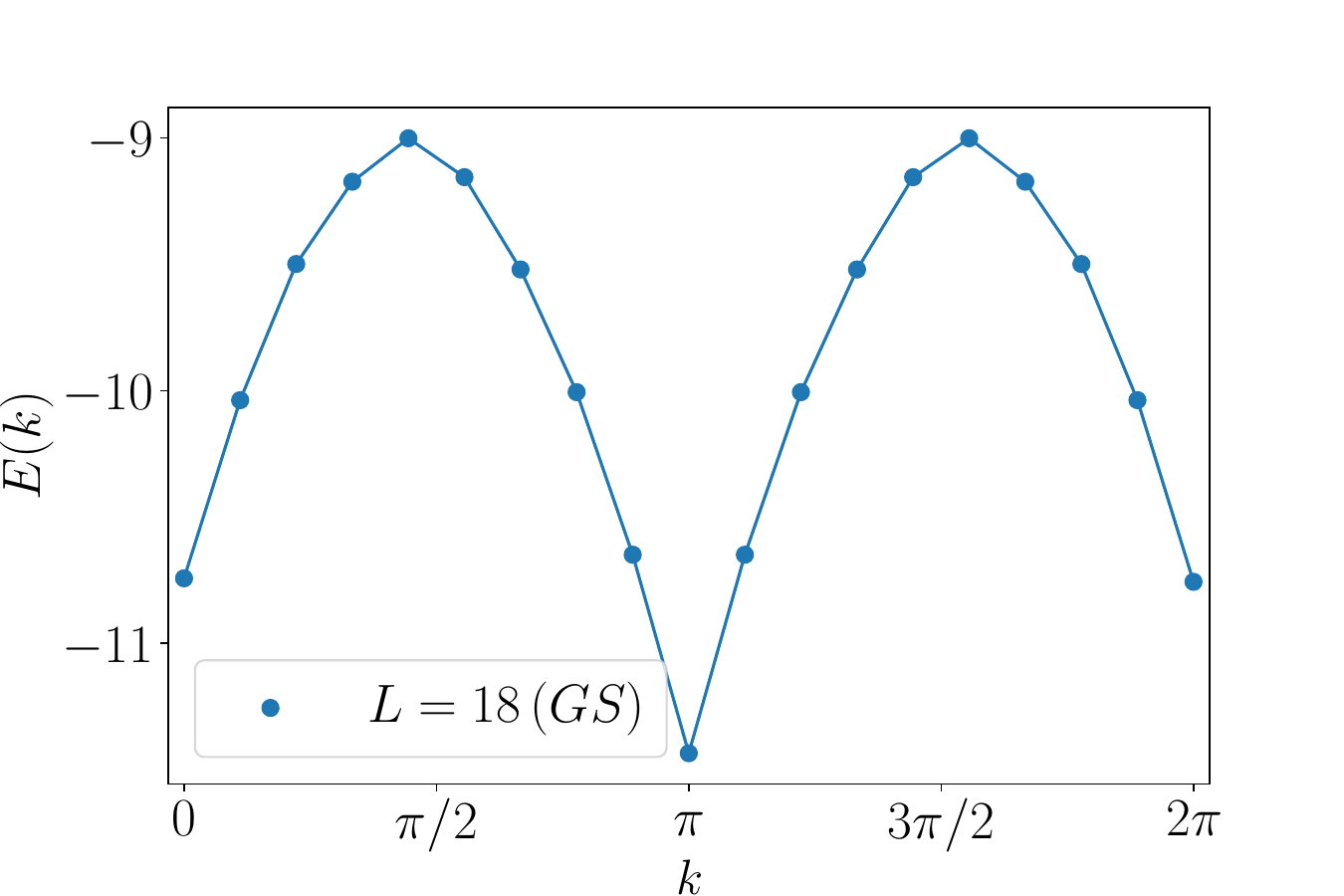}}%
\subfigure[]{\includegraphics[width=0.261\hsize,height=3.7cm]{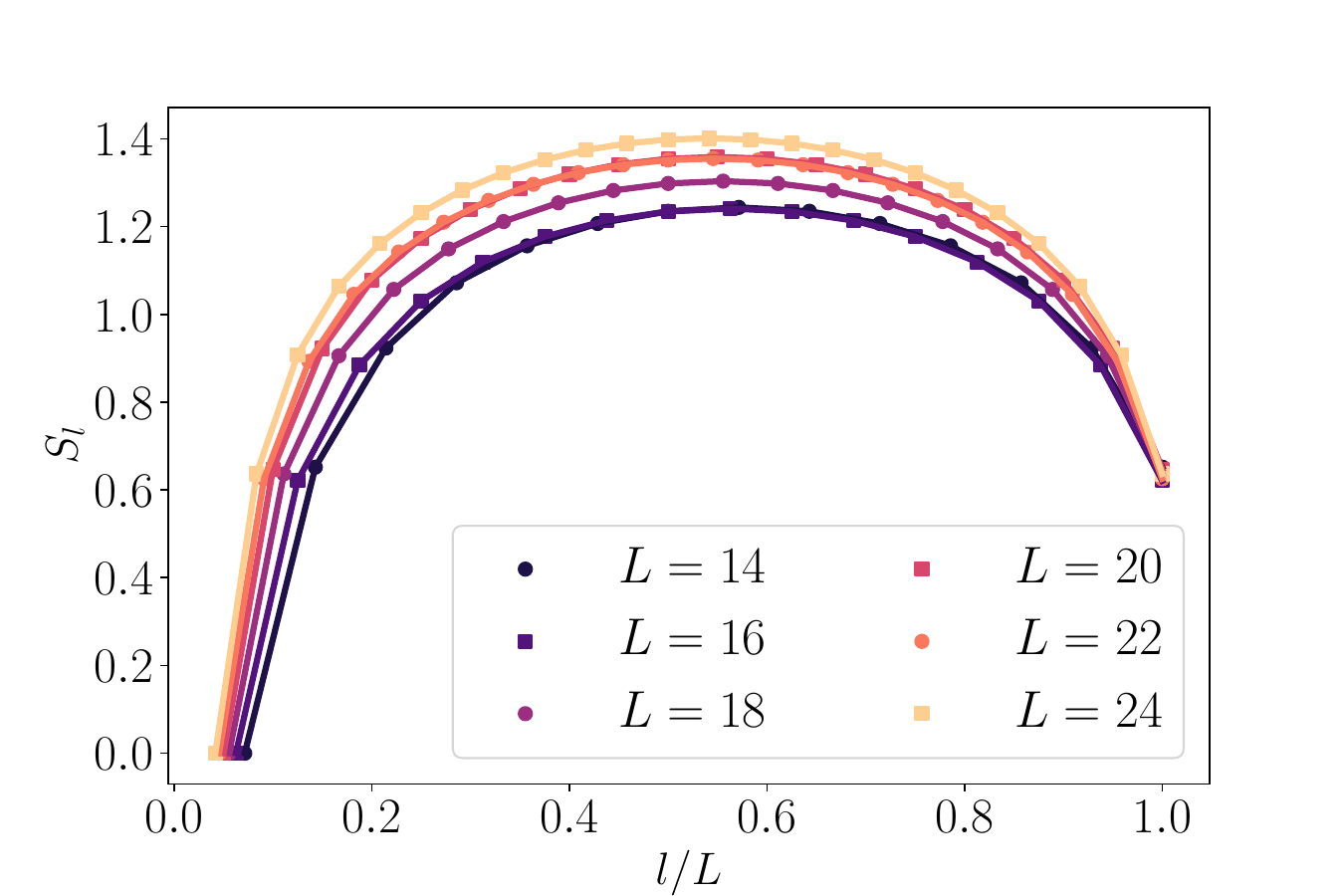}}%
\subfigure[]{\includegraphics[width=0.261\hsize,height=3.7cm]{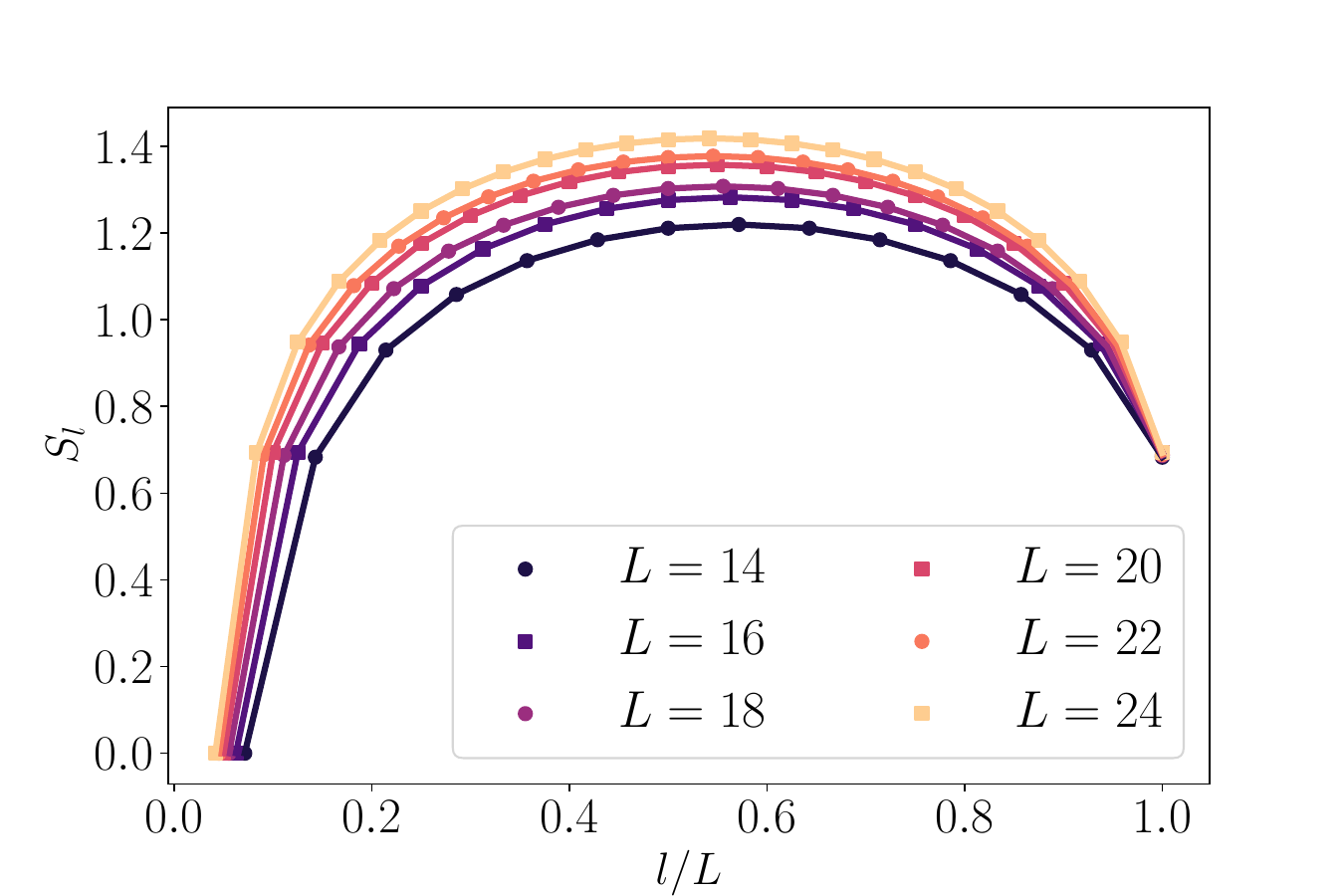}}
\caption{(a-b) $E(k)$ vs $k$ in the range-1 and range-3 cases for $(L,N_{f})=(16,L/2+3)$ and $(18,L/2)$, respectively, (where lies $E_{Gs}$'s in PBCs). (c-d) Bipartite entanglement entropy, $S_{l}$ profile for various cuts, $l$ in the same two cases for different $L$'s in PBCs. (a-b) In both cases, low-lying $E(k)$ exhibits a linear-$k$ dispersion near $k=\pi$ in the BZ, indicating dynamical exponent $z\sim1$. (c-d) $S_{l}$ also exhibits a semicircular profile (excluding edges due to the lack of inversion symmetry), which is typically observed in critical ground states in 1D. This further indicates $log(L)$ scaling of entanglement entropy as we also observed (plots not shown here).}
\label{groundstater2r3}
\end{figure*}

In Fig. \ref{groundstater2r3} (a-b), we exhibit the low-energy dispersion at ground state filling for the range-1 and range-3 cases, which shows linear-$k$ dispersion near certain $k$-points (low-lying excitations coming from only one $k$-point at $k=\pi$ for both cases), like the range-2 case, as shown in Fig. \ref{gsr2_PBC} (c). In addition, the variation of bipartite entanglement entropy $S_{l}$ with various cuts, $l$ displays a semicircular profile (except near the edges due to the lack of inversion symmetry); this is typically observed in one-dimensional local Hamiltonians embodying critical ground states.
\section{Krylov-restricted thermalization in longer-range models}
\begin{figure*}
\subfigure[]{\includegraphics[width=0.5\linewidth,height=4.8cm]{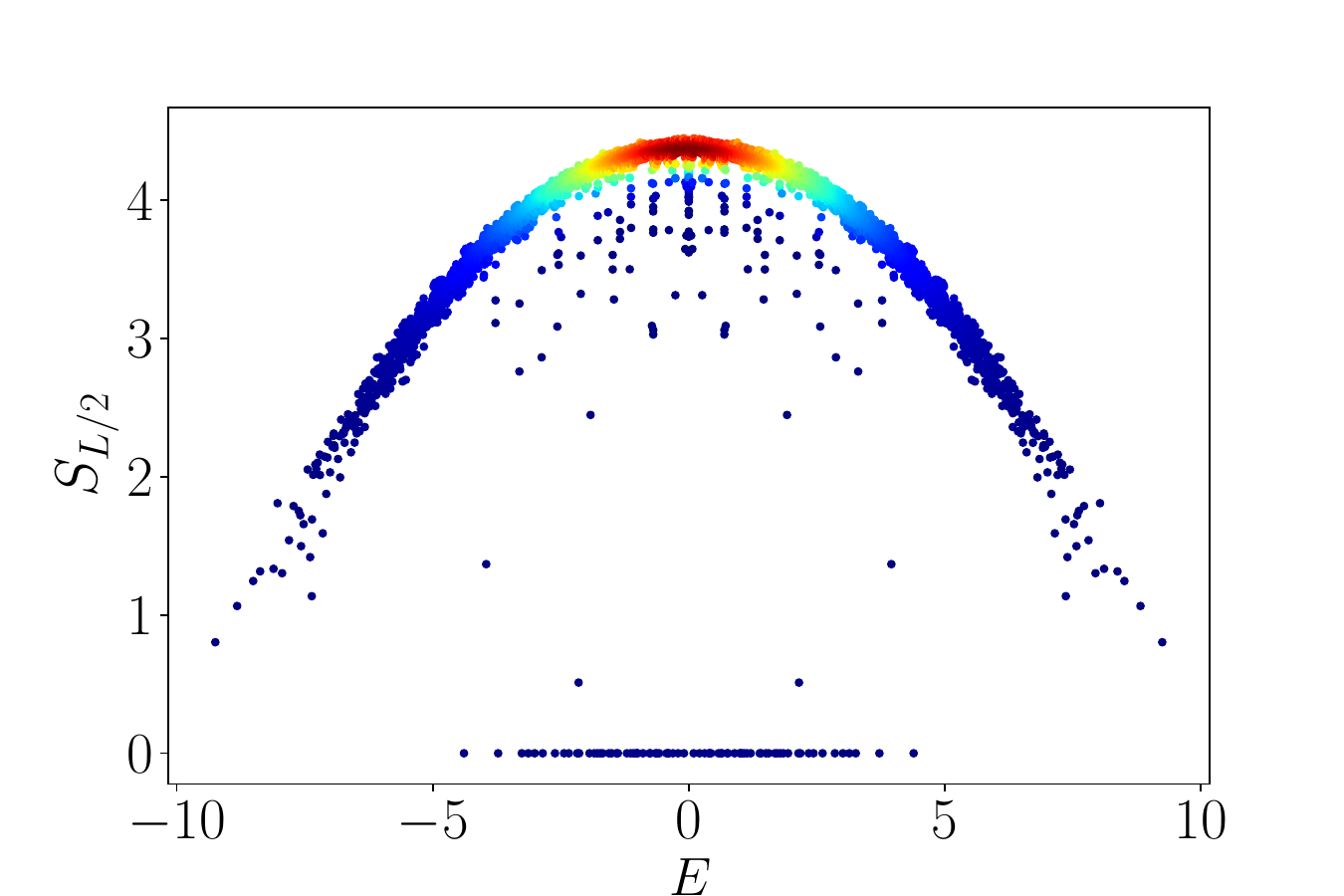}}%
\subfigure[]{\includegraphics[width=0.5\linewidth,height=4.8cm]{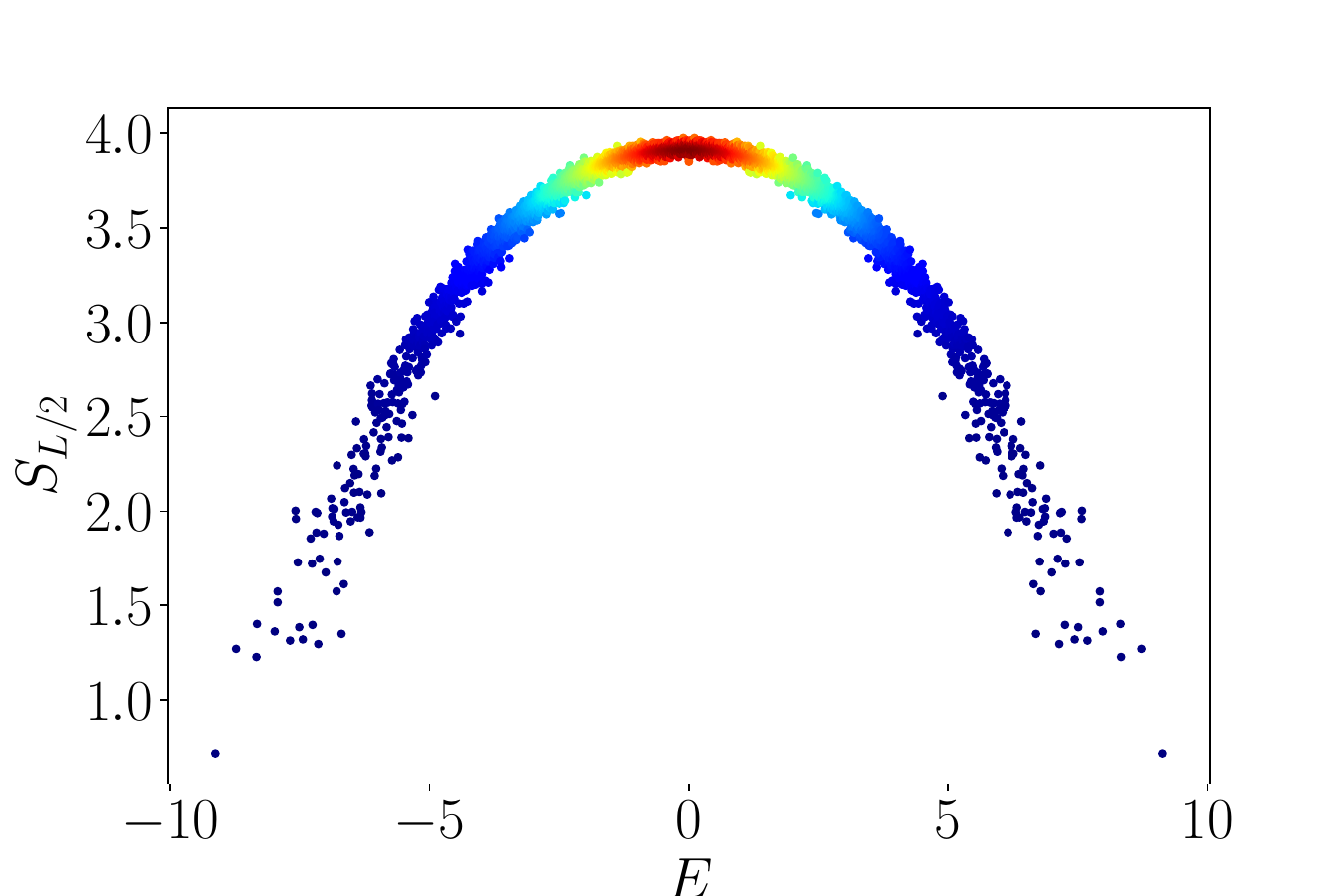}}\\
\subfigure[]{\includegraphics[width=0.5\linewidth,height=4.8cm]{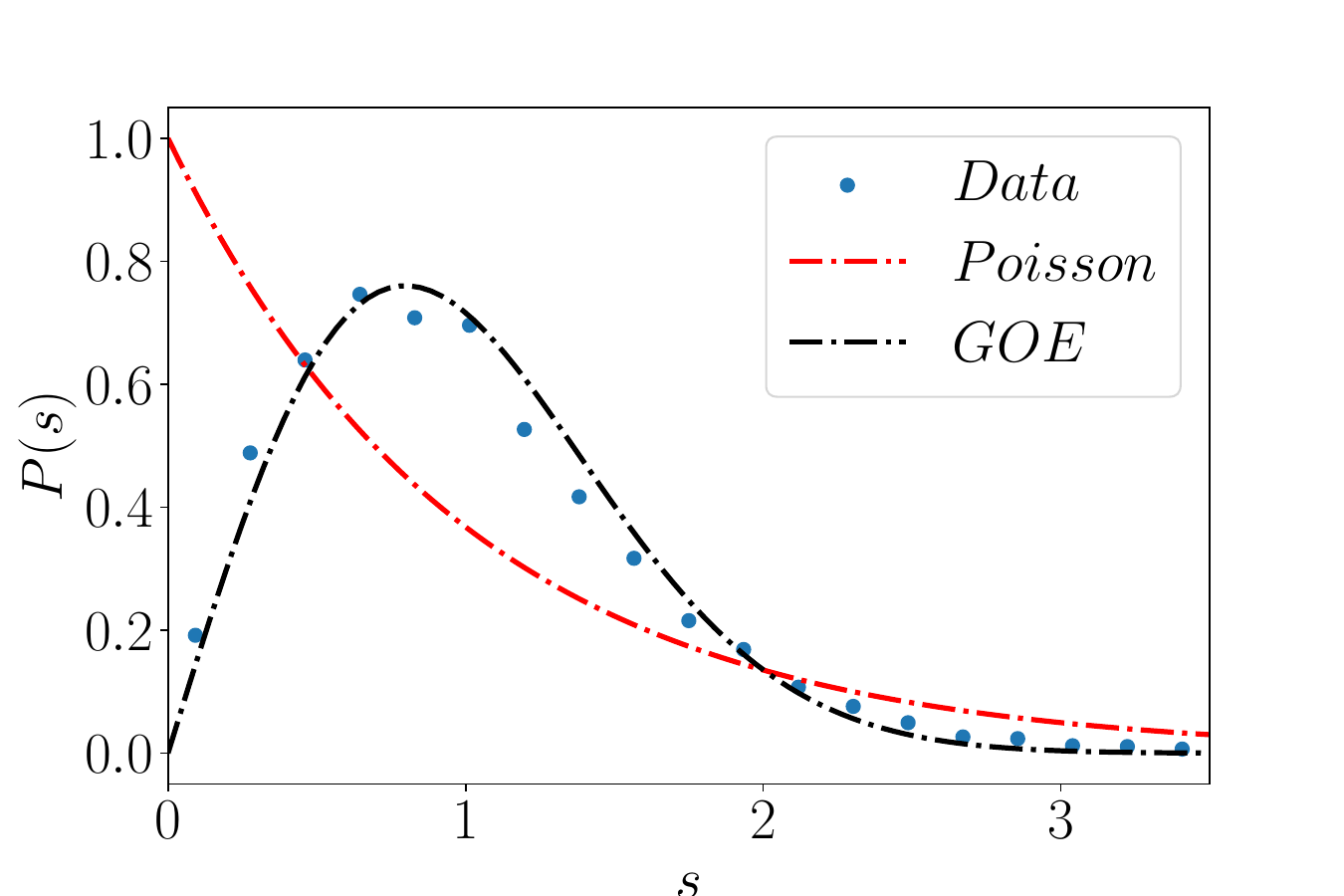}}%
\subfigure[]{\includegraphics[width=0.5\linewidth,height=4.8cm]{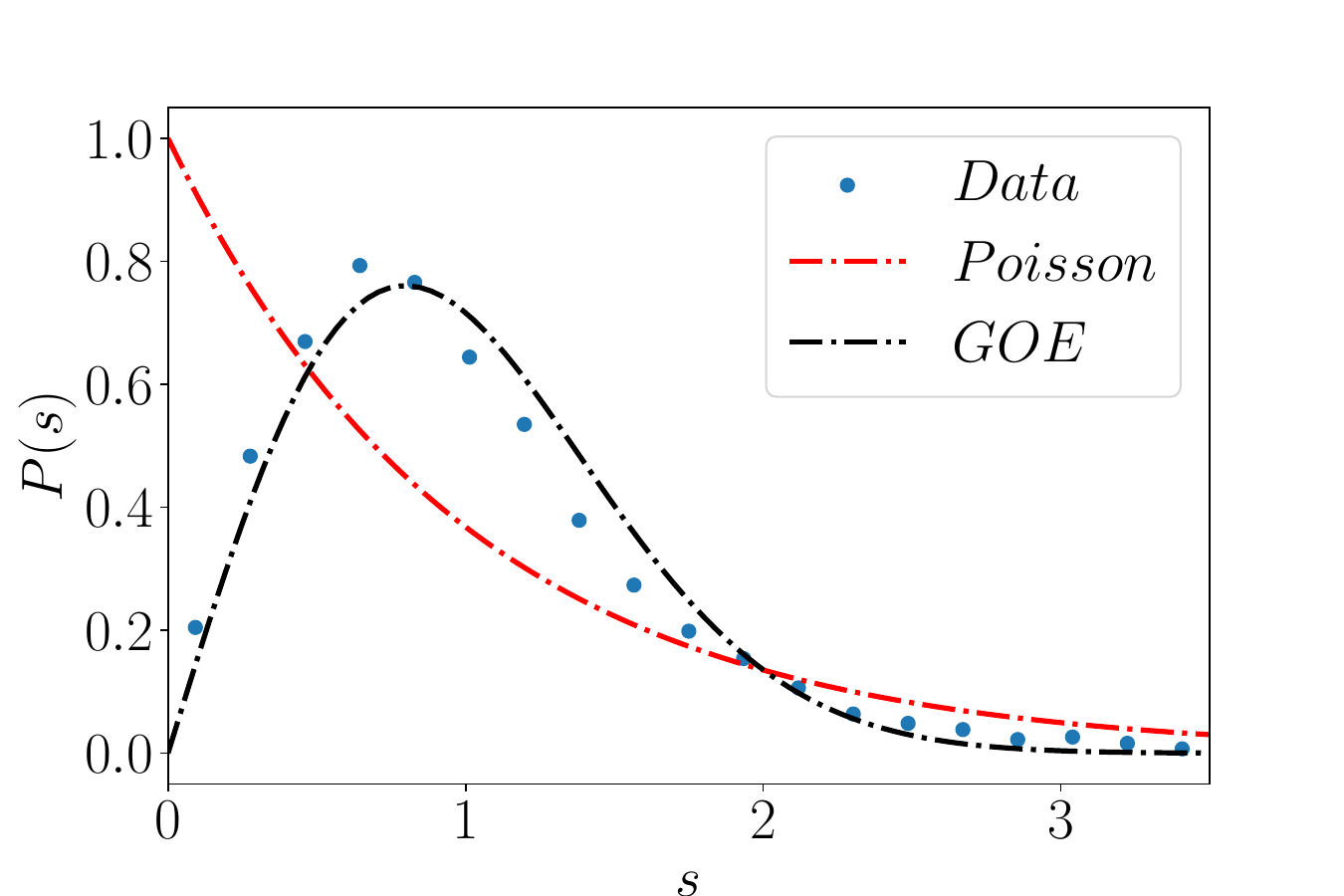}}
\caption{(a-b): Entanglement spectrum $S_{L/2}$ vs $E$ for range-2 and range-3 models within fragments labeling by the root states, $0100100100100111111$ and $001000100010111111$, respectively, for $L=18$. (c-d) The level spacing analysis within these given fragments in the presence of a small diagonal disorder with $w=0.01$ to avoid any accidental degeneracy and limit finite-size effects. However, this diagonal disorder preserves the fragmentation structure of this model. (a) In this case, the entanglement spectrum exhibits a sharp rainbow structure with some athermal eigenstates in the middle of the spectrum, which points toward Krylov-restricted thermalization in a weaker sense. (b) The same holds for the range-3 case, which again demonstrates a sharp rainbow pattern with no athermal states in the middle of the spectrum (at least for the central entanglement cut). (c-d) The level spacing analysis within the fragments follows the GOE statistics, thus revealing the non-integrable nature of these models with these fragments. }
\label{therma}
\end{figure*}

Here, we show the largest fragments in longer-range models are non-integrable~\cite{atas_2013,huse_2013,Wigner_1955,berry_1977} and entanglement spectrum displaying Krylov-restricted thermalization~\cite{Moudgalya_review_2022,brian_thermal,East_Sreemayee}. To elucidate this, we consider range-2 and range-3 models with $L=18$ in OBCs and further consider two fragments represented by root states, $0100100100100111111$ and $001000100010111111$, respectively, which describe the largest fragment at $N_{f}=L/2+1$ and $N_f=L/2$ for range-2 and range-3 models, respectively. In Figs. \ref{therma} (a-b), we demonstrate the entanglement entropy spectrum as a function of $E$ for these two cases, which exhibit a sharp rainbow-like spectrum. This typically symbolizes thermal behavior within the fragment, although there are some athermal states in the middle of the spectrum for the first case. In Figs. \ref{therma} (c-d), we perform a level spacing analysis of the sorted energy spectrum, where $s=(E_{n+1}-E_{n})/\delta$, $E_{n}$ and $\delta$ stand for $n$-th energy eigenvalue and mean level spacing, respectively. In both cases, we note the level spacing distribution follows the GOE statistics instead of the Poisson statistics~\cite{atas_2013,huse_2013,Wigner_1955,berry_1977}, indicating the non-integrability of these models within their respective largest fragments. 

\section{Bulk and boundary autocorrelators for range-1 and range-3 cases}
\label{mazurmore}
\begin{figure*}
\subfigure[]{\includegraphics[width=0.33\hsize,height=4.7cm]{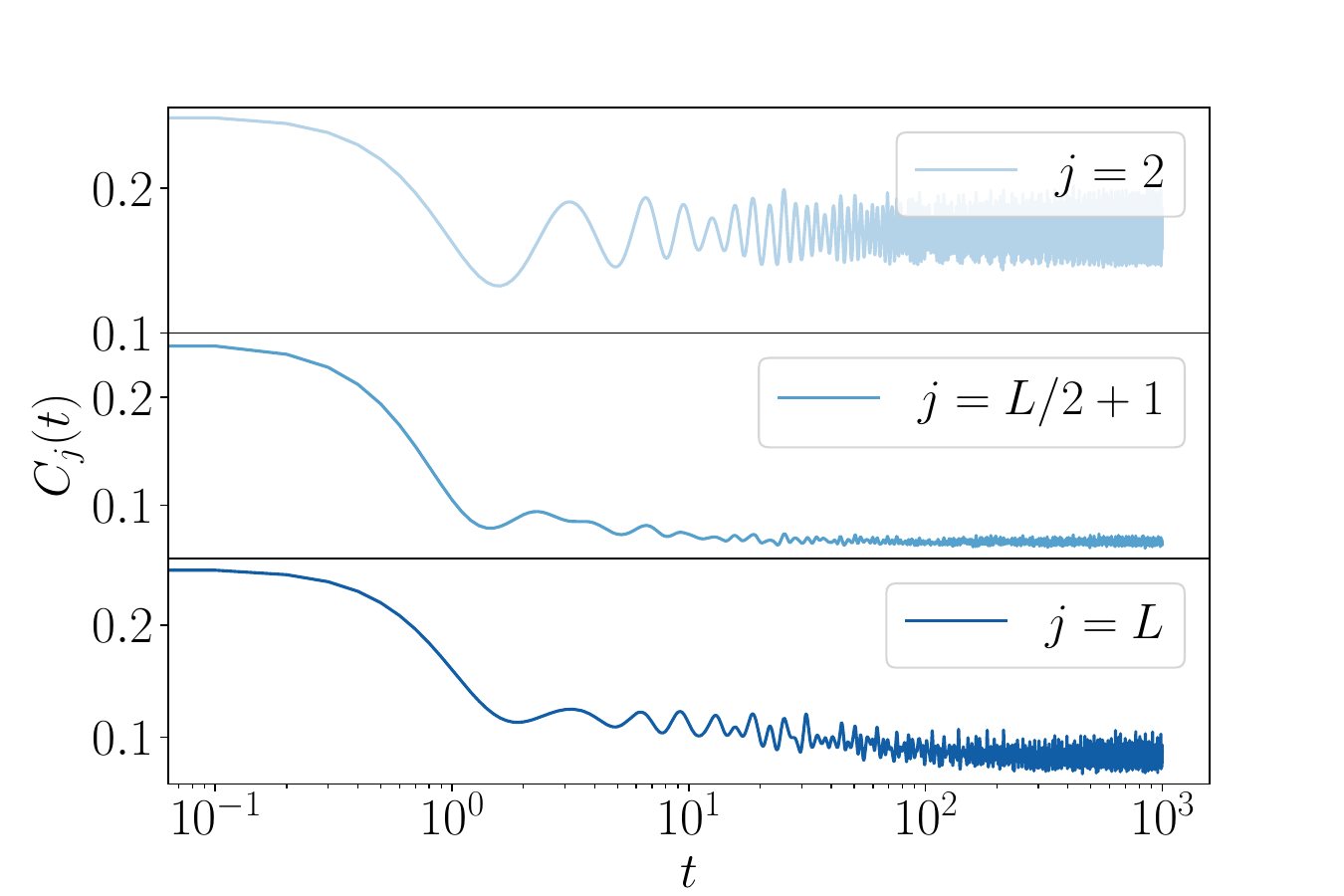}}%
\vspace{-0.1cm}
\subfigure[]{\includegraphics[width=0.33\hsize,height=4.7cm]{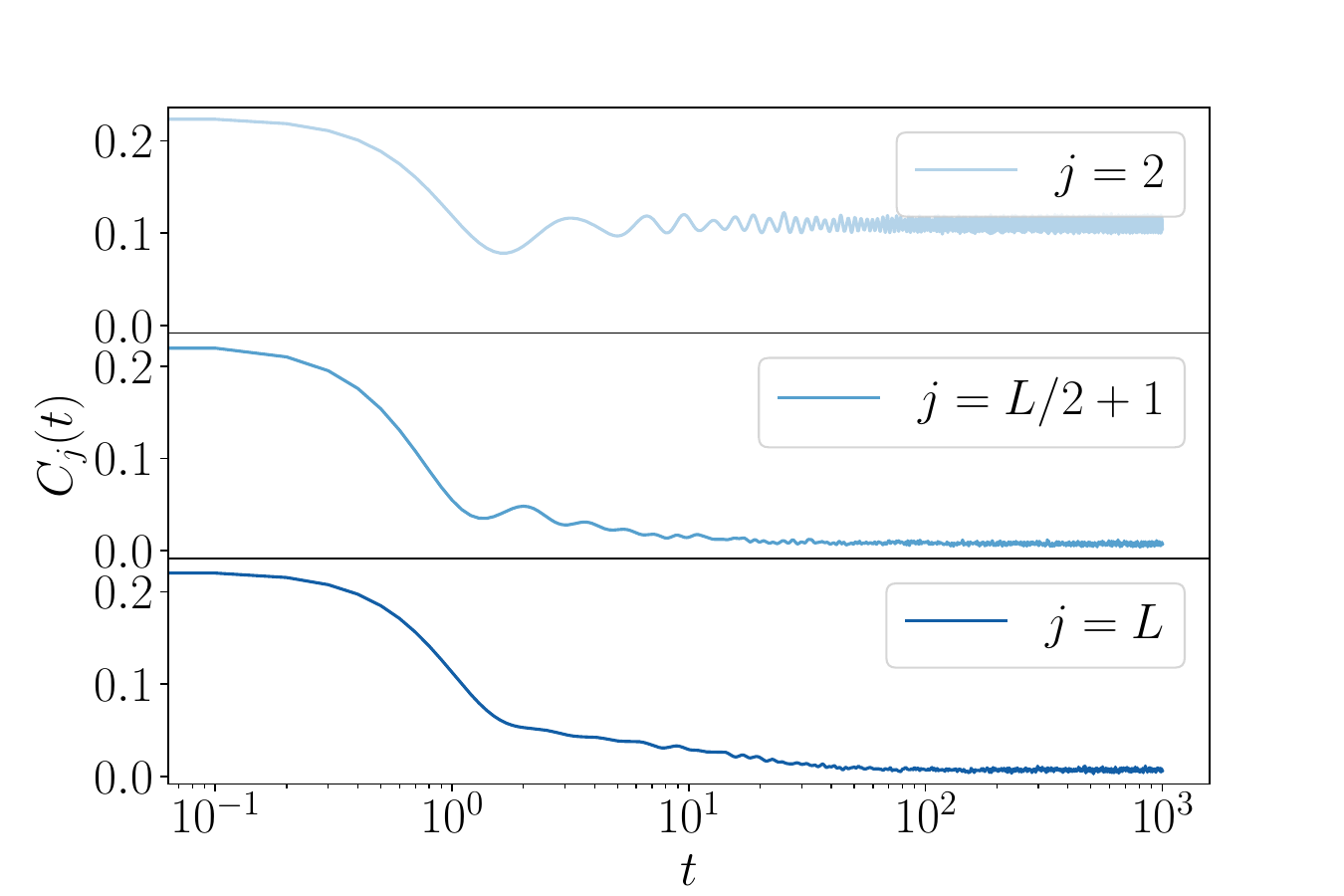}}%
\vspace{-0.1cm}
\subfigure[]{\includegraphics[width=0.33\hsize,height=4.7cm]{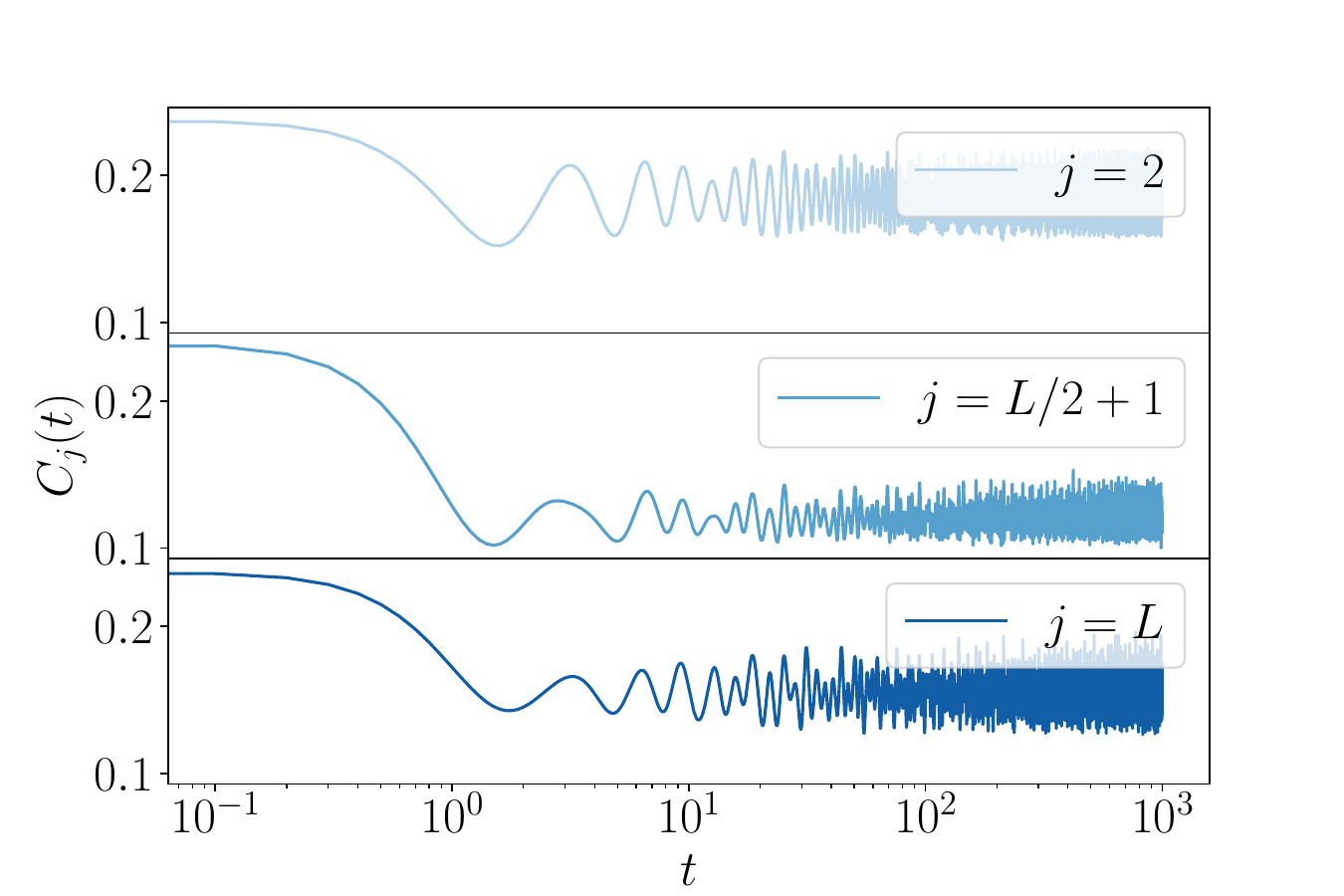}}\\
\caption{(a-c) Plots showing $C_{j}(t)$ vs $t$ at the leftmost active site, in the bulk of the chain, at the rightmost boundary, for $L=18$ and $N_{f}=L/2$ (freezing transition), $N_{f}=L/2+3$ (weakly fragmented) and $N_{f}=L/2-2$ (strongly fragmented), respectively. In all three cases, we observe an identical behavior as seen in the range-2 constraints, as depicted in Fig. \ref{autor2} (a-c). }
\label{freezingr1}
\end{figure*}
\begin{figure*}
\subfigure[]{\includegraphics[width=0.5\hsize]{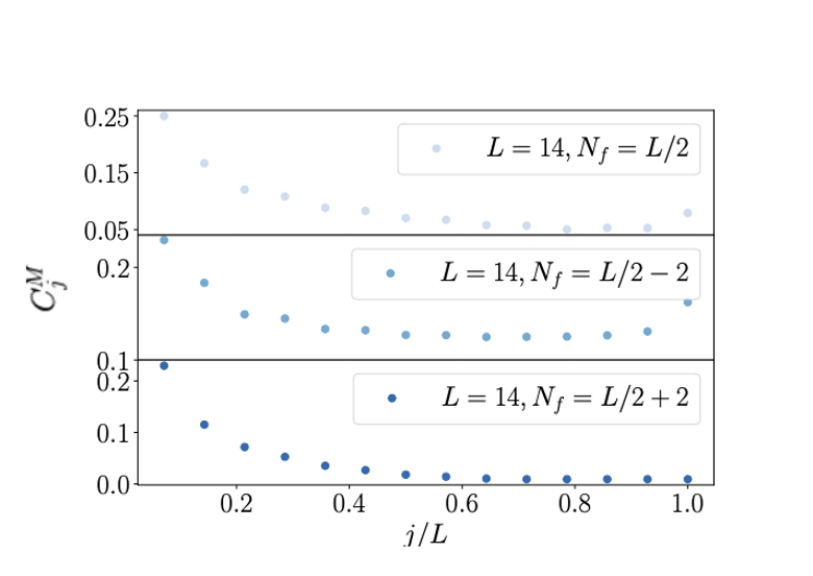}}%
\subfigure[]{\includegraphics[width=0.5\hsize,height=6cm]{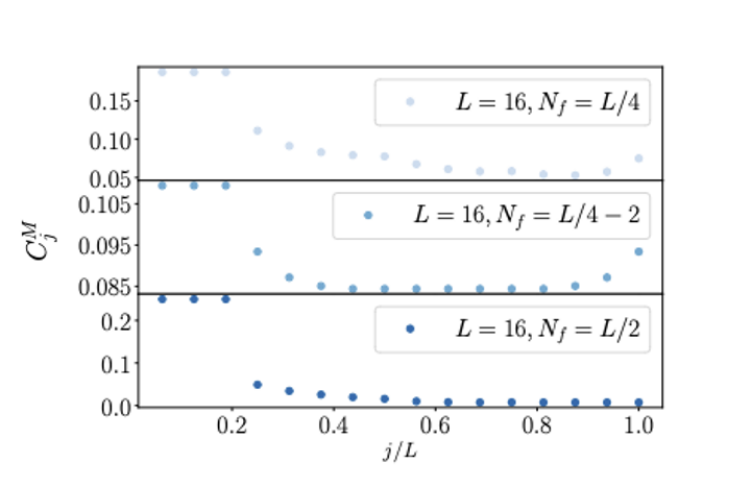}}
\caption{(a-b): The Mazur-bound predicted saturation value, $C_{j}^{M}$ across the chain in the range-1 and range-3 cases with $L=14$ and $L=16$, respectively. In both cases, we examine the profile at the critical filling fraction ($N_{f}=L/2$ and $L/4$ for range-1 and range-3 cases, respectively), below the critical filling (strongly fragmented) and above the critical filling (weakly fragmented). Again, we note identical behaviors to that observed in the range-2 case, as shown in Fig. \ref{autor2} (d).}
\label{mazur23}
\end{figure*}
Here, we also present the characteristics of the autocorrelation functions in time for the range-1 constraint, as illustrated in Figs. \ref{freezingr1} (a-c), for $L=18$ and at ($N_{f}=L/2$), below  ($N_{f}=L/2+3$) and above critical filling ($N_{f}=L/2-2$), respectively. As elucidated in Figs. \ref{autor2} (a-c), the autocorrelation functions, in this case, again demonstrate identical behaviors to the range-2 model depending on whether the filling fraction is at, below, or above the critical filling. While the leftmost boundary for all cases saturates at a finite value, which reveals signatures of non-thermal behavior. The rightmost boundary, on the other hand, behaves quite differently depending on the filling fraction under consideration. It shows non-thermal behaviors below and at critical filling and close to thermal behavior above the critical filling. Moreover, the bulk autocorrelation also hosts thermalization properties similar to the rightmost boundary, as can be seen from Figs. \ref{freezingr1} (a-c). To get a further generalized understanding, we also demonstrate the lower bound of the saturation value of long-time autocorrelation functions, $C_{j}^{M}$ across the chain for range-1 and range-3 cases at various filling fractions, i.e.,  at, below or above critical filling, respectively, employing Eq. \ref{Mazur}~\cite{MAZUR1969533,SUZUKI1971277}, as can be seen in Figs. \ref{mazur23} (a-b). In both cases, the profile of $C_{j}^{M}$ demonstrates exactly analogous behavior to the range-2 case, as shown in Fig. \ref{autor2} (d).

\section{More details on dynamics}
\begin{figure}
%\stackon{\includegraphics[width=\columnwidth]{spreading_modelr1_freezing.pdf}}{(a)}\\
%\stackon{\includegraphics[width=\columnwidth]{spreading_modelr2_freezing.pdf}}{(b)}\\
%\stackon{\includegraphics[width=\columnwidth]{spreading_modelr3_freezing.pdf}}{(c)}\\
\includegraphics[width=\columnwidth]{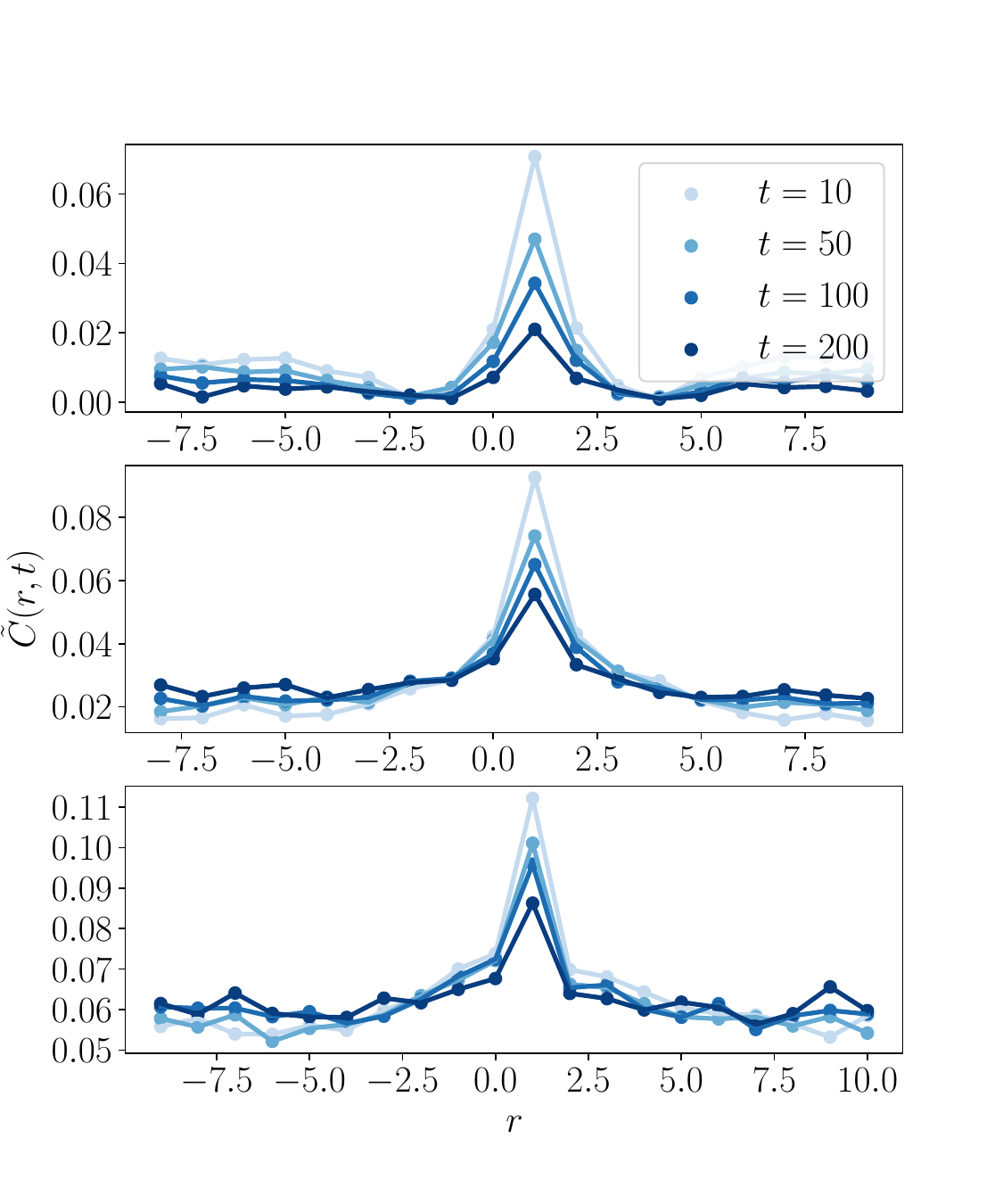}
\caption{Plots showing the spreading of $\tilde{C}(r,t)$ with $r$ at $N_{f}=L/2, L/3$ and $L/4$ in range-1 (top panel) and range-2 cases (middle panel) with $L=18$ and for the range-3 case (bottom panel) with $L=20$, respectively. In all three cases, we see that the spreading of $\tilde{C}(r,t)$ does not exhibit the Gaussian spreading (diffusive behavior), thus implying anomalous dynamics in this family of models. }
\label{spreading}
\end{figure}
To get further insights into the anomalous transport in this family of models, we also plot the behaviors of $\tilde{C}(r,t)$ vs $r$ at critical filling fractions, i.e., $N_{f}=L/2,L/3$ and $L/4$ for the case with range-1 (top panel), range-2 (middle panel) and range-3 constraints (bottom panel) at multiple intermediate times, $t=10,50,100$ and $200$, respectively, as demonstrated in Fig. \ref{spreading}. In all three cases, the spreading of the correlation function fails to exhibit a Gaussian shape. This is not typically a signature of diffusive transport~\cite{transport_review}, most commonly observed in thermalizing generic many-body systems, thus revealing anomalous dynamics in this family of models. In addition, we see that the correlators display a slow spreading of the spatial support at various intermediate times for all three cases, where it appears to be comparatively faster in the range-1 case (top panel) and the slowest in the range-3 case (bottom panel), as can be seen from Fig. \ref{spreading}.

\begin{comment}

\section{Freezing transition with PBCs}
\begin{figure}[htb]
\subfigure[]{\includegraphics[width=\columnwidth]{plots/freezing_transition_r1_PBC.png}}\\
\subfigure[]{\includegraphics[width=\columnwidth]{plots/freezing_transition_r2_model_PBC.png}}\\
\subfigure[]{\includegraphics[width=\columnwidth]{plots/freezing_transition_r3_model_PBC.png}}
\label{freezPBC}
\end{figure}
\end{comment}

\bibliography{ref}

\end{document}